\documentclass{article}

\usepackage{arxiv}

\usepackage{subfigure}
\usepackage[utf8]{inputenc} 
\usepackage[T1]{fontenc}    
\usepackage{hyperref}       
\usepackage{url}            
\usepackage{booktabs}       
\usepackage{subcaption}
\usepackage{amsfonts}       
\usepackage{nicefrac}       
\usepackage{microtype}      
\usepackage{gensymb}
\usepackage{amsmath}
\usepackage{cleveref}       
\usepackage{lipsum}         
\usepackage{graphicx}
\usepackage{natbib}
\usepackage{doi}
\usepackage{tikz} 
\usepackage{calc}
\usepackage[table]{xcolor}
\newlength{\figAwidth} 
\newlength{\figBwidth}
\usepackage{multirow}
\usepackage{array}
\usepackage{booktabs, multirow, threeparttable}
\usepackage{tablefootnote, adjustbox}

\usepackage{graphicx,subfigure}
\newsavebox{\refbox}

\usepackage{graphicx}
\usepackage{subcaption} 

\title{Learning Coupled Earth System Dynamics with GraphDOP}

\newbox{\orcid}\sbox{\orcid}{\includegraphics[scale=0.06]{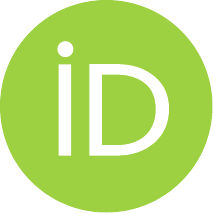}} 

\author{
    {\href{https://orcid.org/0000-0002-6070-2544}{\usebox{\orcid} Eulalie Boucher}}
    \And {\href{https://orcid.org/0009-0007-7798-6524}
    {\usebox{\orcid} Mihai Alexe}}
    \And {\href{https://orcid.org/0000-0002-3662-5382}{\usebox{\orcid} Peter Lean}}
    \AND {\href{https://orcid.org/0000-0003-1869-3426}{\usebox{\orcid} Ewan Pinnington}} 
    \And {\href{https://orcid.org/0000-0003-3952-586X}{\usebox{\orcid} Simon Lang}}
    \And {\href{https://orcid.org/0000-0003-2808-0463}{\usebox{\orcid} Patrick Laloyaux}}
    \And {\href{https://orcid.org/0000-0003-1703-4162}{\usebox{\orcid} Lorenzo Zampieri}} 
    \AND {\href{https://orcid.org/0000-0002-7374-3820}{\usebox{\orcid} Patricia de Rosnay}}
    \And {\href{https://orcid.org/0000-0001-5302-6093}{\usebox{\orcid} Niels Bormann}}
    \And {\href{https://orcid.org/0000-0003-2808-0463}{\usebox{\orcid} Anthony McNally}} \\
    \And
    European Centre for Medium-Range Weather Forecasts (ECMWF)
}

\begin{document}
\maketitle
\begin{abstract}
Interactions between different components of the Earth System (e.g. ocean,  atmosphere, land and cryosphere) are a crucial driver of global weather patterns. Modern Numerical Weather Prediction (NWP) systems typically run separate models of the different components, explicitly coupled across their interfaces to additionally model exchanges between the different components. Accurately representing these coupled interactions remains a major scientific and technical challenge of weather forecasting. GraphDOP is a graph-based machine learning model that learns to forecast weather directly from raw satellite and in-situ observations, without reliance on reanalysis products or traditional physics-based NWP models. GraphDOP simultaneously embeds information from diverse observation sources spanning the full Earth system into a shared latent space. This enables predictions that implicitly capture cross-domain interactions in a single model without the need for any explicit coupling. Here we present a selection of case studies which illustrate the capability of GraphDOP to forecast events where coupled processes play a particularly key role. These include rapid sea-ice freezing in the Arctic, mixing-induced ocean surface cooling during Hurricane Ian and the severe European heat wave of 2022. The results suggest that learning directly from Earth System observations can successfully characterise and propagate cross-component interactions, offering a promising path towards physically consistent end-to-end data-driven Earth System prediction with a single model.
\end{abstract}


\section{Introduction}

In recent years, data-driven approaches have reshaped numerical weather prediction (NWP) \citep{pathak2022fourcastnet,lam2022graphcast,bi2023accurate,bodnar2024aurora,lang2024aifs}, yet a major challenge remains: achieving full coupling across Earth System components in a way that reflects their real-world interactions. Coupling in this context refers to how different subsystems of the Earth, such as the atmosphere, ocean, land, and sea ice exchange energy, mass, and momentum. These interactions are crucial because they strongly influence weather and climate. For example, sea-surface temperatures affect atmospheric circulation patterns, while soil moisture regulates surface fluxes that impact precipitation and near surface temperature. Without adequate representation of such linkages, forecasts risk missing key feedbacks that drive large-scale variability.

In traditional NWP systems, it is important to distinguish between coupled modelling and coupled data assimilation (DA): the former can be implemented without the latter, but not vice versa. Coupling is typically handled through distinct physical models for each component \citep{coupled_da_1,coupled_da_2,coupled_da_3}. These models interact by exchanging boundary conditions (e.g., ocean surface temperature provided to the atmosphere model, or atmospheric winds driving ocean currents). Each component is typically initialized using its own DA scheme, which blends observations with prior model forecasts to generate the best estimate of the current state. Several coupled DA strategies exist \citep{penny_coupled} to improve consistency across these systems. 
Weakly coupled DA assimilates each component separately, but their forecasts are cycled together so that information gradually propagates between them.
Strongly coupled DA, by contrast, adjusts multiple components simultaneously in a single assimilation step, allowing observational information from one domain (e.g., sea ice extent) to directly influence another (e.g., atmospheric circulation). These approaches have led to significant advances, but they are computationally demanding and rely on hand-crafted parametrizations, error covariance tuning, and reconciliation of disparate model components. 

The Integrated Forecasting System (IFS) at the European Centre for Medium-Range Weather Forecasts (ECMWF) has gradually evolved into a comprehensive Earth System Model (ESM) and DA system, incorporating components such as land, ocean, and sea ice. This evolution reflects a growing emphasis on the consistent treatment of interactions across domains, particularly in the context of coupled data assimilation. By leveraging interface observations — measurements that provide information about multiple subsystems simultaneously — coupled 4D-Var \citep{rs11030234,Rosnay22} aims to improve the coherence of initial conditions and reduce the risk of cross-component error propagation. 
While such interface observations are difficult to exploit in physics-based systems due to complex forward modelling and uncertain parameters (e.g., \cite{geer2024}), ECMWF's coupled DA developments, particularly in the pre-operational cycle 50r1, represent major progress in exploiting this valuable source of cross-domain information.

In parallel with advances in coupled physical models, machine learning (ML)–based forecasting systems (e.g. \cite{keisler2022forecasting}, \cite{lam2022graphcast},\cite{lang2024aifs}) have recently demonstrated skill comparable to traditional NWP at a fraction of the computational cost. However, most existing models remain limited to atmospheric prediction, which has motivated growing interest in extending ML methods to coupled Earth System forecasting. For instance, \citet{wang2024coupledoceanatmospheredynamicsmachine} present a data-driven model that learns atmosphere–ocean dynamics from coupled model data, demonstrating the feasibility of representing cross-sphere interactions within an ML framework. At ECMWF, the Artificial Intelligence Forecasting System (AIFS; \cite{lang2024aifs,lang2024aifscrpsensembleforecastingusing}) is trained on ERA5 reanalysis and initialized from operational analysis. It already incorporates land variables operationally \citep{aaifssingle2025}, and research is underway to develop a joint atmosphere–wave prototype by combining ERA5 atmospheric fields with wave variables from the operational wave model (ecWAM; \cite{ecmwf2025aifswaves}). Such joint modelling allows the ML system to internalize atmosphere–wave interactions, and has shown the potential to capture coupling even when not explicitly encoded in training data; for example by attenuating wave heights beneath sea ice despite the absence of an explicit sea ice representation \cite{ecmwf2025aifswaves}. Nevertheless, the realism of these learned couplings still depends on the fidelity of the training datasets rather than on new interactions empirically constrained by observations.

A more radical approach is being pursued at ECMWF through Artificial Intelligence Direct Observation Prediction (AI-DOP) \citep{mcnally2024, alexe2024graphdopskilfuldatadrivenmediumrange}. Unlike reanalysis-trained systems, AI-DOP seeks to learn forecast dynamics directly and only from raw observations. In this framework, satellite and \textit{in-situ} measurements from different Earth System components are projected into a shared latent space via a joint graph representation. This allows the model to naturally learn a coupled latent representation, since many observation types are sensitive to multiple components simultaneously. If successful, this approach could open the path toward a new generation of fully coupled forecasting systems where Earth System interactions are learned directly from observation data rather than imposed through separate model components.



In this paper, we extend the analysis of GraphDOP, a graph based AI-DOP model developed at ECMWF, first presented in \cite{alexe2024graphdopskilfuldatadrivenmediumrange}, by examining its forecasts across different Earth System components. Three case studies spanning the atmosphere, ocean, cryosphere and land surfaces show that GraphDOP learns a coherent internal representation of the coupled Earth System state that allows it to produce skilful forecasts of weather parameters. These findings indicate that GraphDOP captures physically meaningful relationships within the coupled Earth System, rather than merely interpolating observed correlations. 


\section{The GraphDOP Model}
\label{sec:model}
\begin{figure}[htbp]
    \centering
    \includegraphics[width=0.99\textwidth]{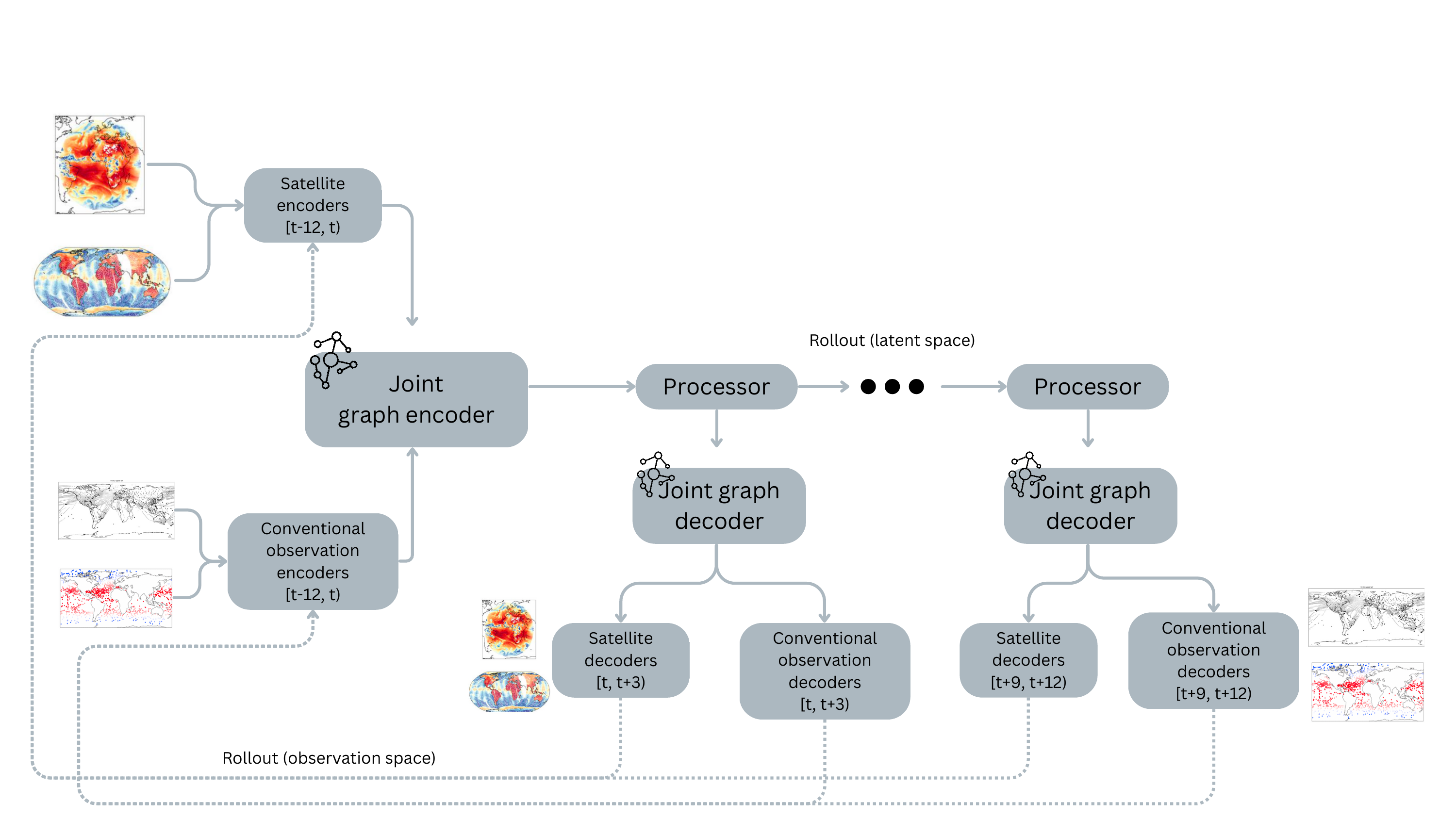}
    \caption{GraphDOP \citep{alexe2024graphdopskilfuldatadrivenmediumrange} with 3-hour autoregressive time-stepping (rollout) in the latent space. During fine-tuning, the forecasted observations are fed back as inputs to produce the next forecast (rollout in observation space).} 
    \label{fig:dop_model_schematic}
\end{figure}

The GraphDOP model used in this study largely follows the encoder-processor-decoder architecture described in \citep{alexe2024graphdopskilfuldatadrivenmediumrange} and Figure \ref{fig:dop_model_schematic}.  Its design shares many aspects with AIFS \citep{lang2024aifs}, though here the architecture is adapted specifically for observation prediction. A graph-based encoder projects the 12 hours of input observations onto a latent space. Subsequently, the processor (a sliding-window transformer) evolves the latent space representation throughout a 12-hour target observation window. Finally, a decoder projects the latent representation back into observation space to produce the model forecasts. The latent representation is defined on an o96 reduced Gaussian grid \citep{malardel2016} (a spatial resolution of ca. 1 degree), with 1024 features. The encoder graph mapping is constructed on the fly from the input observations available in each training and validation batch to their nearest grid points in the latent space representation, a $k$ nearest-neighbour graph, with $k=1$. Conversely, the decoder graphs connect each target observation to its three nearest neighbour nodes on the latent mesh.

Departing from the original training recipe, GraphDOP is asked to forecast 12-hour observation windows by auto-regressively processing and then decoding the target observations across four consecutive three-hour chunks, a procedure referred to in Figure \ref{fig:dop_model_schematic} and \citep{alexe2024graphdopskilfuldatadrivenmediumrange} as latent-space rollout. This has led to improvements in the sharpness and accuracy of the GraphDOP forecasts compared to the simpler, "single-shot" decoding of all the observations inside the 12-hour target window. The training objective is the weighted mean square error, same as that used by \cite{alexe2024graphdopskilfuldatadrivenmediumrange}.

GraphDOP is trained entirely using Earth System observations from satellite platforms and sparse \textit{in-situ} sources, without direct dependency on NWP (re)analysis fields or forecast model outputs. Each observation type is associated with its own encoder–decoder pair (the satellite and conventional observation encoders shown in Figure \ref{fig:dop_model_schematic}). This enables the model to learn modality-specific representations, akin to observation operators, and a common latent space for cross-variable interaction. At inference time, these trained decoders can be used to produce gridded forecasts for any observed quantity at any chosen time inside the target window, and on grids of arbitrary spatial resolution. In practice, however, we find that the effective resolution of the forecasts depends on several factors, such as the spatial density and resolution of the available observations, and the dimensionality of the latent representation.

\section{Observation Datasets}
\label{sec:data}

A key advantage of a purely observation-driven system like GraphDOP is its ability to ingest a wide range of observational data, including those that may not be currently utilized in traditional DA systems. DA systems, which integrate observations with a model-based background state, require predefined observation error covariances and observation operators to assimilate new data effectively. Another important aspect is that they also rely on the physical model to accurately represent the processes to which the observations are sensitive. 
In contrast, GraphDOP only requires an appropriate encoder and decoder to incorporate an observation type, making it inherently more flexible and adaptable to diverse observational datasets. This naturally extends to the inclusion of interface observations, which bridge multiple Earth System components but entail careful design of multi-domain observation operators and error covariances to represent the interactions between coupled model components in traditional systems \citep{geer22,Rosnay22}.

In this study, we focus on introducing as many interface observations as possible, whilst keeping a focus on medium-range forecasting, and therefore using the same data-selection strategy as presented in \cite{alexe2024graphdopskilfuldatadrivenmediumrange}. This means prioritizing observations that capture the large-scale thermodynamic structure of the atmosphere (such as from polar orbiting satellite sounding instruments), as these characteristics fundamentally govern atmospheric dynamics at the medium range. Alongside this, we include diverse sources of \textit{in-situ} observations of physically meaningful variables that could then be predicted on a regular grid to provide meaningful forecasts. 

Table \ref{table:dop-observations} gives an overview of the datasets used for the model presented in this paper. Instruments in bold are new additions that were not used in \cite{alexe2024graphdopskilfuldatadrivenmediumrange}. Improved quality control was performed for several of the datasets used in the original presentation of the GraphDOP model. 

The model was trained on 20 years of data (2000–2020), validated on 2021, and evaluated in case studies from 2022, shown in this paper. Raw observations (Level 1) available from multiple channels are used from microwave and infrared sensors. Physical observed parameters are used from conventional observations. Although a majority of the observations used give valuable information on the atmosphere, many of them also play a role in other components of the Earth System. For instance, radar altimeters like SARAL RALT provide insights into significant wave height and surface wind speed over the ocean. Microwave imagers such as AMSR-2 and SMOS are crucial for monitoring sea ice concentration and snow cover, and are sensitive to thin sea ice thickness, ocean sea surface temperature and salinity, wind speed over ocean surface, and soil moisture over land surfaces, \textit{in-situ} snow reports help track changes in the cryosphere, while drifting buoys (e.g., BUFR Drifting Buoys) provide valuable data on sea surface temperature. Additionally, visible and infrared sensors, such as MetOp AVHRR and Meteosat SEVIRI, support land surface and vegetation monitoring, etc...

\begin{table}[!htbp]
\centering
\small
\begin{threeparttable}

\begin{tabular}{p{.2\textwidth}p{.2\textwidth}p{.22\textwidth}p{.35\textwidth}}
\toprule
\textbf{Category} & \textbf{Instrument} & \textbf{Platform} & \textbf{Parameters} \\
\midrule
\multirow{10}{*}{Microwave}
    & AMSR-2    & GCOM-W1 & Channels 1-14 \\
    
    & \textbf{AMSR-E}    & AQUA      & Channels 1-10 \\
    & AMSU-A    & NOAA-15/18, MetOp-A/B/C & Channels 1–15 \\
    
    & \textbf{AMSU-B }   & NOAA-16/17 & Channels 1-5 \\
    & ATMS      & NPP, NOAA-20 & Channels 1–22 \\
    
    & \textbf{MHS}       & NOAA-18/19, MetOp-B/C & Channels 1-5 \\
    
    & \textbf{MSU}       & NOAA-14  & Channels 1-4 \\
    
    & \textbf{SSMI}      & DMSP F13  & Channels 1-13 \\
    
    & \textbf{SMOS}      & SMOS & V and H polarizations \\
    
    & \textbf{GMI}       & GPM & Channels 1-13 \\
    
    & \textbf{WindSat} & Coriolis & Channels 3-16 \\
\addlinespace
\addlinespace

\multirow{5}{*}{Infrared}
    & SEVIRI    & Meteosat 8/8 IODC/9/10/11   & Channels 4-11 \\
    
    & \textbf{AHI}       & Himawari 8    & Channels 2,3,4,8 \\
    & IASI      & MetOp-B       & 17 channels \\
    & AVHRR      & MetOp-B       & visible \\
    
    & \textbf{GOES ABI}      & GOES 16-18 &  Channels 2,3,4,8 \\
    
    & \textbf{GOES Imager}  & GOES 12-14  & Channels 1-4 \\
    
\addlinespace
\addlinespace

Radar Altimeter
    & SRAL  & Sentinel-3  & Significant wave height, wind speed\\

\addlinespace
\addlinespace

Atmospheric Motion Vectors & \textbf{A range of geostationary and polar satellites} & Full list of satellites is presented in \cite{hersbach2020era5} & u and v derived from cloud motions in long-wave infrared, water-vapour and visible channels \\

\addlinespace   
\addlinespace

\multirow{2}{*}{Lidar Winds}
    
    & \textbf{ALADIN (Mie) }     & Aeolus & Line-of-sight wind (Mie) \\
    
    & \textbf{ALADIN (Rayleigh)} & Aeolus & Line-of-sight wind (Rayleigh) \\

\addlinespace
\addlinespace

\multirow{4}{*}{Conventional Surface}
    & SYNOP (Land)      &  & ps, T2m, rh2m, u10m, v10m\\
    & Ship Observations &  & ps, T2m, rh2m, u10m, v10m \\
    & DRIBU / Buoys     &  & ps, sst \\
    
    & \textbf{SYNOP Visibility}  &  & Visibility  \\
\addlinespace
\addlinespace

\multirow{4}{*}{Conventional Upper-Air}
    & Radiosondes       & & t, u, v, z \\
    & PILOT Reports     &  & u, v \\
    & Dropsondes        &  & t,u,v,z \\
    & Aircraft          & & t,u,v \\
    
\bottomrule
\end{tabular}
\end{threeparttable}

\vspace{.3cm}
\caption{Observations used in GraphDOP, grouped by instrument category. New observations that are included here compared to \cite{alexe2024graphdopskilfuldatadrivenmediumrange} are shown in \textbf{bold}.}
\label{table:dop-observations}

\end{table}

\section{Sea ice rapid freezing case study: cryosphere and ocean}
\label{sec:seaice}

The sea ice is a critical component of the coupled climate system \citep{Perovich2016}. Its dynamic and thermodynamic processes, such as freezing, melting, transport, and deformation, directly affect atmospheric circulation \citep{Persson2016}, most strongly in polar regions but also in the mid-latitudes through episodic yet substantial teleconnection patterns \citep{Overland2018}. Beyond its regional influence, sea ice is a key regulator of the Earth’s energy balance and an important driver of ocean circulation. It is also among the most responsive components of the Earth System to both inter-annual climate variability and anthropogenic global warming. Consequently, accurate sea ice forecasts are essential across timescales, from nowcasting to climate monitoring, and are well motivated by the well-documented mechanisms that underpin sea ice predictability \citep{Jung2016}.

Due to Arctic amplification from global warming, sea ice has changed dramatically in recent decades, with declining extent and volume and shorter residence times, progressively shifting the Arctic Ocean toward a first-year ice regime \citep{Sumata2023}. As a result, recent years have seen an increase in the frequency and intensity of extreme sea ice events, such as prolonged melting seasons \citep{Stroeve2018}, but also sharper freezing episodes in the winter months \citep{Cornish2022}. These changes pose substantial challenges for marine operations and for coastal communities that depend on sea ice. At the same time, they expose the limitations of traditional modelling frameworks, which can develop substantial biases in response to a crude representation of the sea ice physics and its coupling with the marine and atmospheric components \citep{Zampieri2018}. In this study, we demonstrate that purely data-driven approaches such as GraphDOP could, in the future, be a viable alternative framework for sea ice predictions, as it can successfully represent several aspects of sea ice dynamics and thermodynamics solely based on observations. To do so, we will focus in particular on a case study: a rapid refreezing event that occurred in the fall of 2022.




Using GraphDOP for sea ice prediction is motivated by the strong, well-characterized sensitivity of satellite observables to sea ice properties and their evolution \citep{Spreen2025}. Passive microwave measurements (e.g., AMSR-2) respond to changes in concentration, phase state, surface temperature, snow cover, and microstructure—the very signals exploited by operational sea ice concentration retrievals from OSI-SAF (EUMETSAT; \citet{Lavergne2019}) and NSIDC (NASA; \citet{Meier2014})–and will be the focus on this study. 

To produce a credible forecast of sea ice extent, an observation-driven framework like GraphDOP must learn a coherent latent representation that links the evolving sea ice state to multiple, complementary sources of information. In practice, this means reconciling the relationships among conventional geophysical predictors (such as sea-surface temperature and near-surface winds) and the multispectral brightness temperature fields available over the region of interest from microwave and infrared sensors. Beyond what is learned during training, the quality and coverage of the observations used for initialization play a decisive role in how the forecast develops, because they anchor the state of the system at lead time zero and shape the subsequent trajectory. It is recognised that microwave brightness temperature may not be as accessible to users as a forecast of sea-ice concentration, but this could be addressed through appropriate post-processing or through including relevant satellite-derived products in the training.

\begin{figure}[htpb]
\centering

\renewcommand{\arraystretch}{1.2}
\begin{adjustbox}{width=\linewidth}
\begin{tabular}{c c c @{}c c @{}c c @{}c}
 & Target & Forecasted & &  Difference & & ORAS6 SIC & \\
\raisebox{1cm}{\rotatebox{90}{21/10/2022}} &
\includegraphics[height=3.5cm]{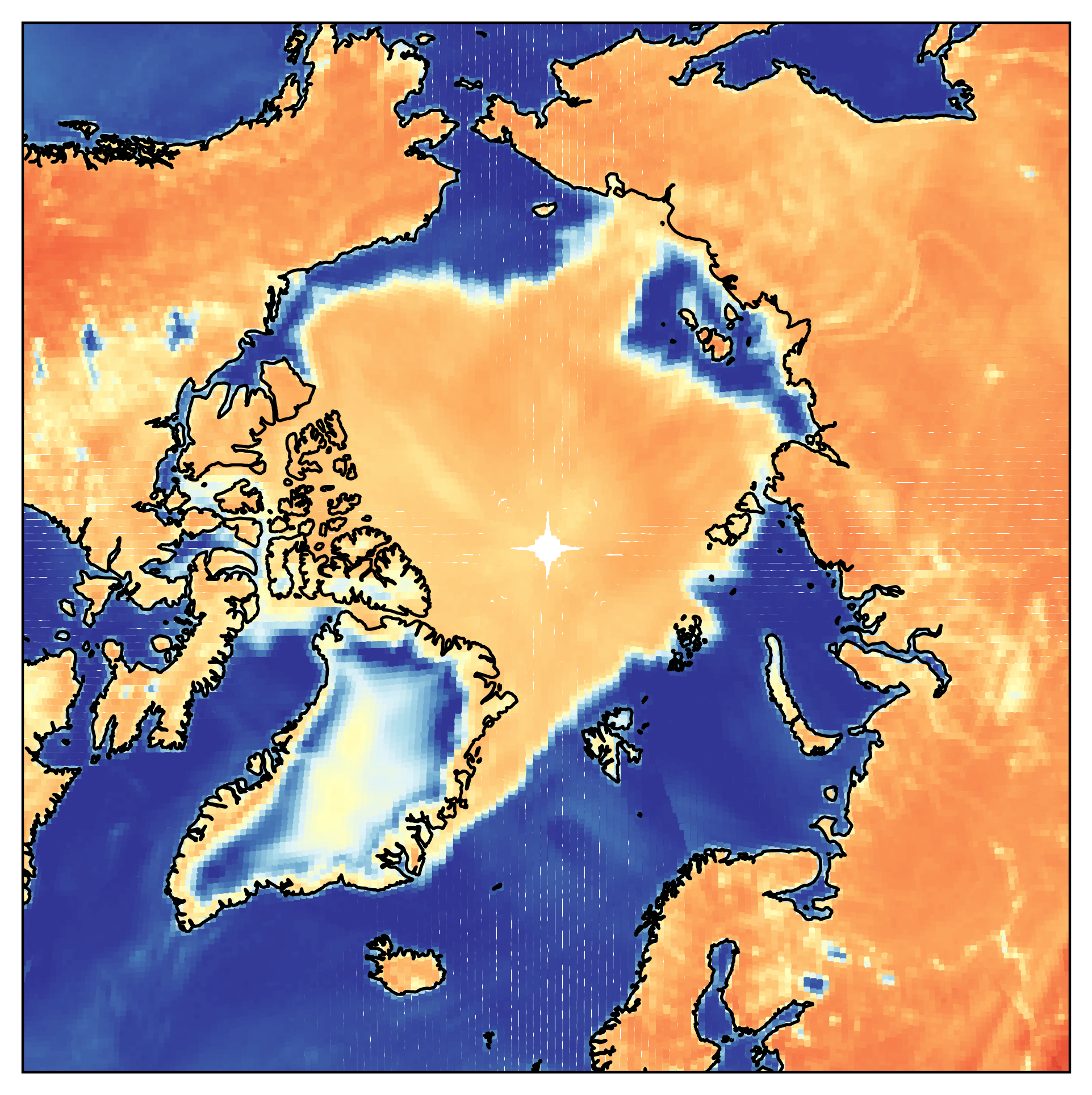} &
\includegraphics[height=3.5cm]{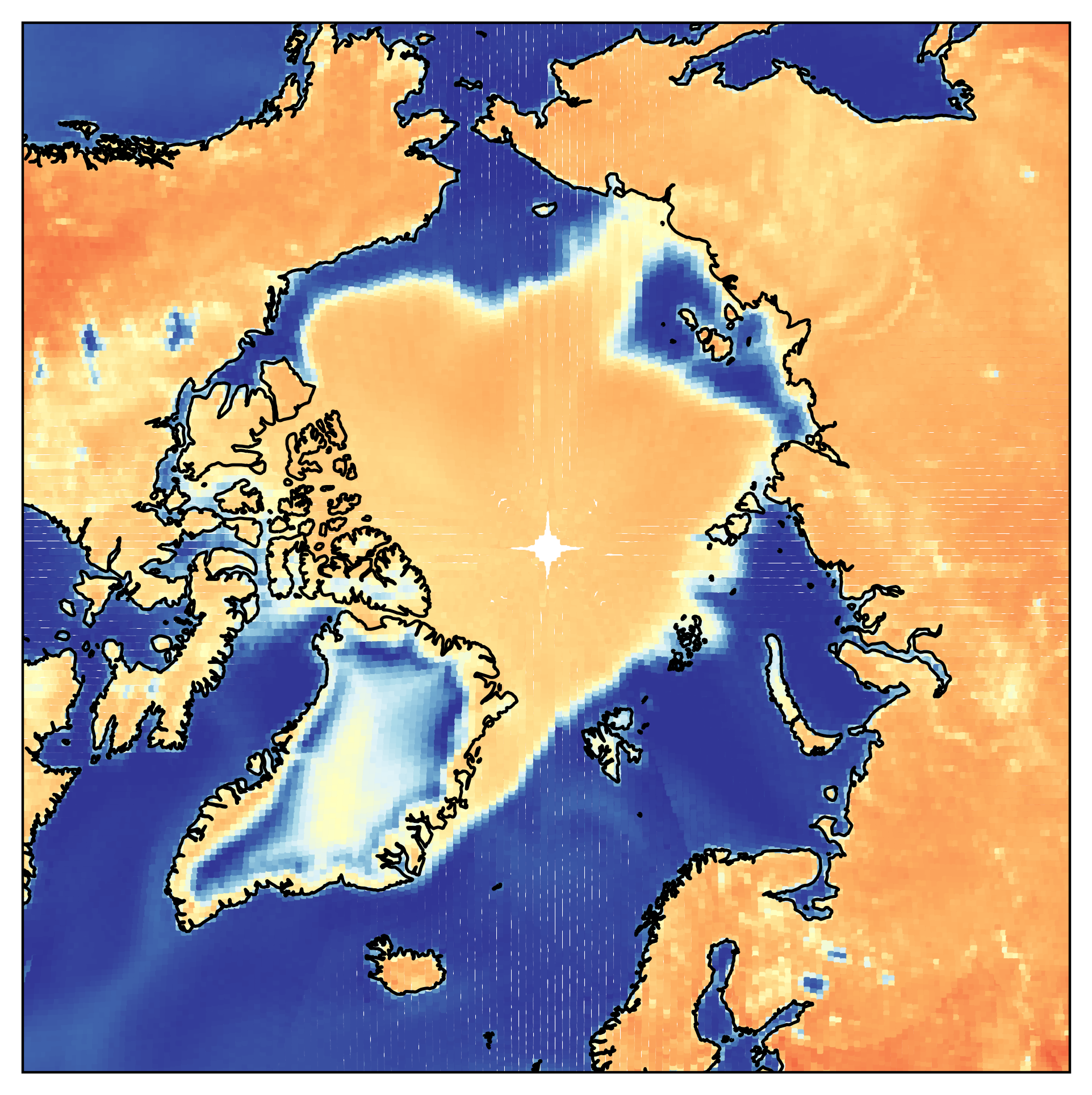} & \multirow{3}{*}{\includegraphics[height=5cm]{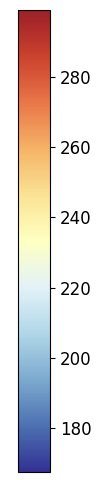}}{} &
\includegraphics[height=3.5cm]{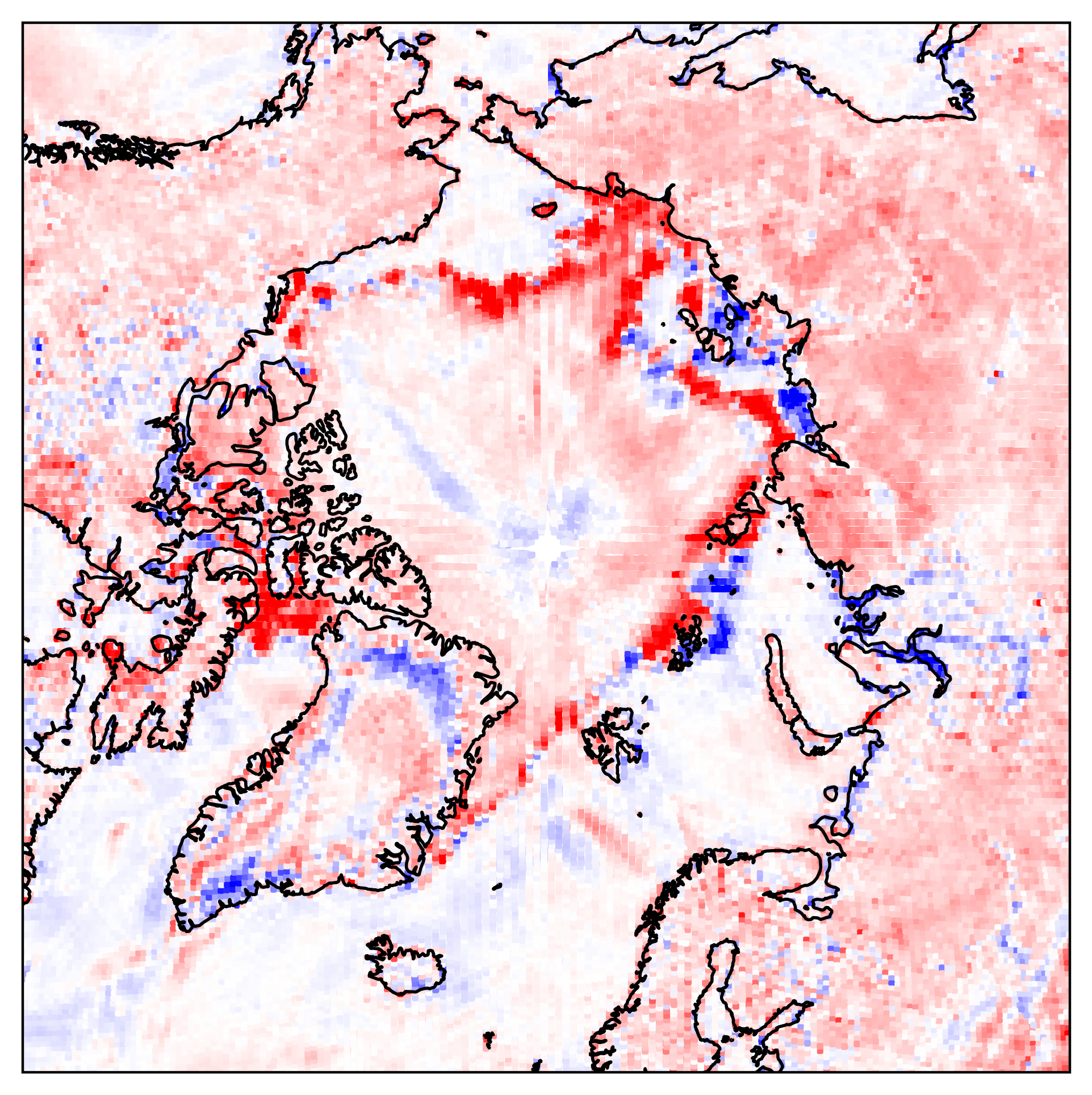} & \multirow{3}{*}{\includegraphics[height=5cm]{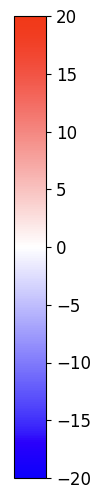}}{} & \includegraphics[height=3.5cm]{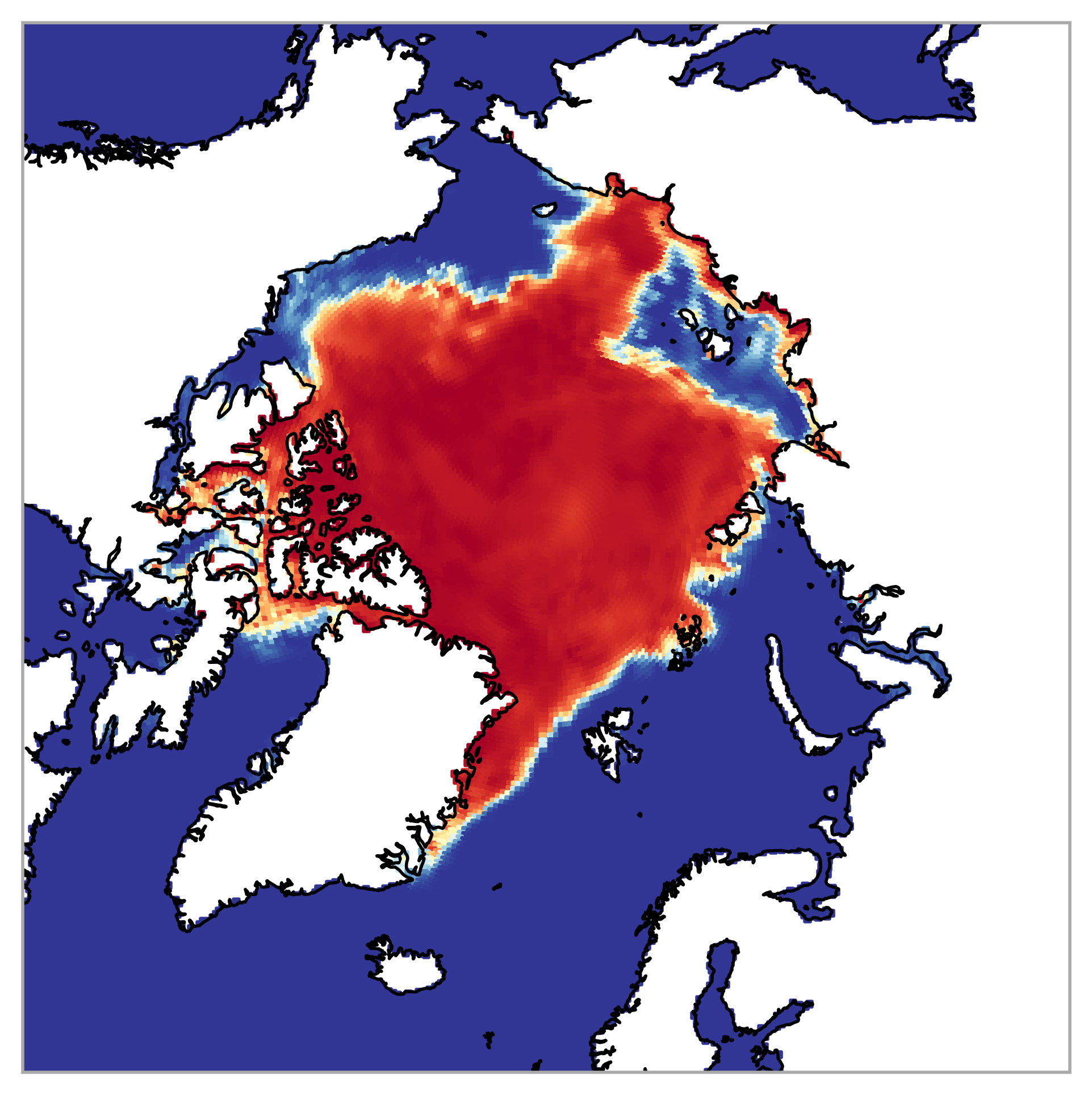} & \multirow{3}{*}{\includegraphics[height=5cm]{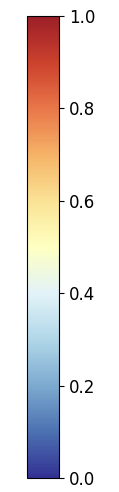}}{} \\
\raisebox{1cm}{\rotatebox{90}{25/10/2022}} &
\includegraphics[height=3.5cm]{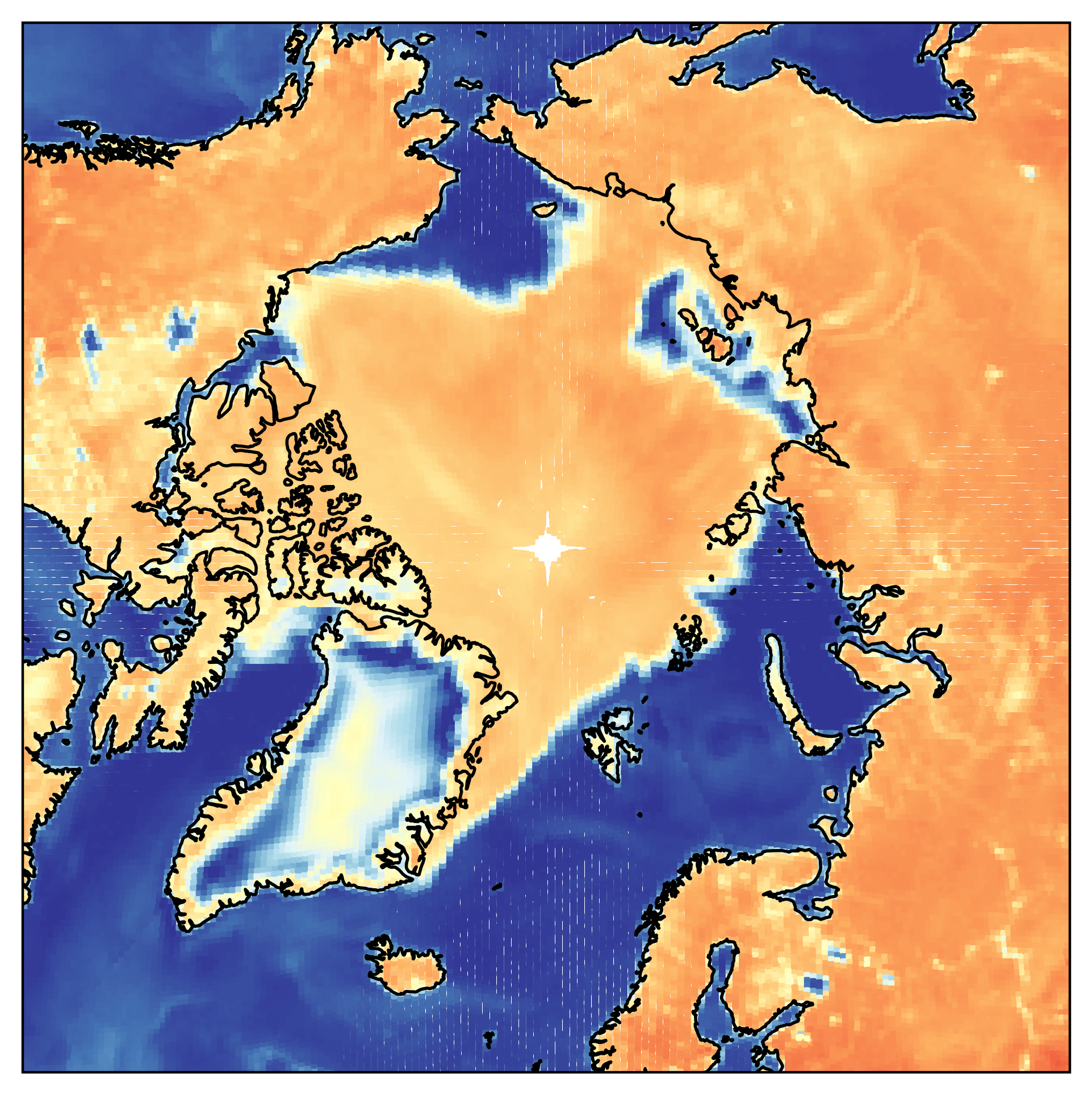} &
\includegraphics[height=3.5cm]{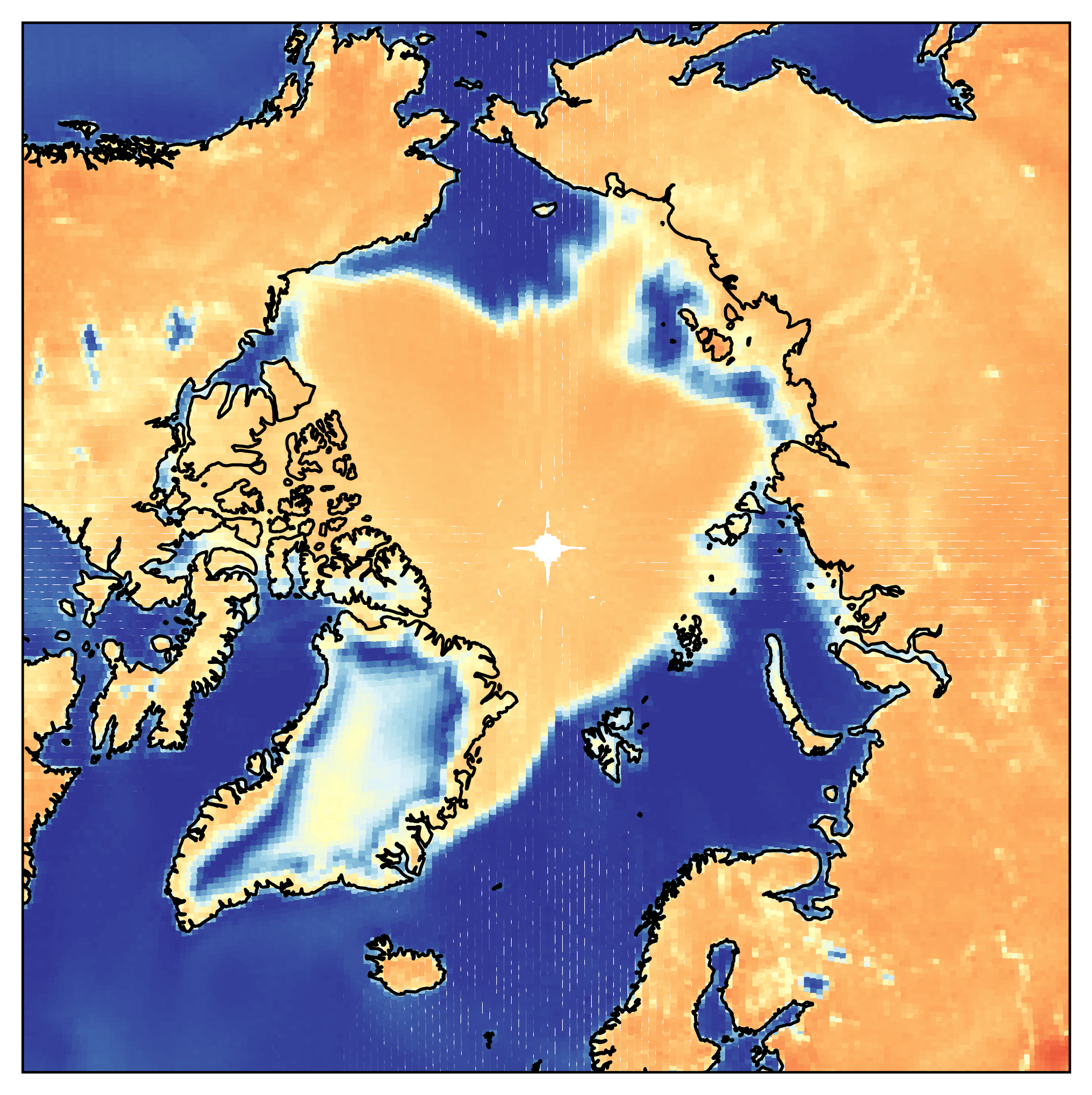} & &
\includegraphics[height=3.5cm]{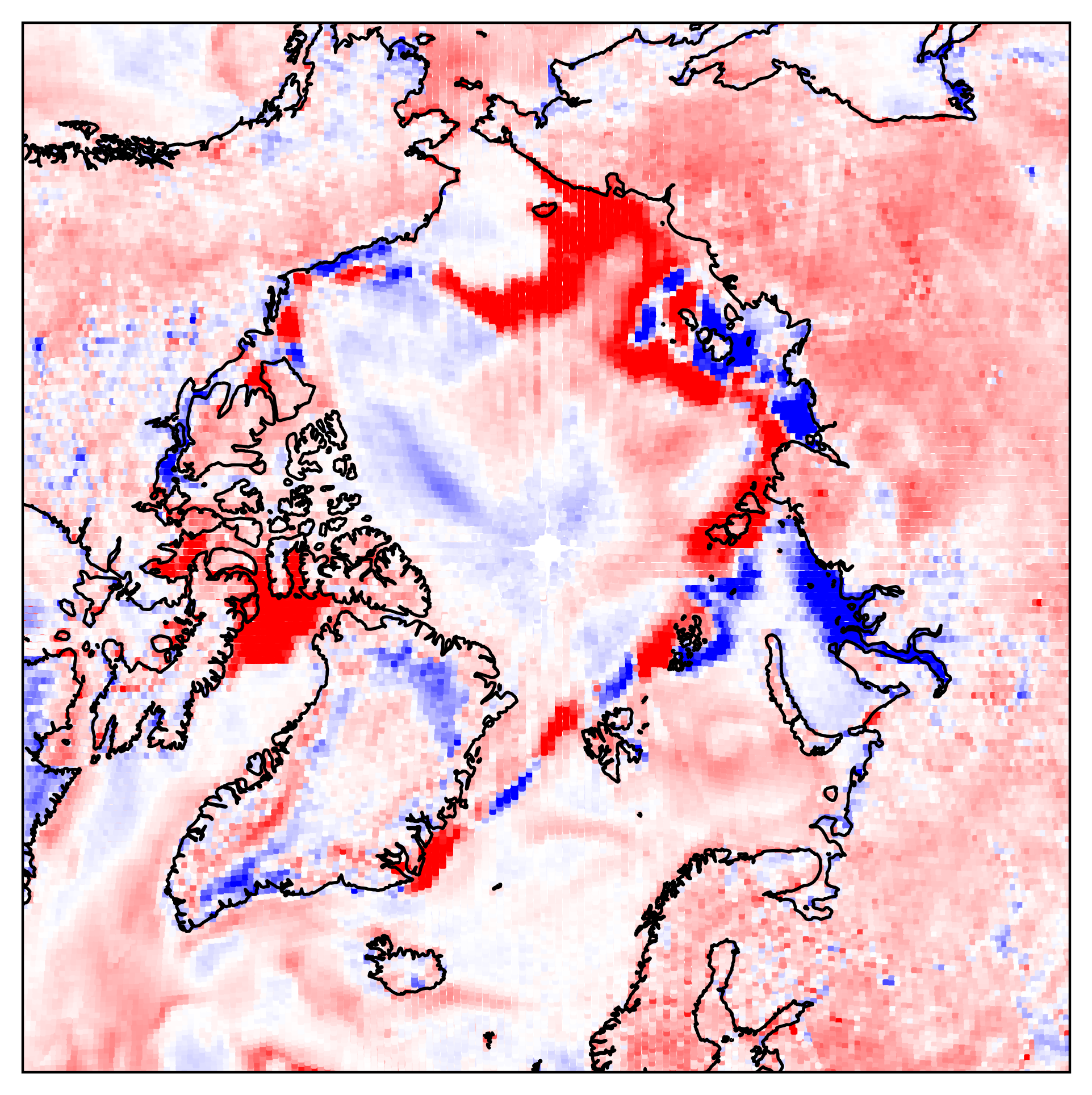} & & \includegraphics[height=3.5cm]{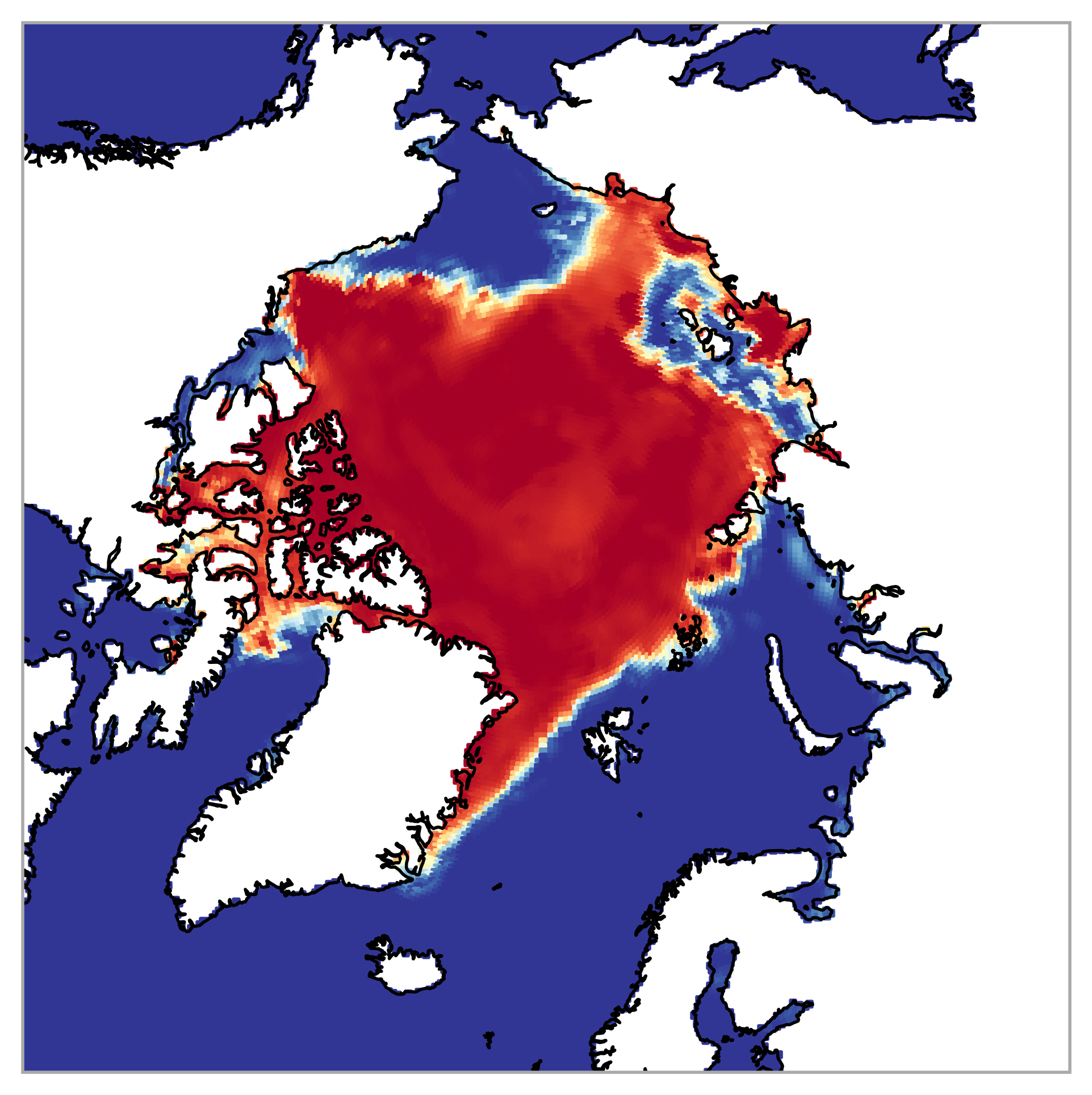} & \\
\raisebox{1cm}{\rotatebox{90}{30/10/2022}} &
\includegraphics[height=3.5cm]{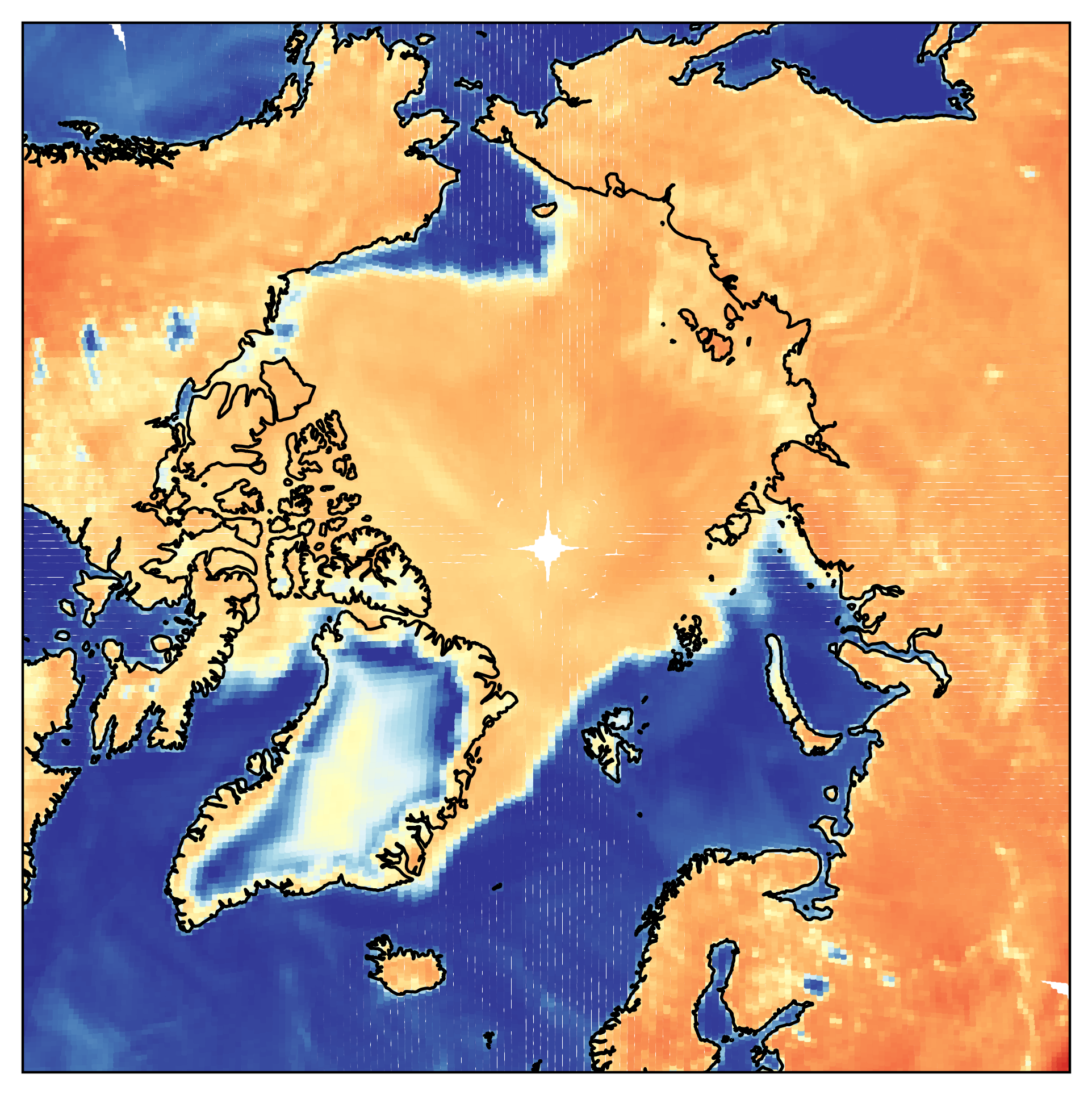} &
\includegraphics[height=3.5cm]{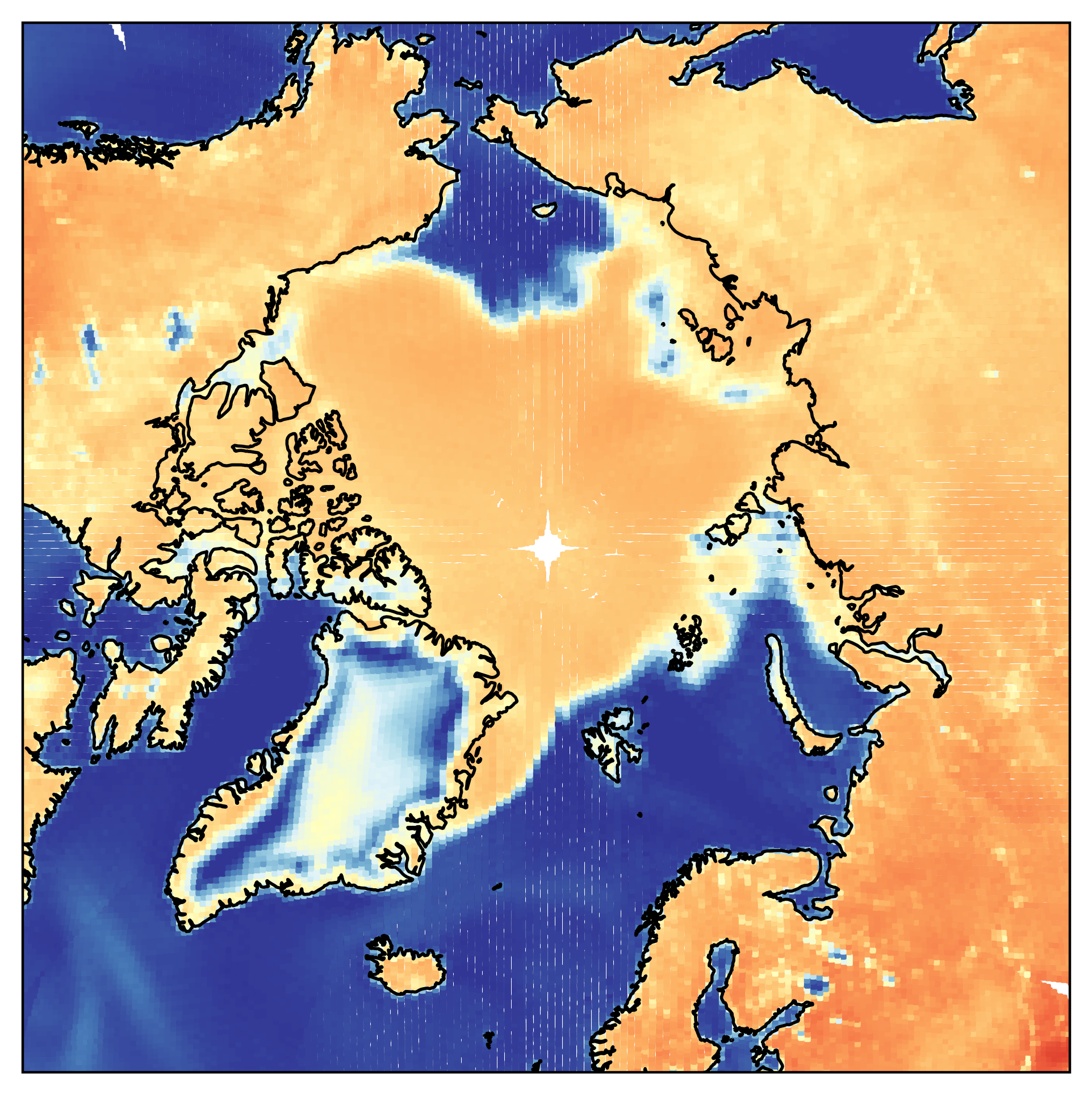} & &
\includegraphics[height=3.5cm]{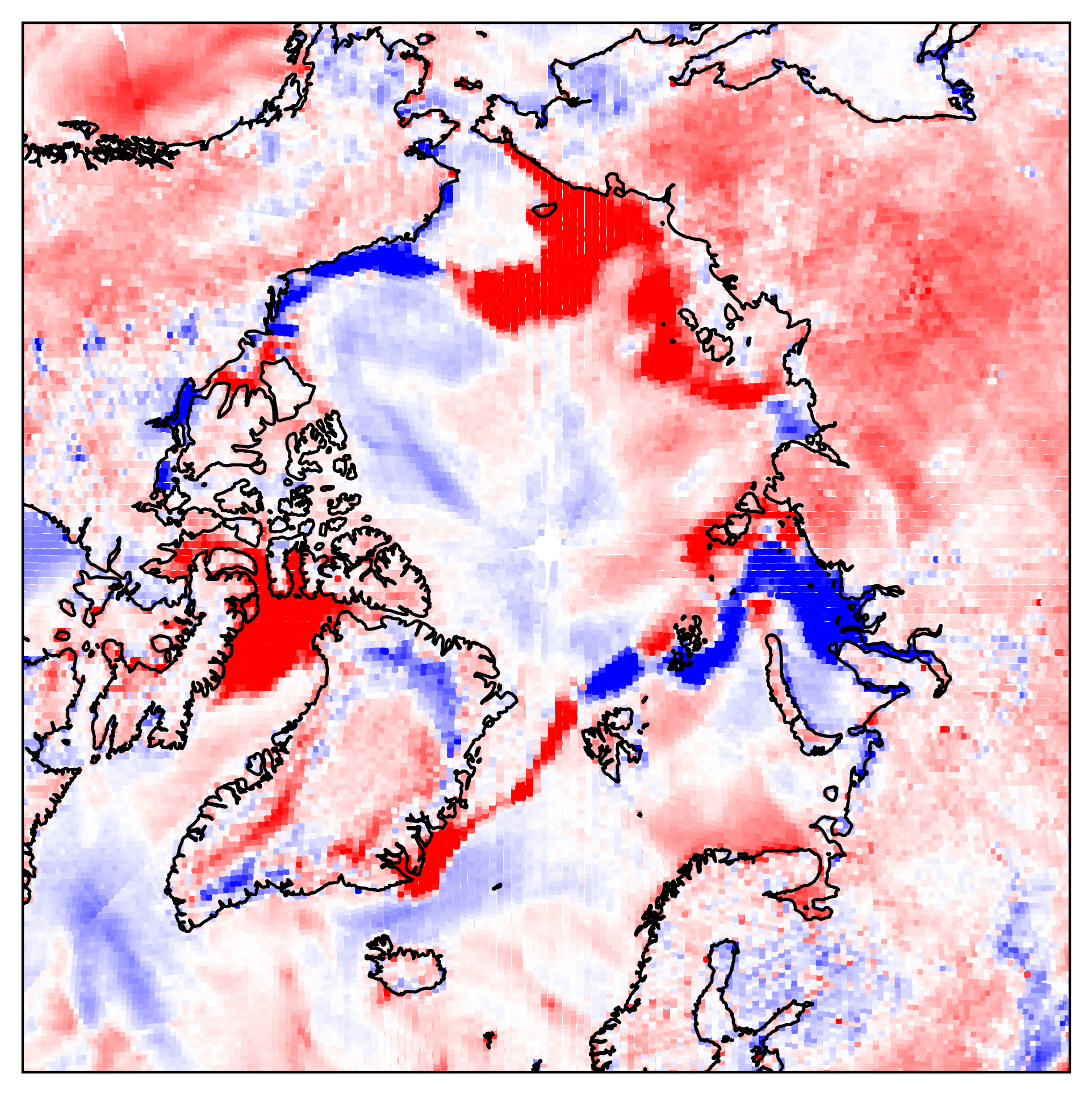}& & \includegraphics[height=3.5cm]{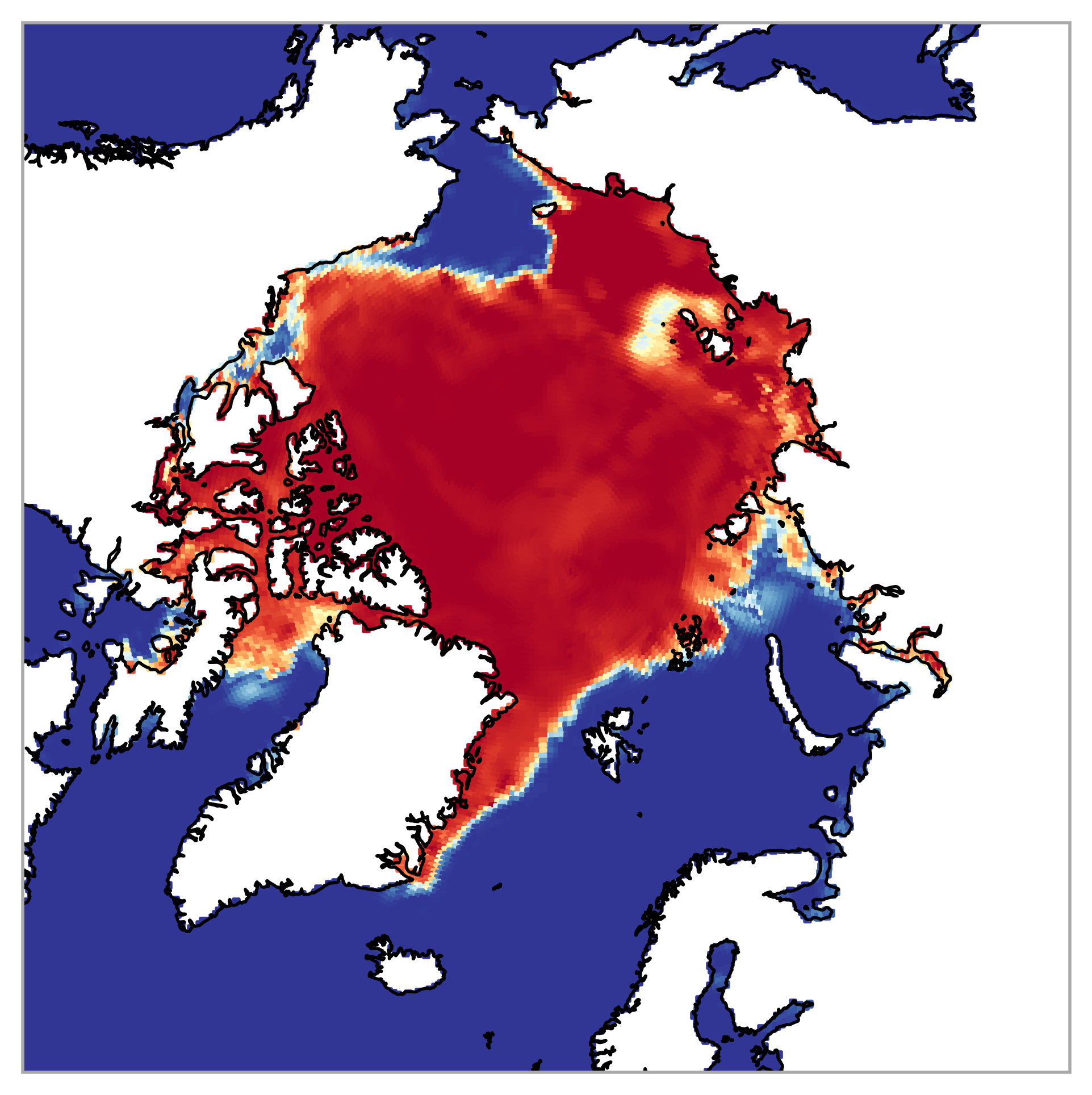} & \\

\end{tabular}
\end{adjustbox}
\caption{AMSR-2 channel 5 (10v) brightness temperatures (K) from left to right: target (observed), forecasted, difference (target minus forecast). The rightmost column shows the sea ice concentration of the ECMWF ORAS6 reanalysis. The top row shows observations at 24-hour lead time (21 October 2022), the middle row shows the predicted observations at 5-day lead time (25 October 2022), whereas the bottom row shows the predicted observations at 10-day lead time (30 October 2022). The 10-day forecast shown here has been chosen because it features a rapid freezing event, and was initialised using 12 hours of observations between  09:00 and 21:00~UTC on 20 October 2022.}
\label{fig:seaice_amsr2}

\end{figure}

The top row of Figure \ref{fig:seaice_amsr2} illustrates this idea by comparing the verifying brightness temperatures (“target”, left) with the corresponding GraphDOP forecasts (second column) for the AMSR-2 (Advanced Microwave Scanning Radiometer 2) 10 GHz, vertically polarized channel (V-pol; channel 5) over the first 24 hours of a 10-day trajectory, spanning 21:00~UTC on 20 October 2022 to 21:00~UTC on 21 October 2022. AMSR-2 channel 5, together with other frequencies not shown here, is notably sensitive to sea ice concentration because sea ice, open ocean, and land exhibit distinct microwave emissivities and scattering behaviours. The short-range forecast reproduces the observed large-scale gradients and much of the mesoscale variability, including the sharp contrast between the land surface (e.g., Greenland) and the adjacent sea ice cover, which appears prominently in warmer hues. This contrast arises because sea ice (and most land surfaces) has a higher effective emissivity at these microwave wavelengths than open water, yielding elevated brightness temperatures over the ice pack and allowing it to stand out against the cooler ocean background.


Since direct \textit{in-situ} observations of sea ice are largely unavailable in this region and time period, the AMSR-2 microwave brightness temperatures serve as a crucial observational proxy for sea ice extent in the Arctic. The ability of GraphDOP to reproduce these radiometrically derived surface signals, despite being trained without access to explicit sea ice labels, underscores its capacity to learn physically grounded representations from indirect observations. This level of agreement serves as a strong indication that GraphDOP has successfully encoded the relationship between surface emissivity and observed brightness temperatures into a coherent and physically consistent latent representation of sea ice dynamics.

The second and third row of Figure \ref{fig:seaice_amsr2} compare the target (left) to the forecasted observations (second column) at a respective lead time of 5 and 10 days. Over the course of these 10-days, a rapid freezing event occurred in the Laptev and East Siberian Seas, as well as in the north of Baffin Bay, between the Canadian Archipelago and Greenland. Remarkably, the overall growth was well captured by the GraphDOP forecast, and the extent of sea ice after 5 and 10 days also compared well to the ECMWF ORAS6 \citep{oras6} sea ice reanalysis shown in the rightmost column of Figure \ref{fig:seaice_amsr2}. 

The sea ice expansion along the Siberian shelf exhibits in particular a recurrent pattern for this time of the year, given the local geography and sea ice processes. The freezing does not only proceed from the ice pack towards the coast but also in the opposite direction. This occurs because the coastal seas are less deep and cool more quickly to freezing temperature compared to deeper ocean domains. Moreover, coastal regions are more affected by intrusions of cold air positioned over land.   

\begin{figure}[htpb]
\centering
\includegraphics[width=\linewidth]{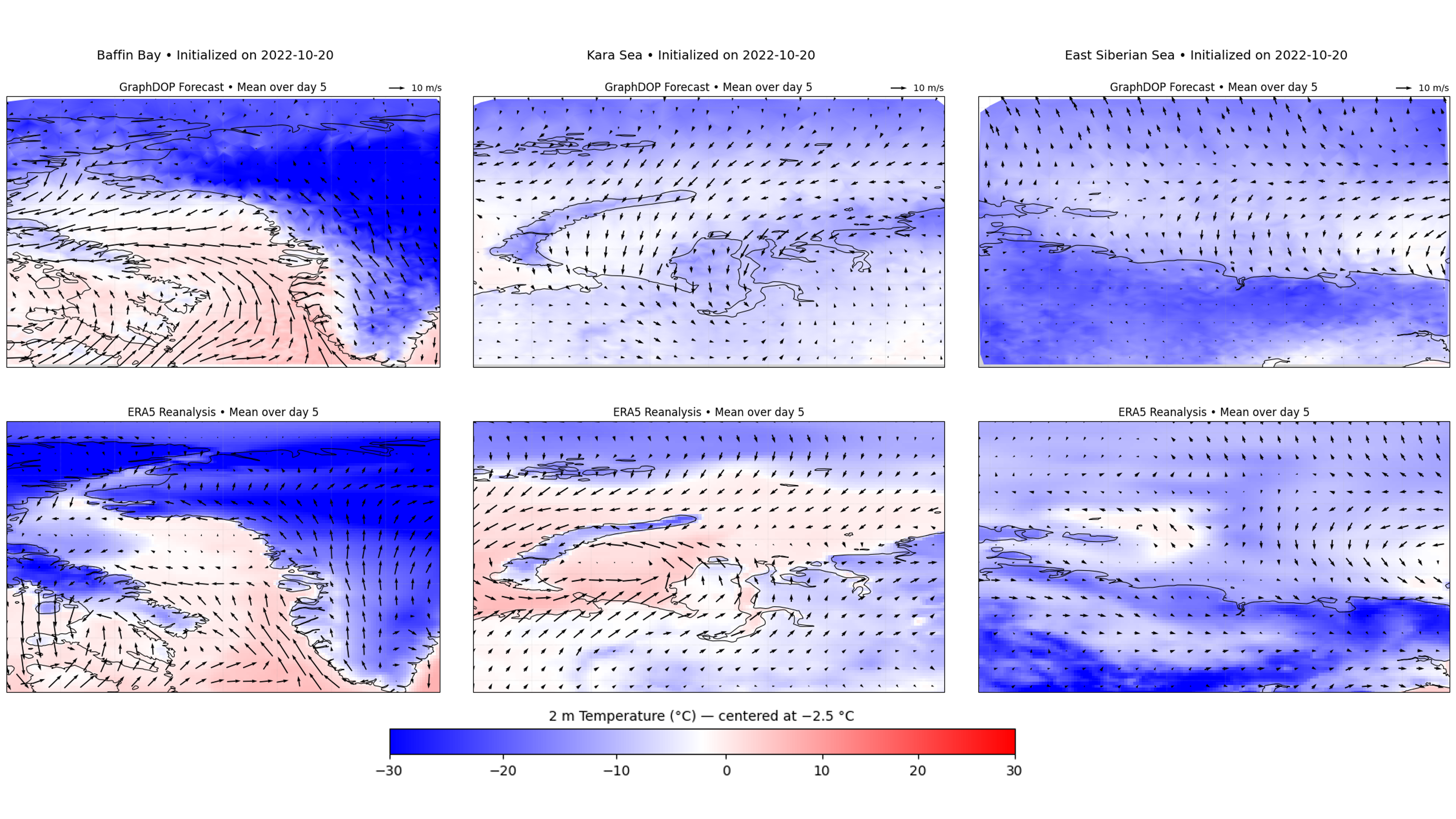}

\caption{Near-surface atmospheric conditions (10\,m wind and 2\,m temperature) from the GraphDOP forecast (top row) and the ERA5 reanalysis (bottom row) for the three regions that exhibit the largest sea-ice extent mismatches: Baffin Bay, the Kara Sea, and the East Siberian Sea (from left to right, respectively). The colour scale of the 2\,m temperature is centred at $-2.5\,^{\circ}\mathrm{C}$ to highlight the likely sea-ice edge position.}
\label{fig:atmo_on_ice}
\end{figure}

However, the 10-day forecast also reveals regions of notable discrepancy, especially in localized areas where the predicted fields depart from the observations, visible as red and blue patches in the difference plots (Figure \ref{fig:seaice_amsr2}). In these areas, GraphDOP has either forecasted too little or too much sea ice growth. Large mismatches are found, in particular, in the easternmost part of the East Siberian Sea (model underestimation), the Kara Sea (model overestimation), and Baffin Bay (model underestimation), indicating a geographically structured error pattern that is most apparent over shelf seas and marginal-ice zones. These errors can be explained by multiple factors, including limitations in the representation of sea ice physics within GraphDOP, insufficient information in the initial conditions, or, finally, inconsistencies in the sea ice drivers between the verifying observations and the forecast. An investigation of the circulation and temperature patterns in ERA5 and in the decoded GraphDOP fields suggests that the last hypothesis could play a substantial role. In particular, errors in the forecast large-scale atmospheric circulation have led to different wind and thermodynamic forcing, resulting in a redistribution or erroneous formation sea ice across the Arctic seas and, consequently, the spatially coherent under- and overestimation signals noted above. Figure~\ref{fig:atmo_on_ice} compiles examples comparing GraphDOP near-surface fields (2\,m temperature and 10\,m wind) with ERA5 on forecast day~5 (26~October~2022). In Baffin Bay, a spurious warm-air intrusion from lower latitudes inhibits sea-ice growth along the Baffin Island coast. In the Kara Sea, slightly stronger southerlies displace the sea-ice edge farther south than in ERA5, contributing to the positive AMSR-2 brightness-temperature bias. Moreover, the absence in the GraphDOP forecast of a low pressure system in the proximity of Novaja Zemlja reduces the advection of warm air from lower latitudes. Finally, over the East Siberian Sea, a more extensive westward circulation in GraphDOP compared to ERA5 could enhance the sea-ice drift, leading to an underestimation of sea-ice extent in that region. Readers should note that a perfect correspondence between the GraphDOP forecast and the observed atmospheric circulation is unlikely at medium-range lead times, especially in the polar regions, where pronounced variability and vigorous weather — arising from local air–ice–ocean coupling and from mid-latitude intrusions —constrain attainable forecast skill. This reflects the general growth of forecast errors with lead time, similar to what is observed in physically based systems.

Overall, these results remain very promising, showing that by leveraging a diverse set of satellite observations, including AMSR-2 data for sea ice and snow, GraphDOP captured the subtle shifts in microwave emissions associated with sea ice refreezing processes. Several other sea ice-sensitive sensors are already available or could be added in the future to the GraphDOP prediction system to increase its skill in simulating sea ice. For example, thermal-infrared instruments can measure the skin temperature under clear-sky conditions; L-band radiometry (e.g., SMOS) is sensitive to thin-ice thickness and brine content; and radar and laser altimeters measure freeboard with complementary penetration characteristics (radar approaching the snow–ice interface, laser sensing the snow surface). By predicting directly in observation space on the satellite-footprint graph, GraphDOP leverages these sensitivities and has the potential to learn an intrinsic representation of the sea ice evolution and deliver useful information to stakeholders. To enhance user relevance, future developments could train GraphDOP directly against Level-2 sea ice concentration (SIC) products, linking its observation-space predictions to a familiar variable.

\section{Hurricane Ian case study: ocean and atmosphere}
\label{sec:hurricane}

Hurricanes are inherently multi-domain phenomena, driven by and impacting the atmosphere, ocean surface, and wave field simultaneously. Thus, successful forecasts of such events offer a compelling indication of the model’s ability to implicitly capture coupled dynamics of the Earth System. Hurricane Ian, a Category 5 storm \citep{osti_10492534} that struck in late September 2022, serves as a challenging and revealing case study for assessing GraphDOP’s ability to forecast this type of extreme, cross-domain weather event. This case study was briefly addressed in \cite{alexe2024graphdopskilfuldatadrivenmediumrange}, but the forecasts presented here are improved, primarily due to the model update described in Section \ref{sec:model}, the incorporation of additional datasets (cf. Table \ref{table:dop-observations}) and also the grid resolution on which the forecasts are output (N320 vs. O96 in \cite{alexe2024graphdopskilfuldatadrivenmediumrange}.

\begin{figure}[h!]
\centering

\subfigure[Mean sea-level pressure (Pa)]{
    \begin{tabular}{c @{} c} 
        \begin{minipage}{0.9\linewidth}
        \centering
        \includegraphics[trim=0 3 0 0, clip, height=.21\linewidth]{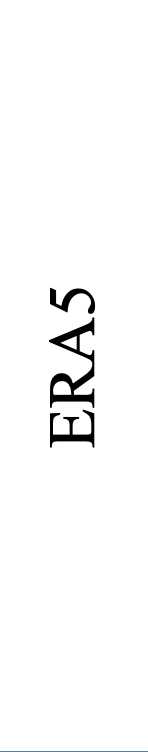}
        \includegraphics[height=.21\linewidth]{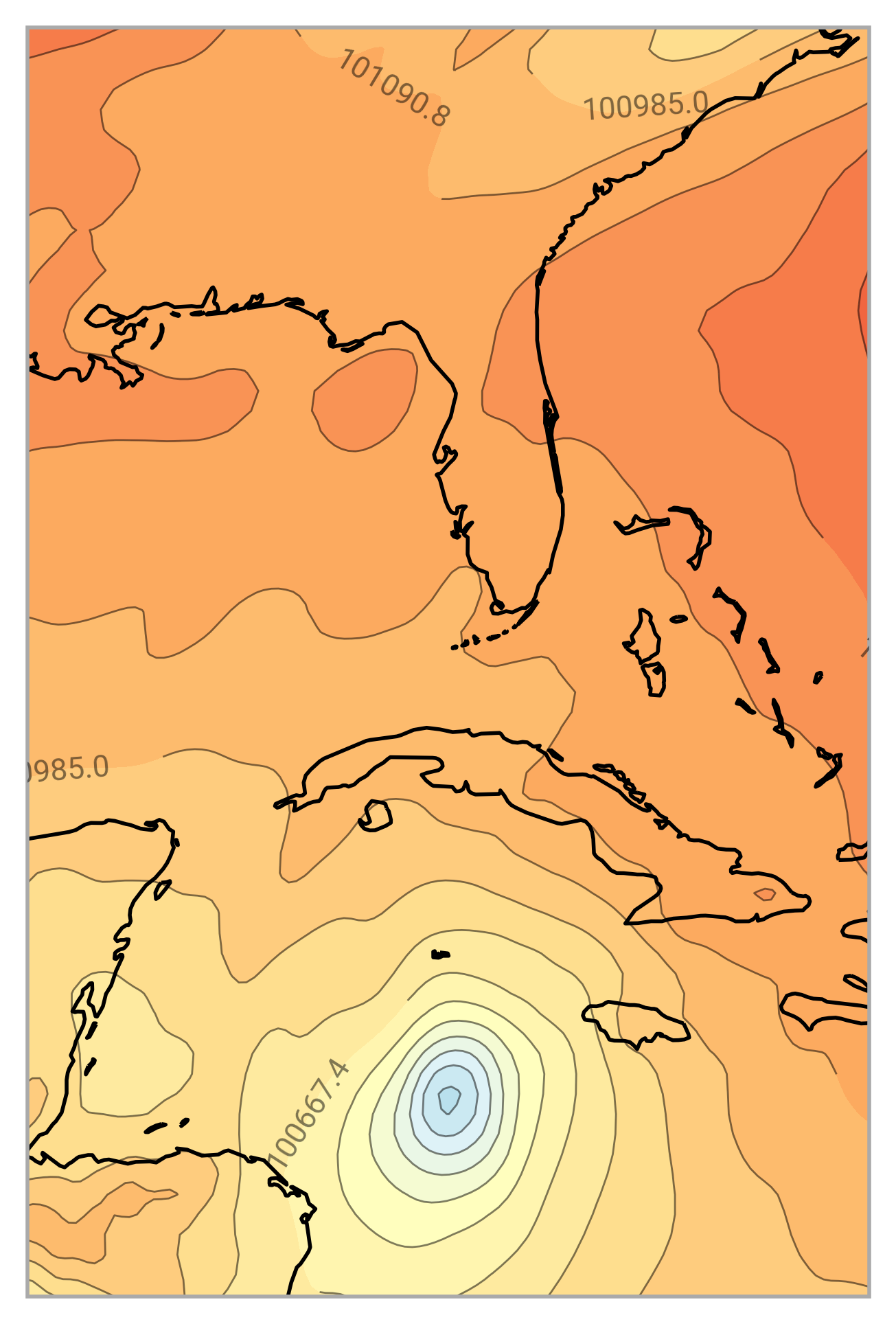}
        \includegraphics[height=.21\linewidth]{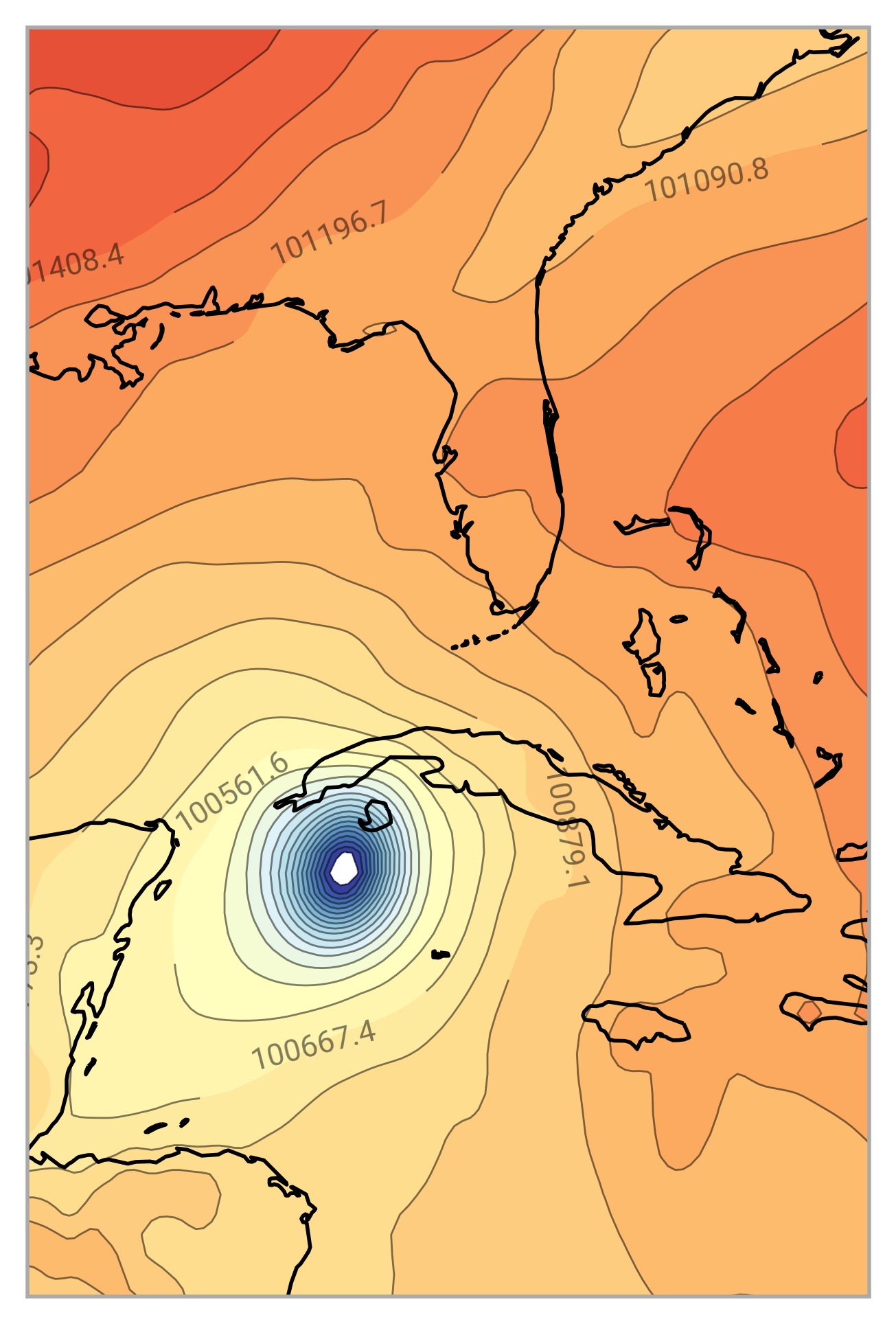}
        \includegraphics[height=.21\linewidth]{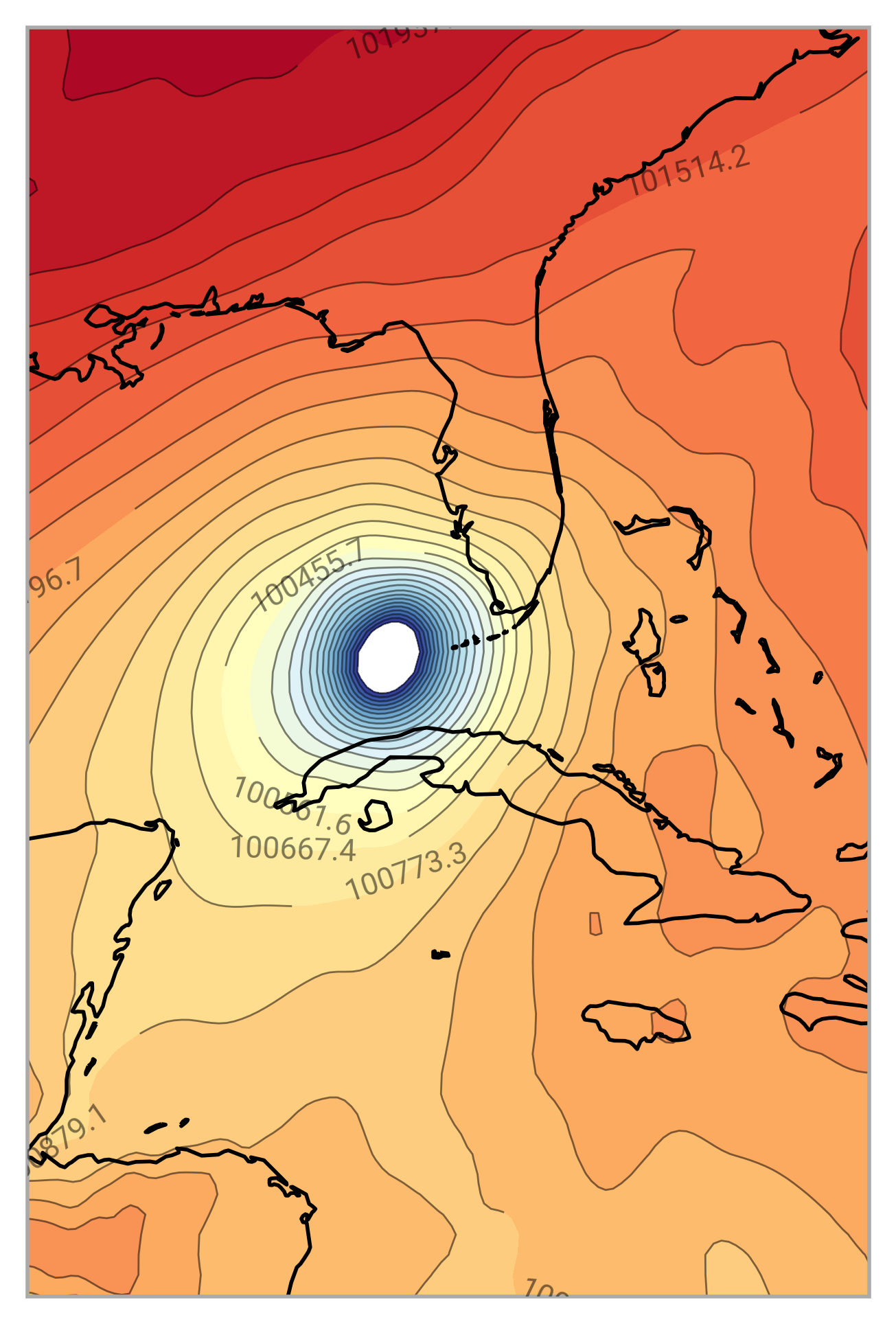}
        \includegraphics[height=.21\linewidth]{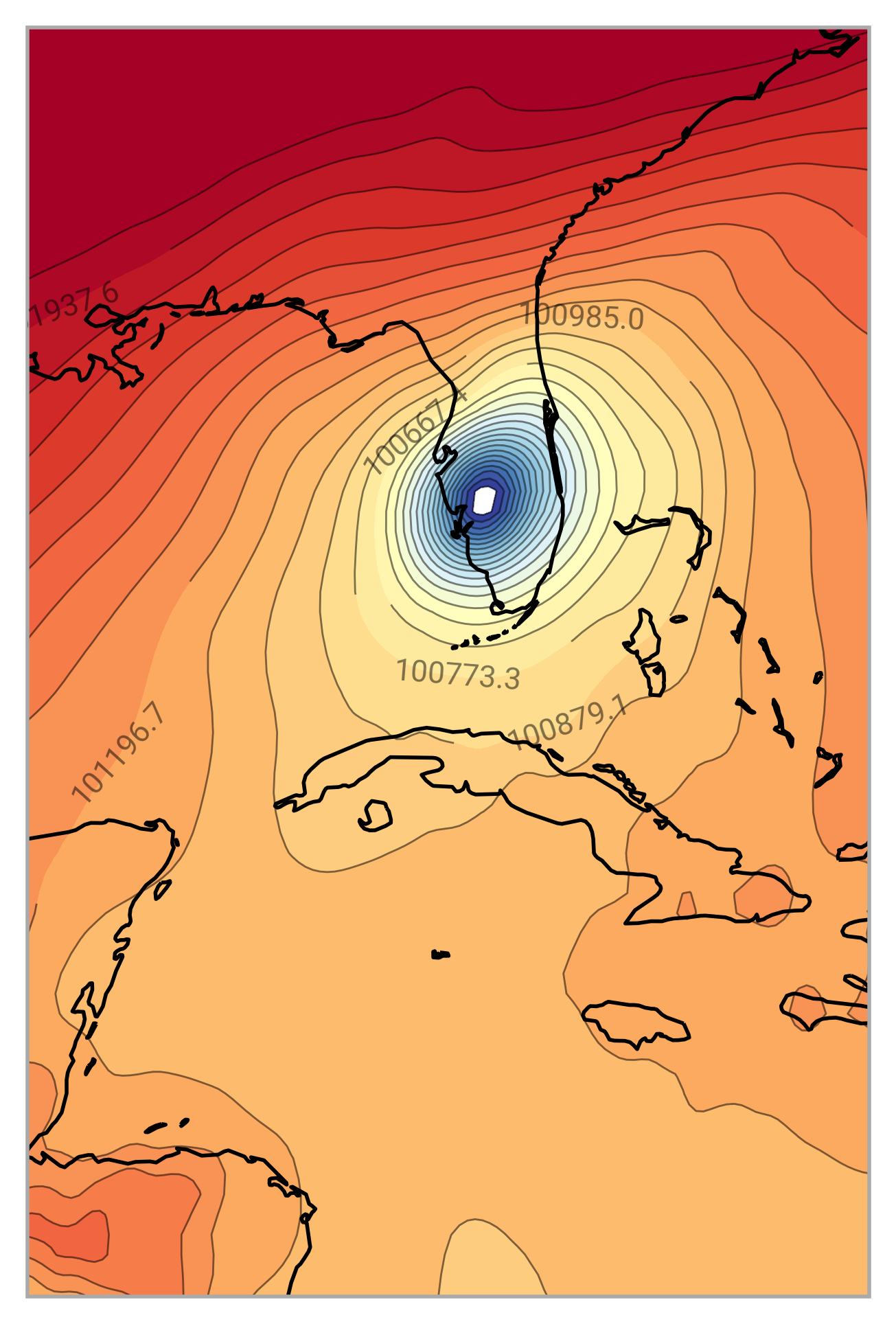}
        \includegraphics[height=.21\linewidth]{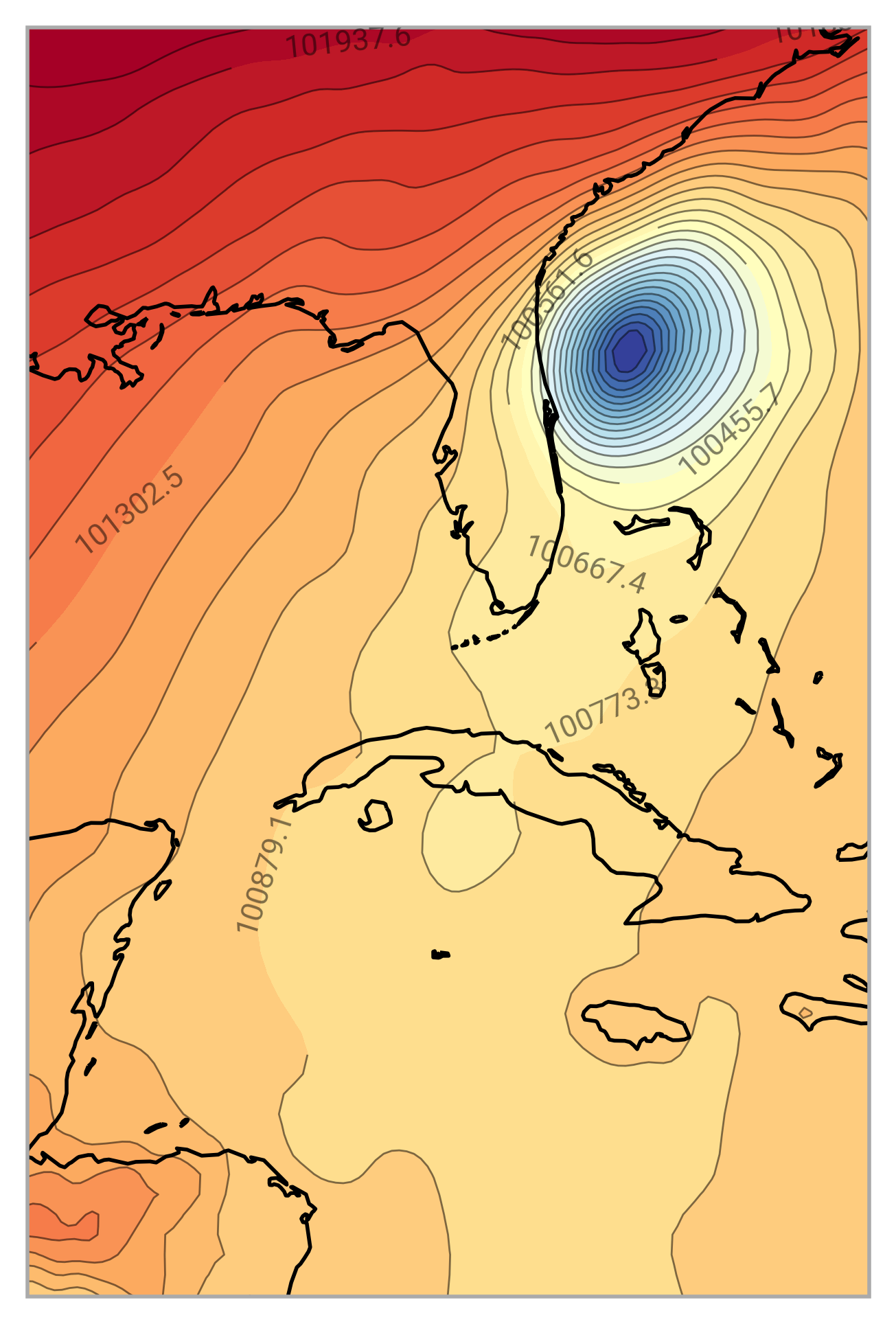}
        \includegraphics[trim=0 6 57 0, clip, height=.21\linewidth]{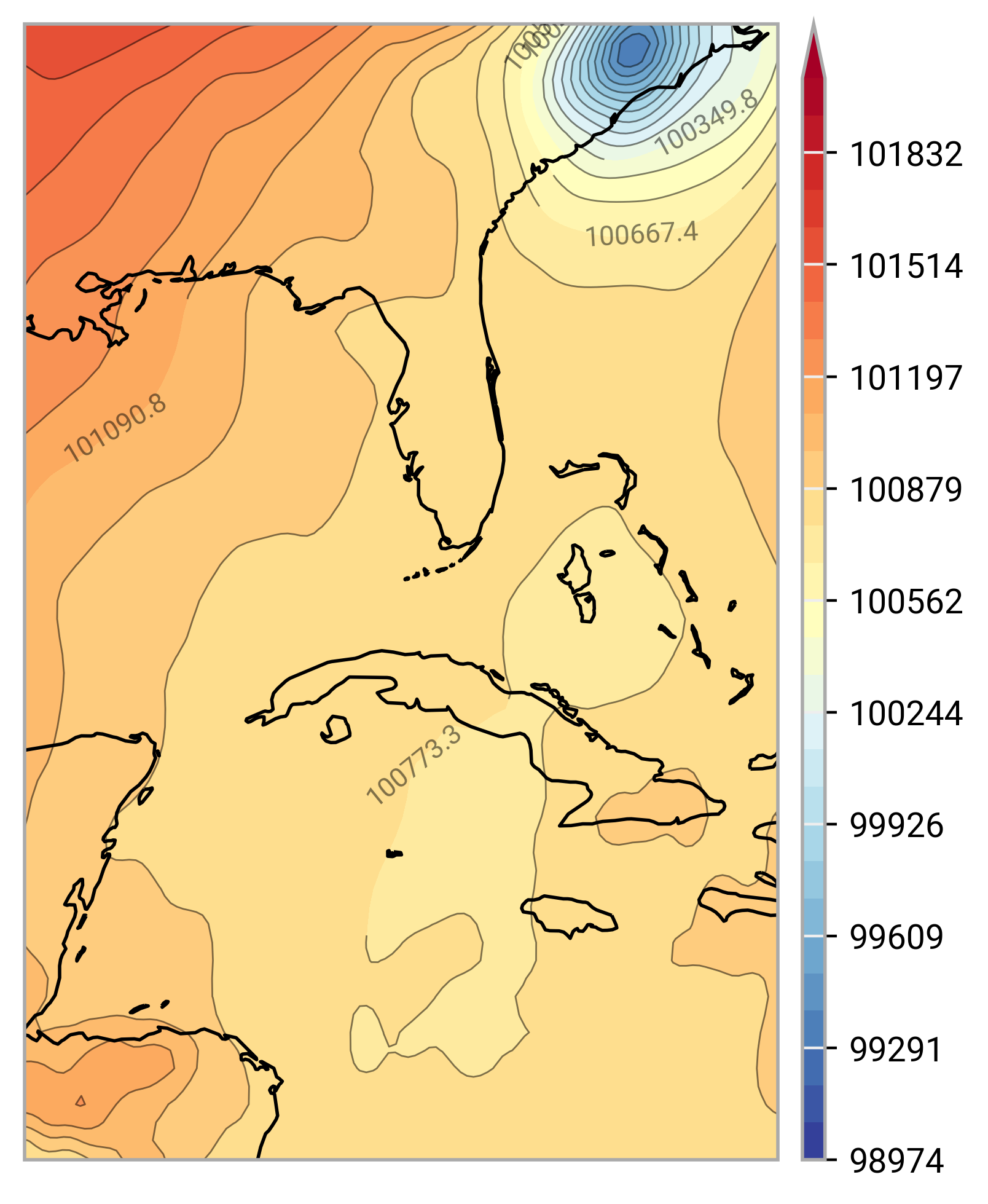}

        \vspace{0.3em}

        \includegraphics[trim=0 3 0 0, clip, height=.21\linewidth]{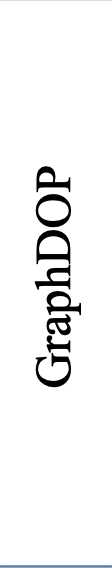}
        \includegraphics[height=.21\linewidth]{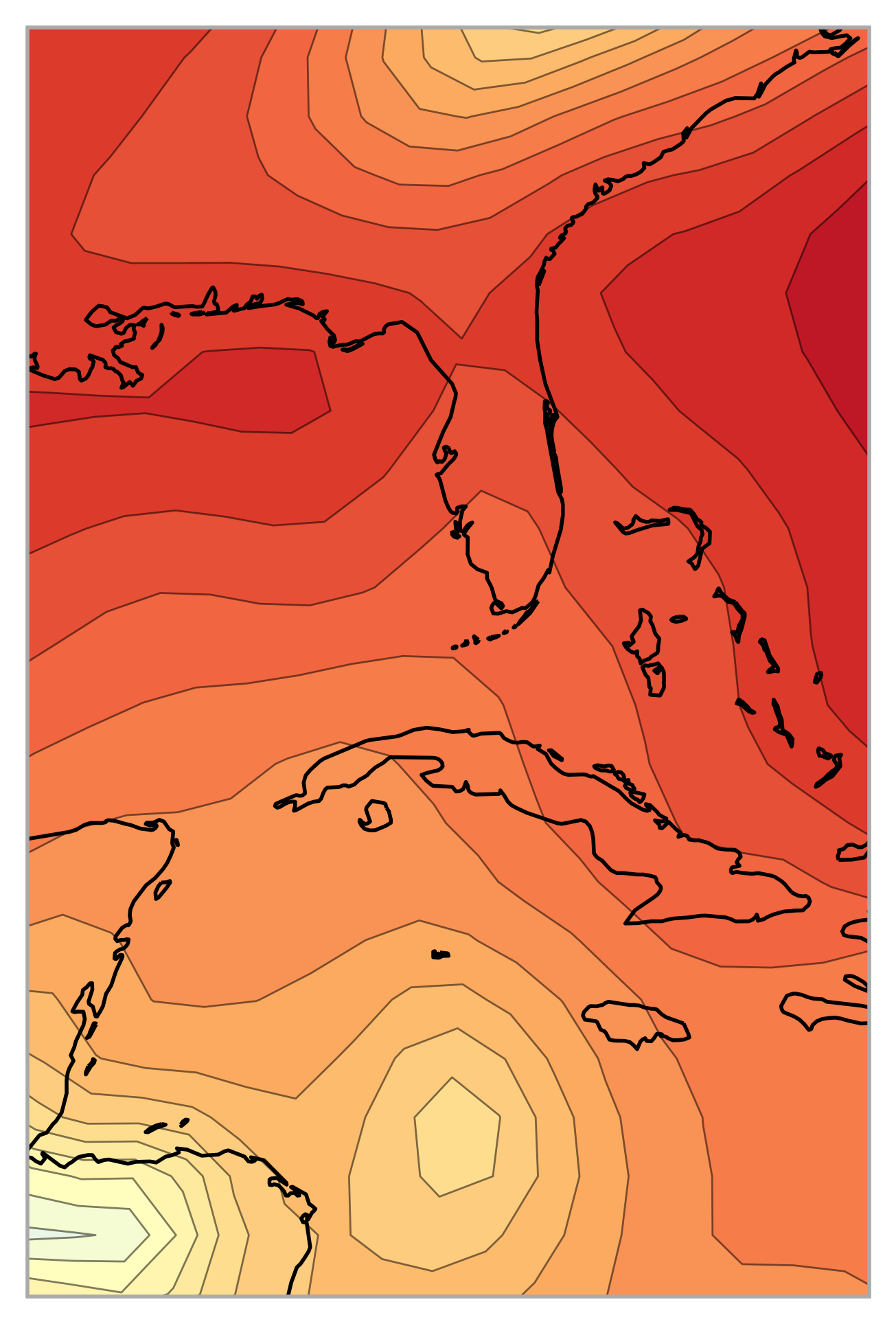}
        \includegraphics[height=.21\linewidth]{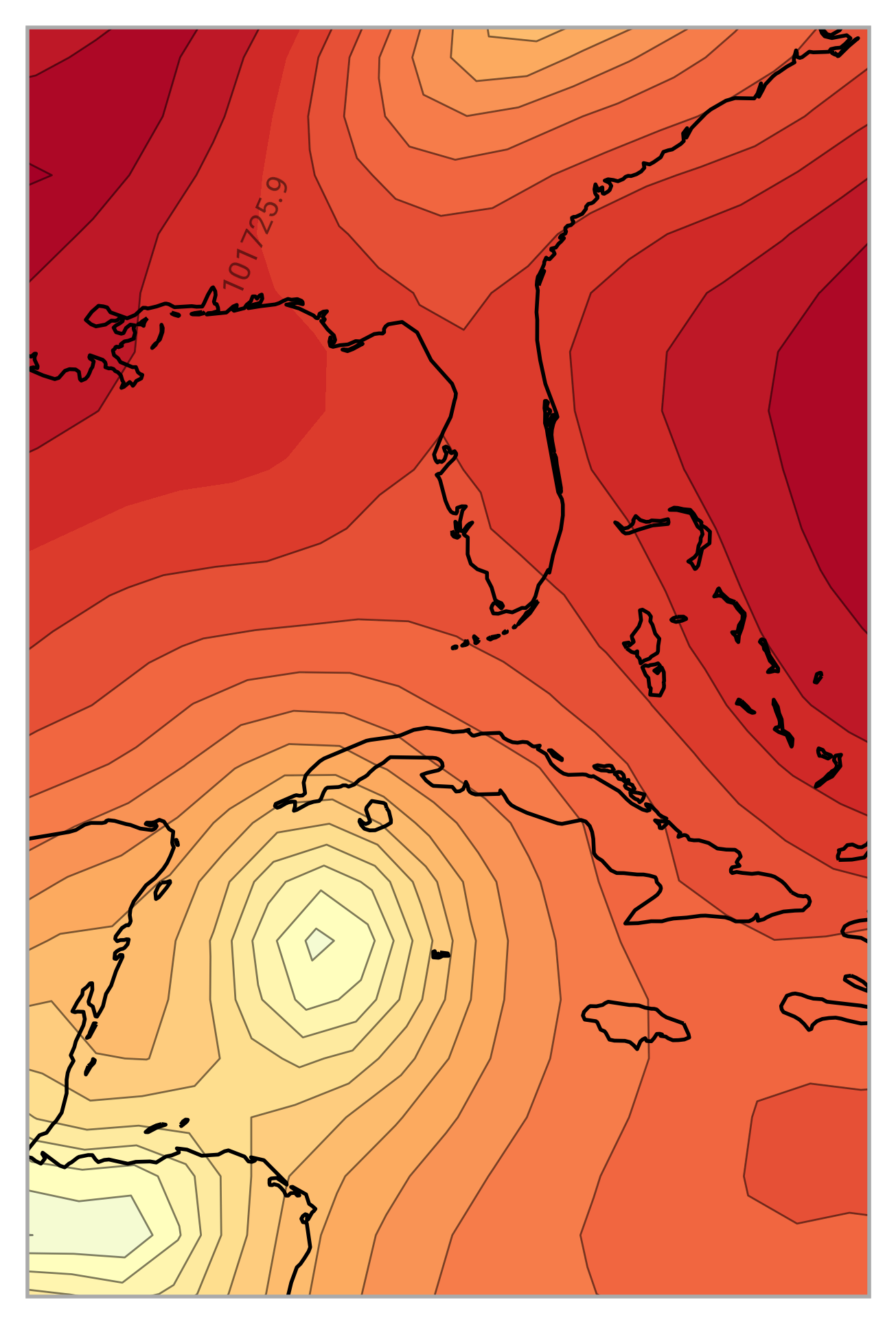}
        \includegraphics[height=.21\linewidth]{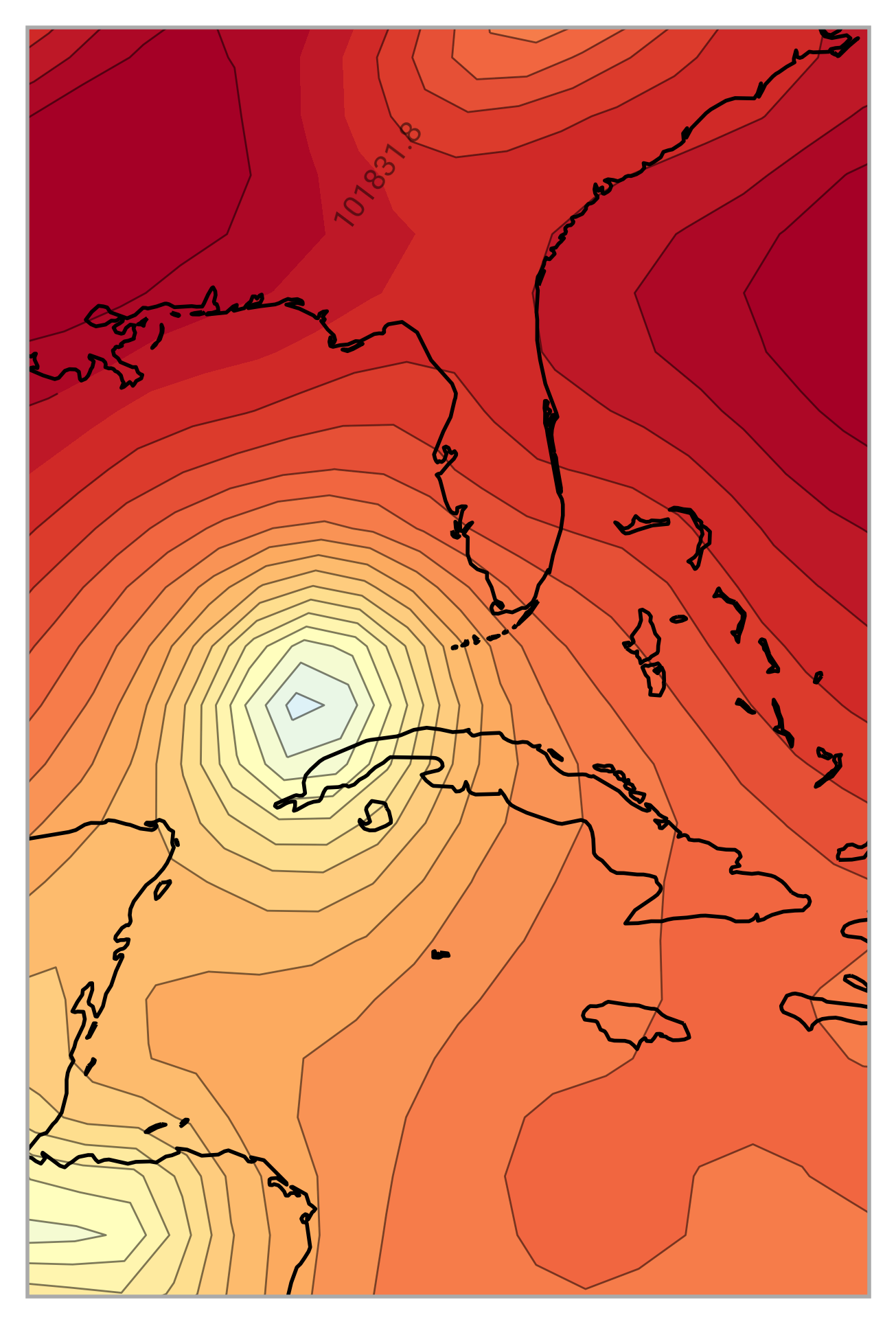}
        \includegraphics[height=.21\linewidth]{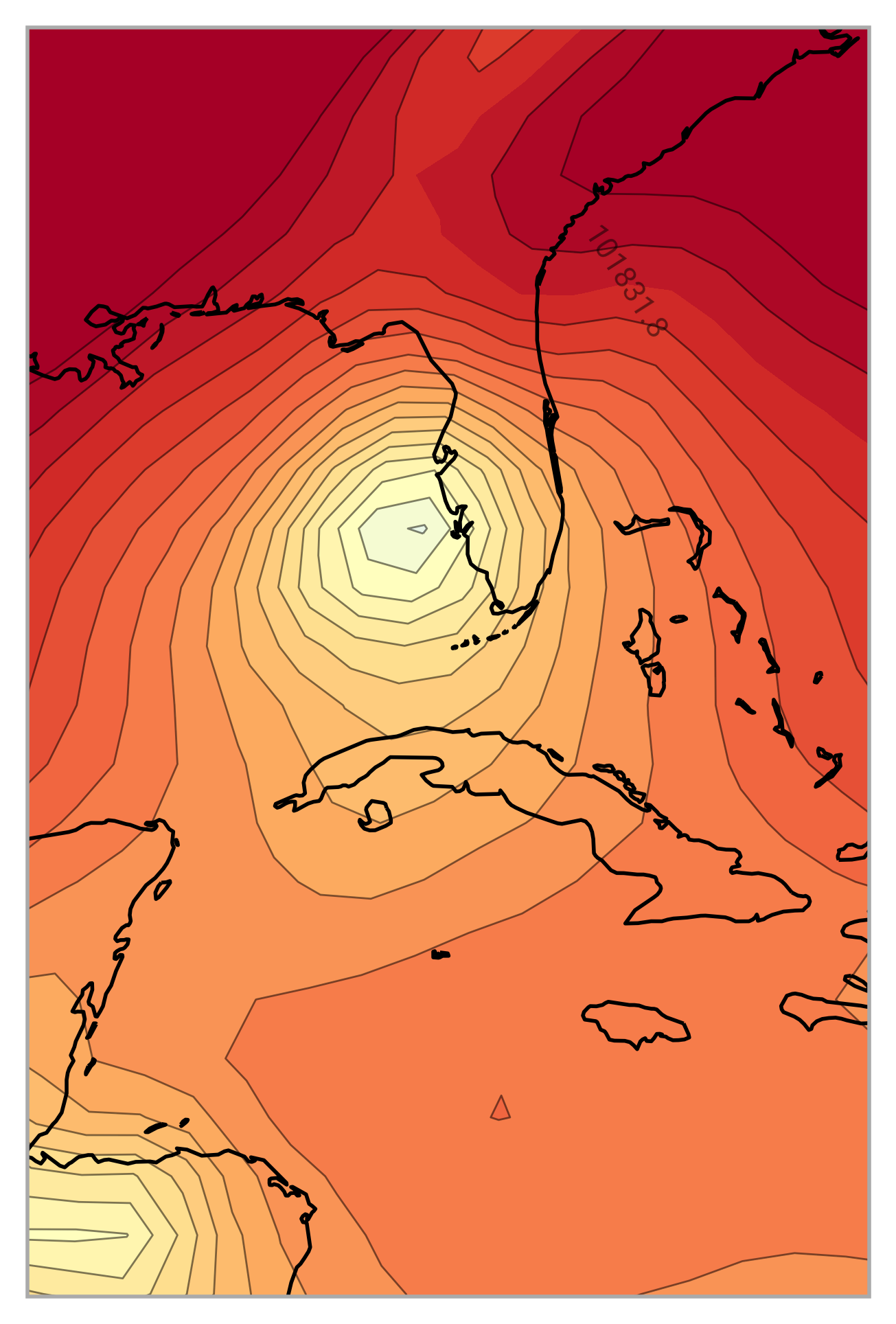}
        \includegraphics[height=.21\linewidth]{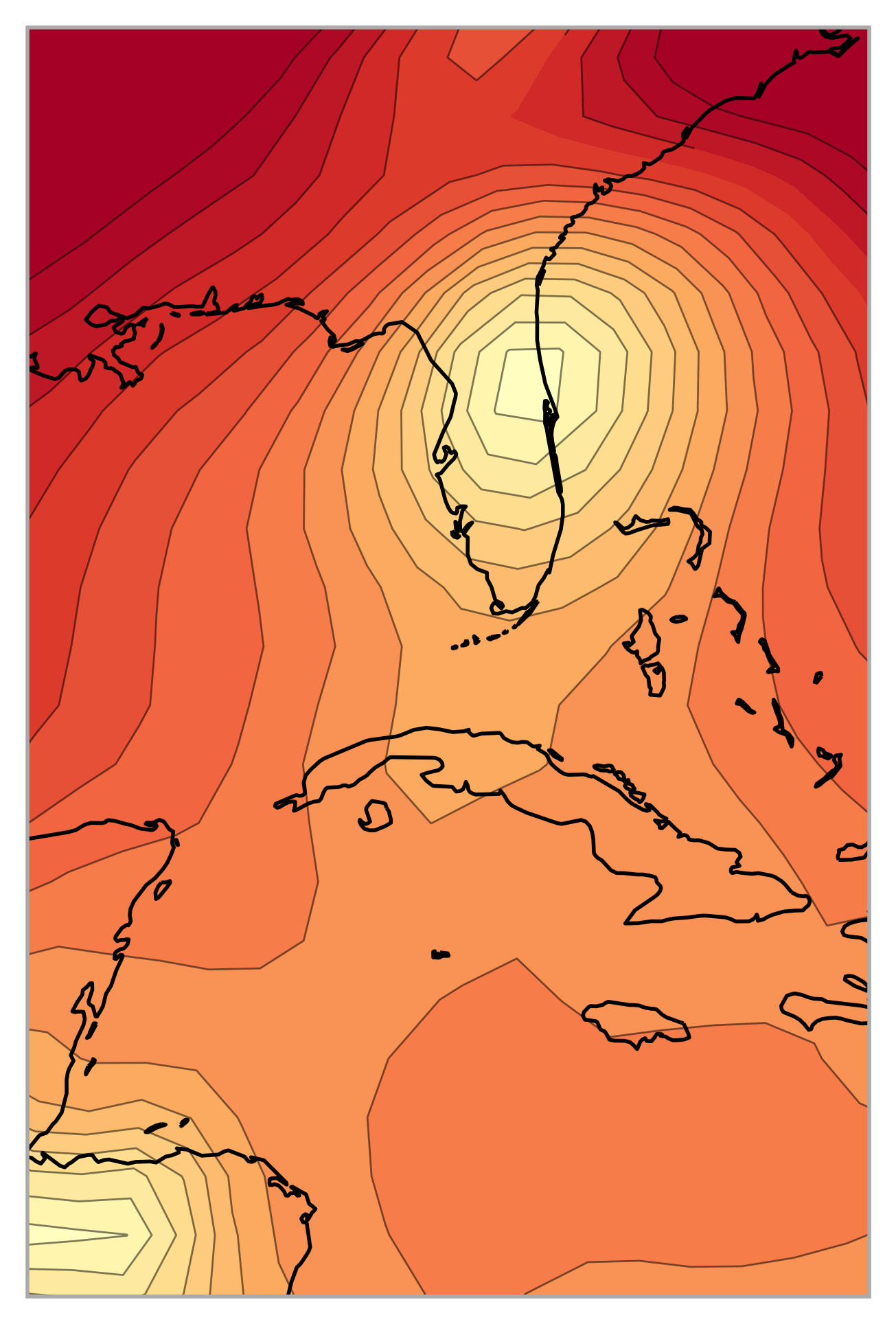}
        \includegraphics[trim=0 5 55 0, clip, height=.21\linewidth]{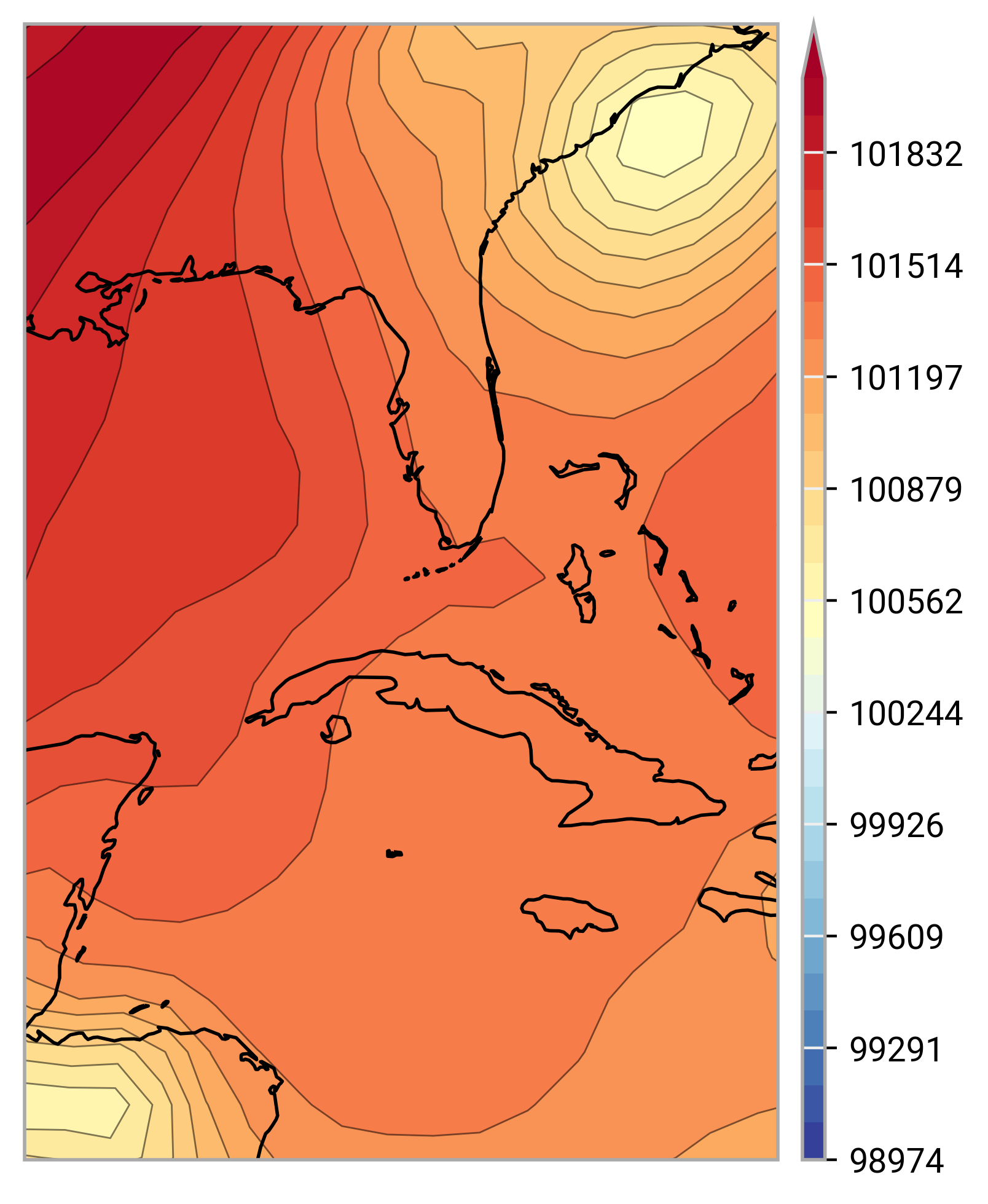}
        \end{minipage}
        &
        \begin{minipage}{0.03\linewidth}
        \includegraphics[height=4.5cm]{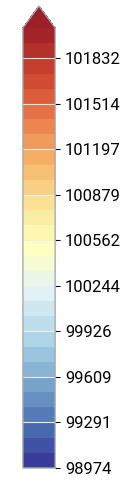}
        \end{minipage}
    \end{tabular}
}

\hspace{1cm} 

\subfigure[Wind speed (m/s)]{
    \begin{tabular}{c @{} c}
        \begin{minipage}{0.9\linewidth}
        \centering
        \includegraphics[trim=0 3 0 0, clip, height=.21\linewidth]{hurricane_ian/era.jpg}
        \includegraphics[height=.21\linewidth]{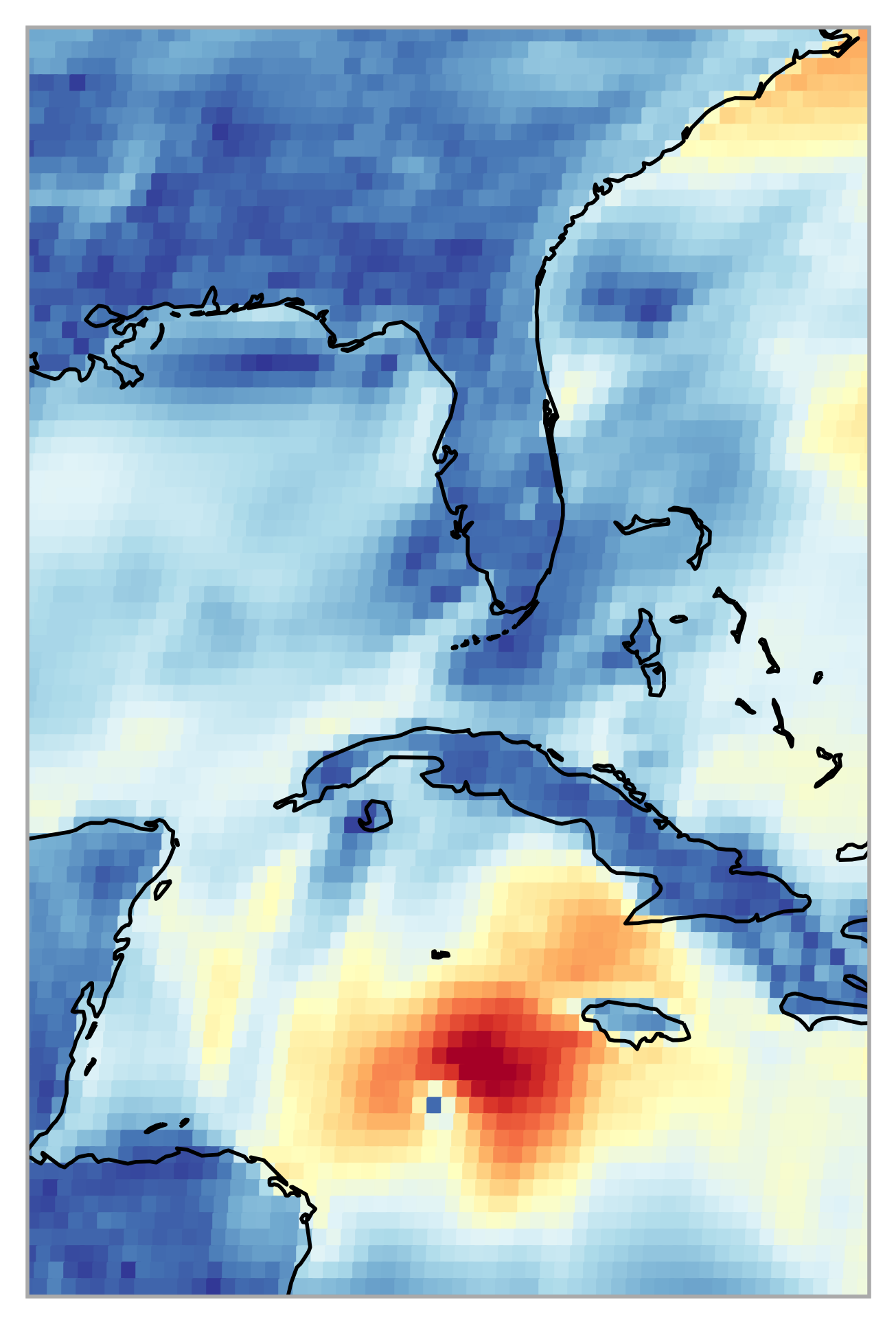}
        \includegraphics[height=.21\linewidth]{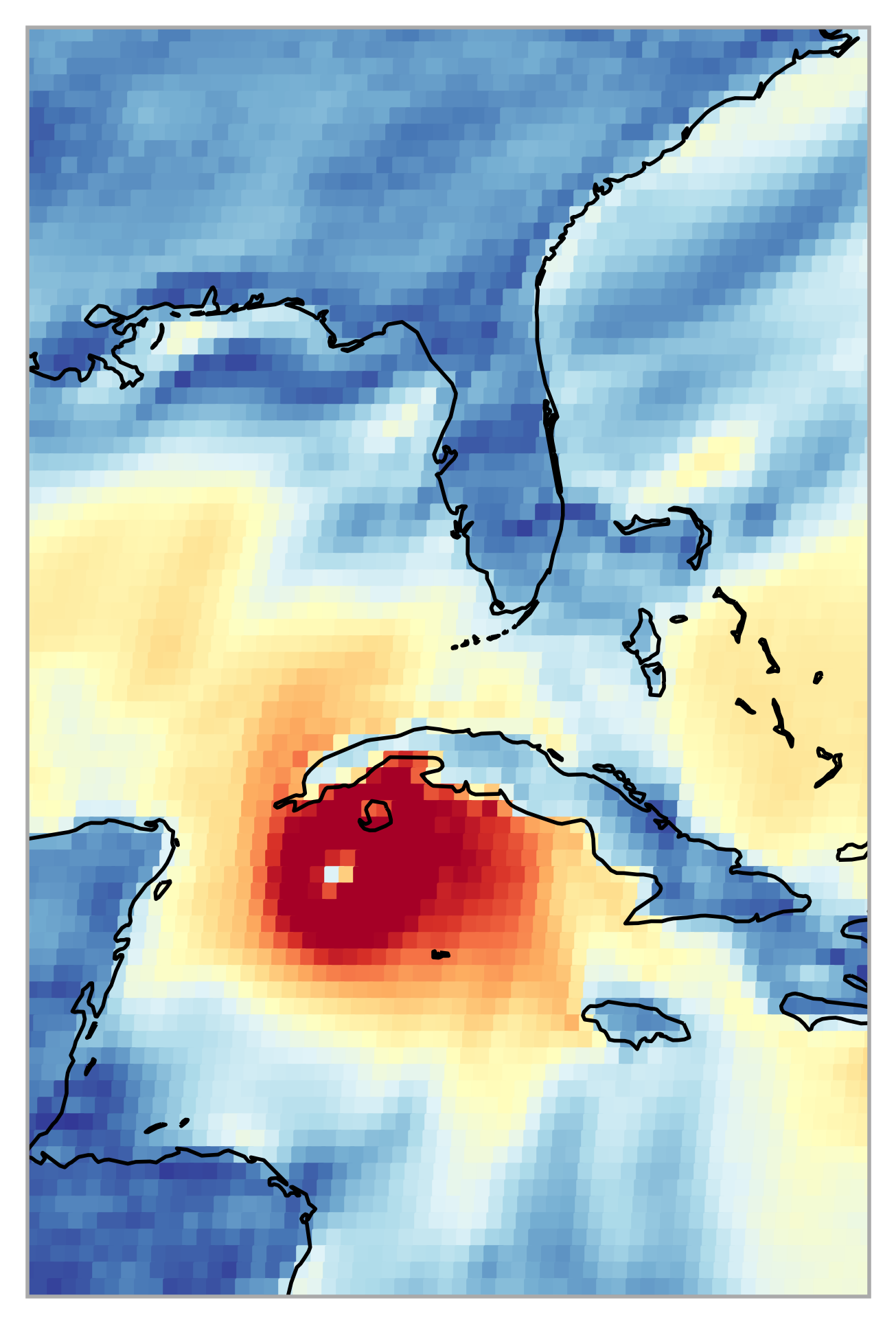}
        \includegraphics[height=.21\linewidth]{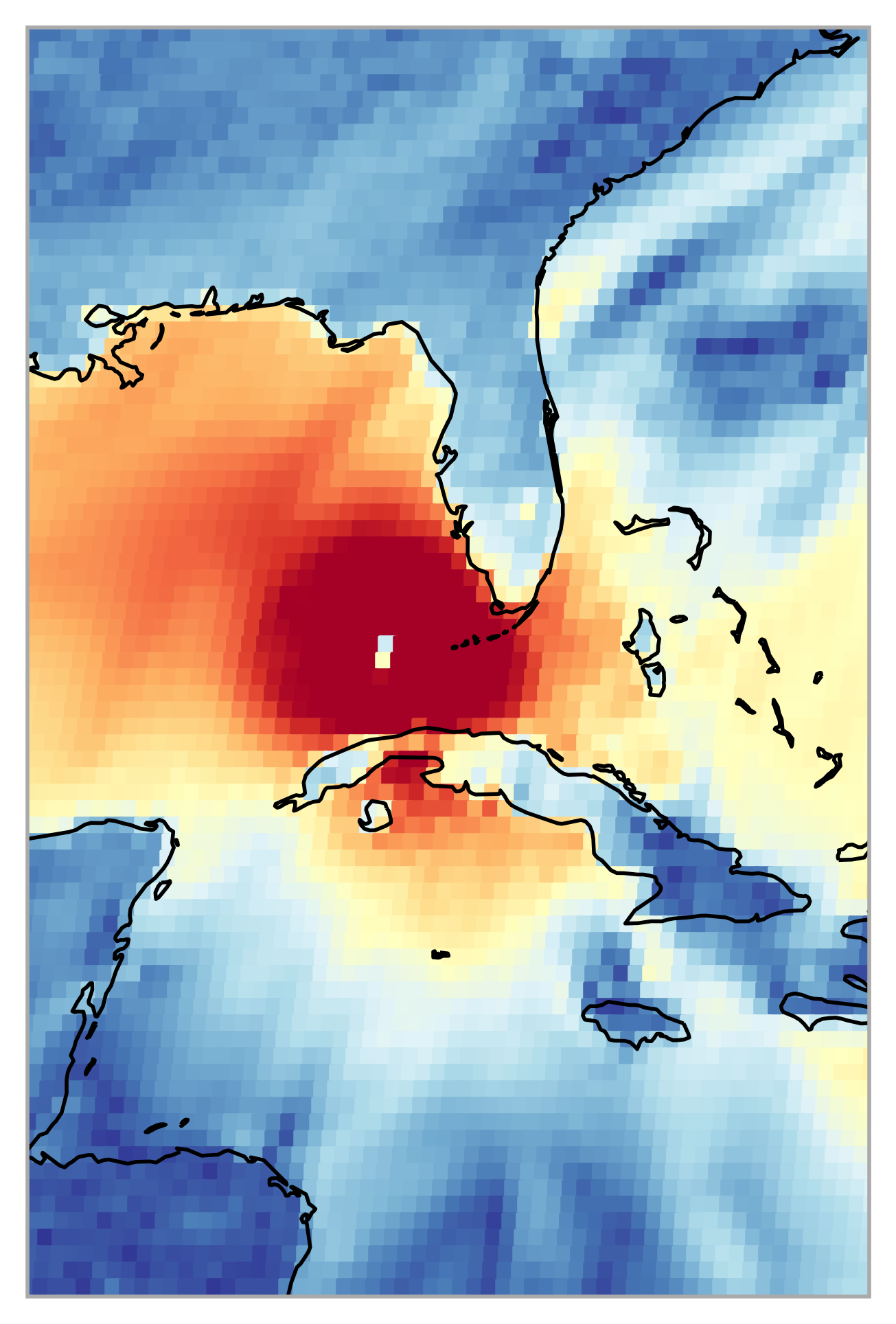}
        \includegraphics[height=.21\linewidth]{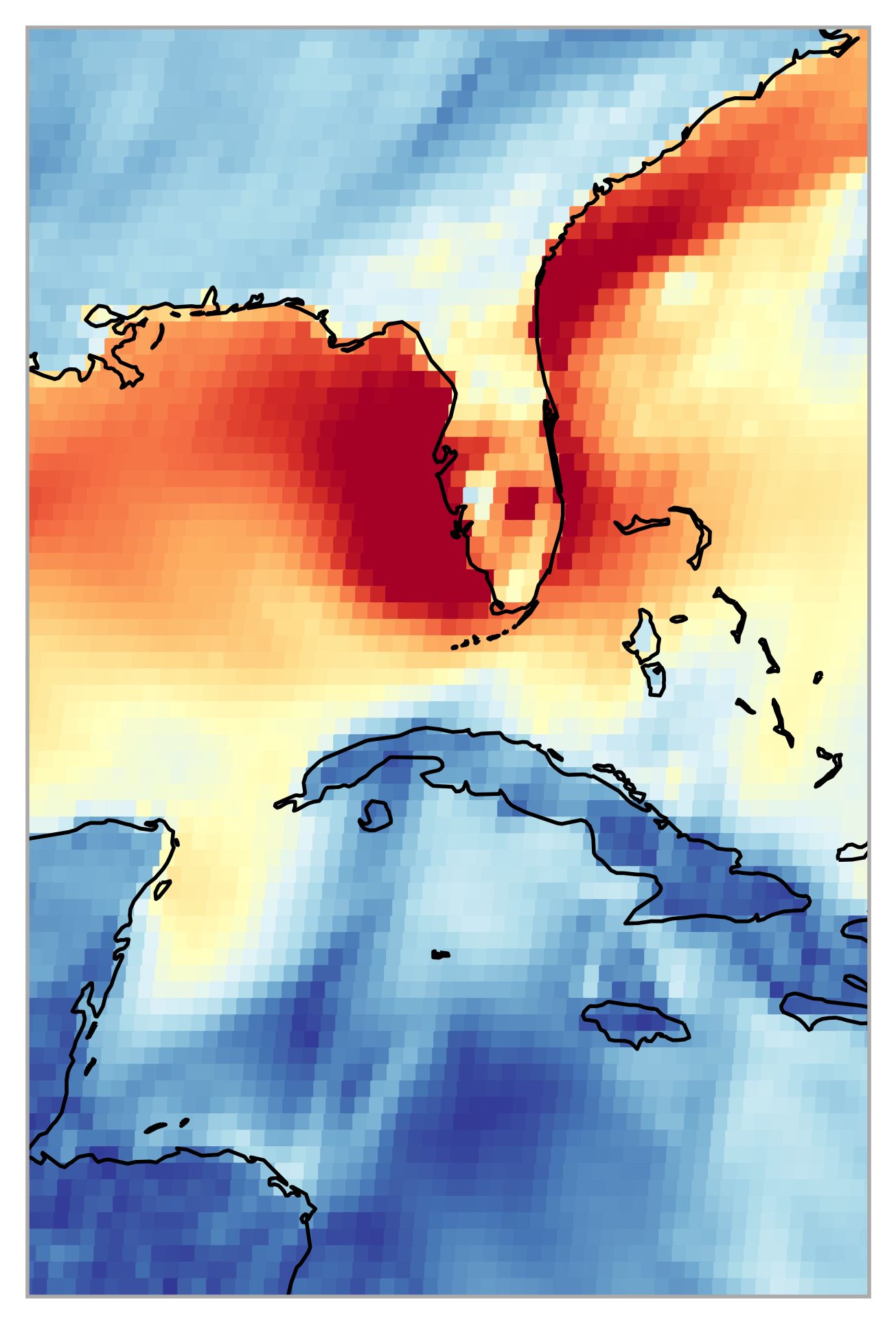}
        \includegraphics[height=.21\linewidth]{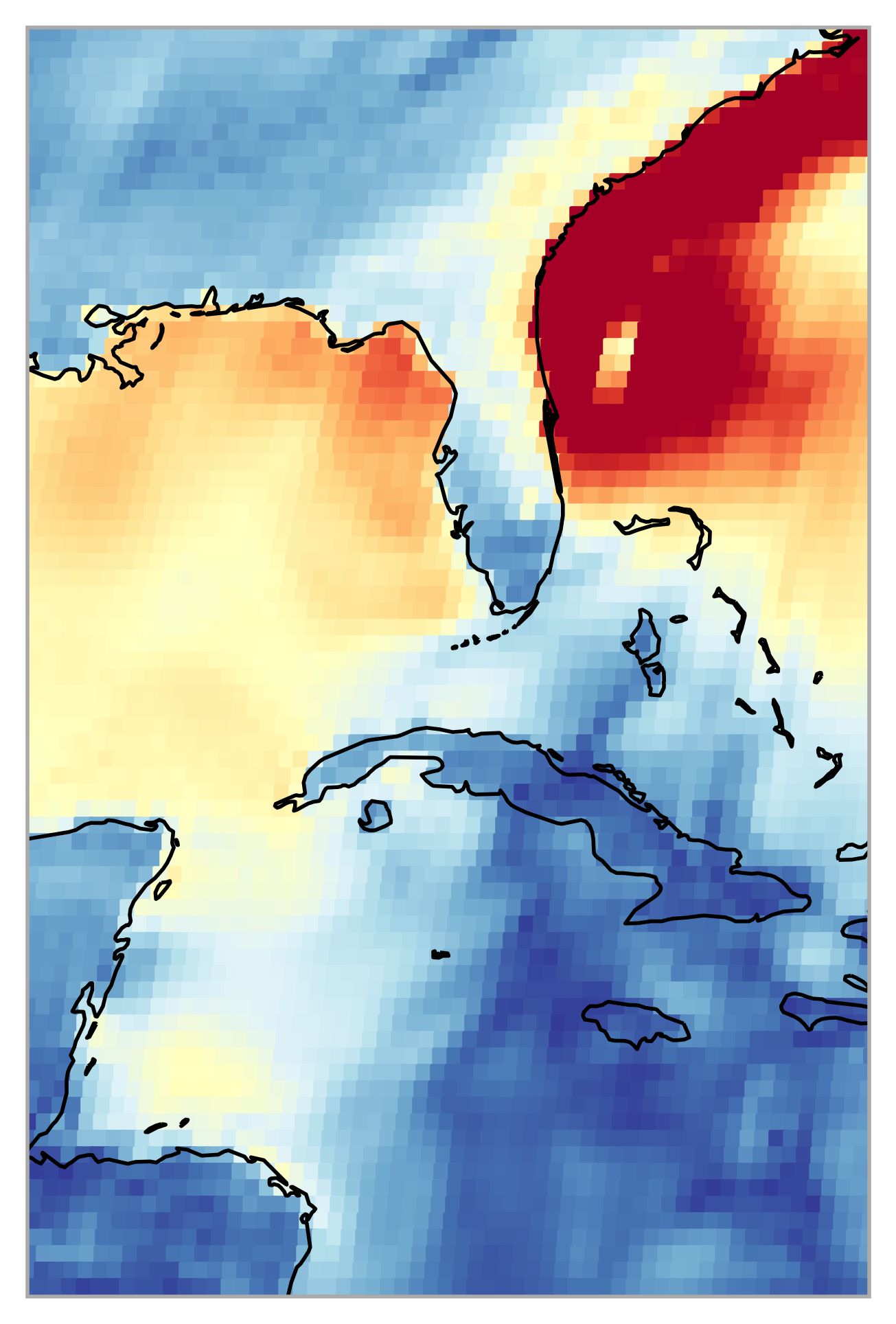}
        \includegraphics[trim=0 6 35 5, clip, height=.21\linewidth]{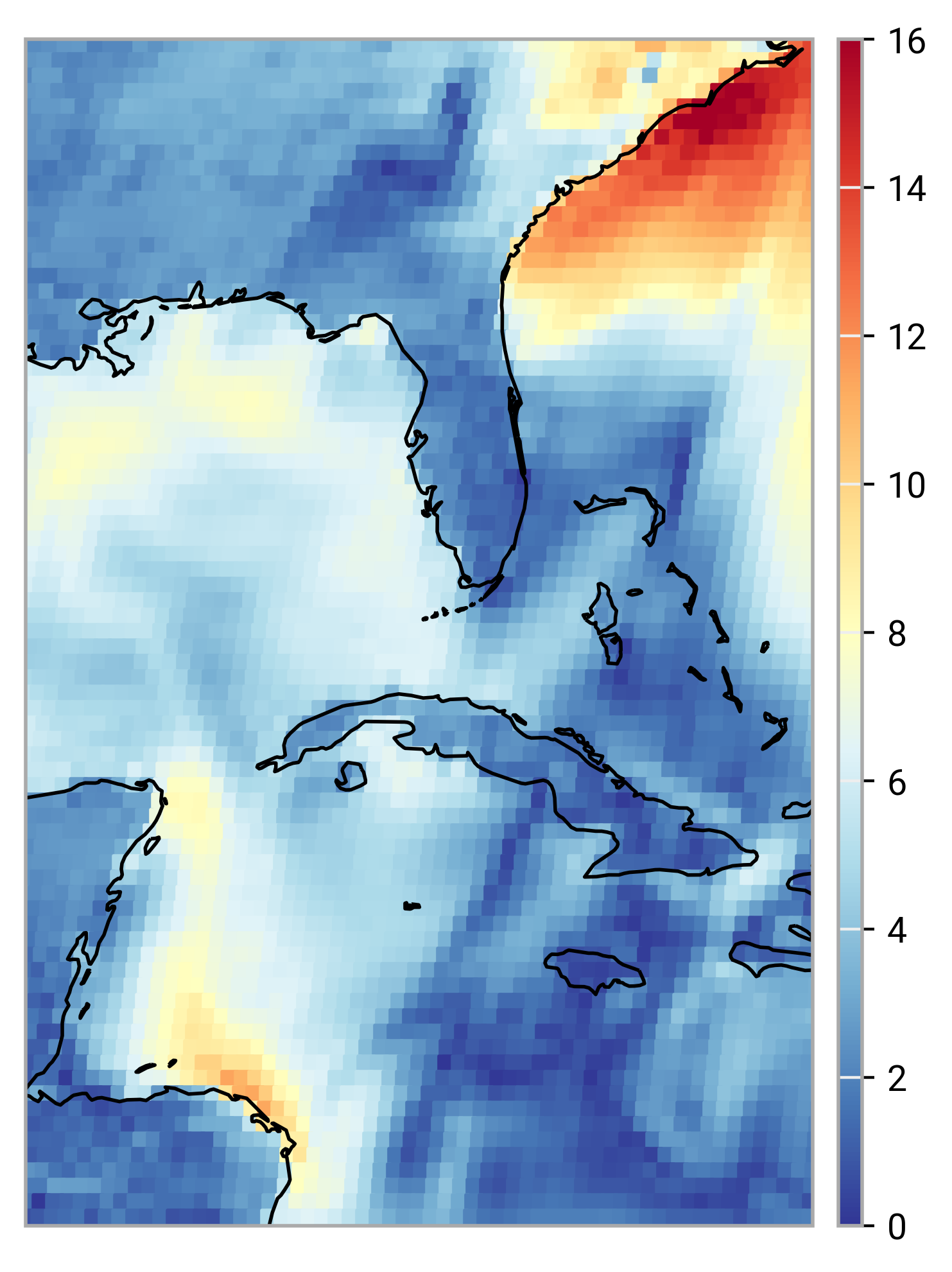}

        \vspace{0.3em}

        \includegraphics[trim=0 3 0 0, clip, height=.21\linewidth]{hurricane_ian/dop.jpg}
        \includegraphics[height=.21\linewidth]{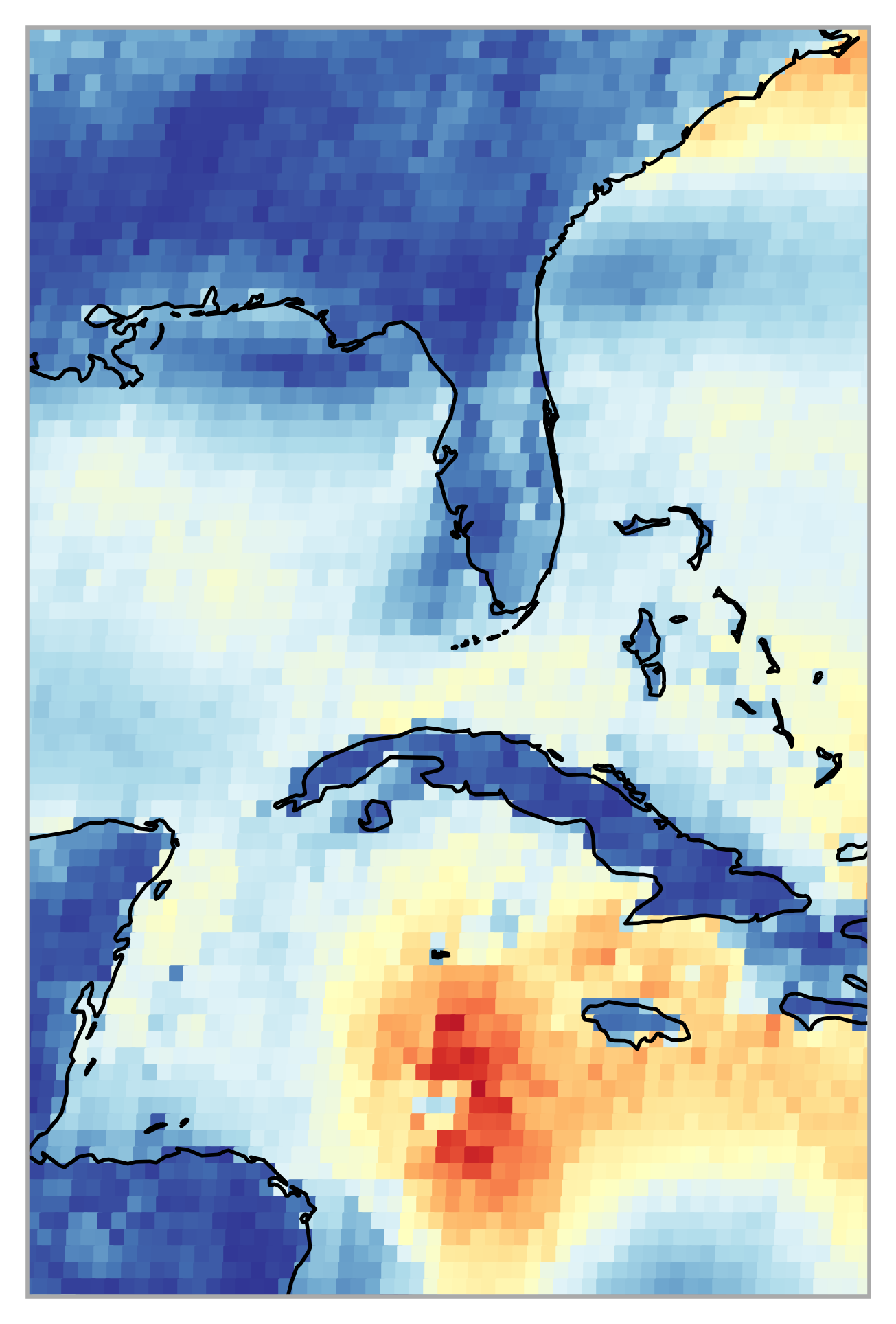}
        \includegraphics[height=.21\linewidth]{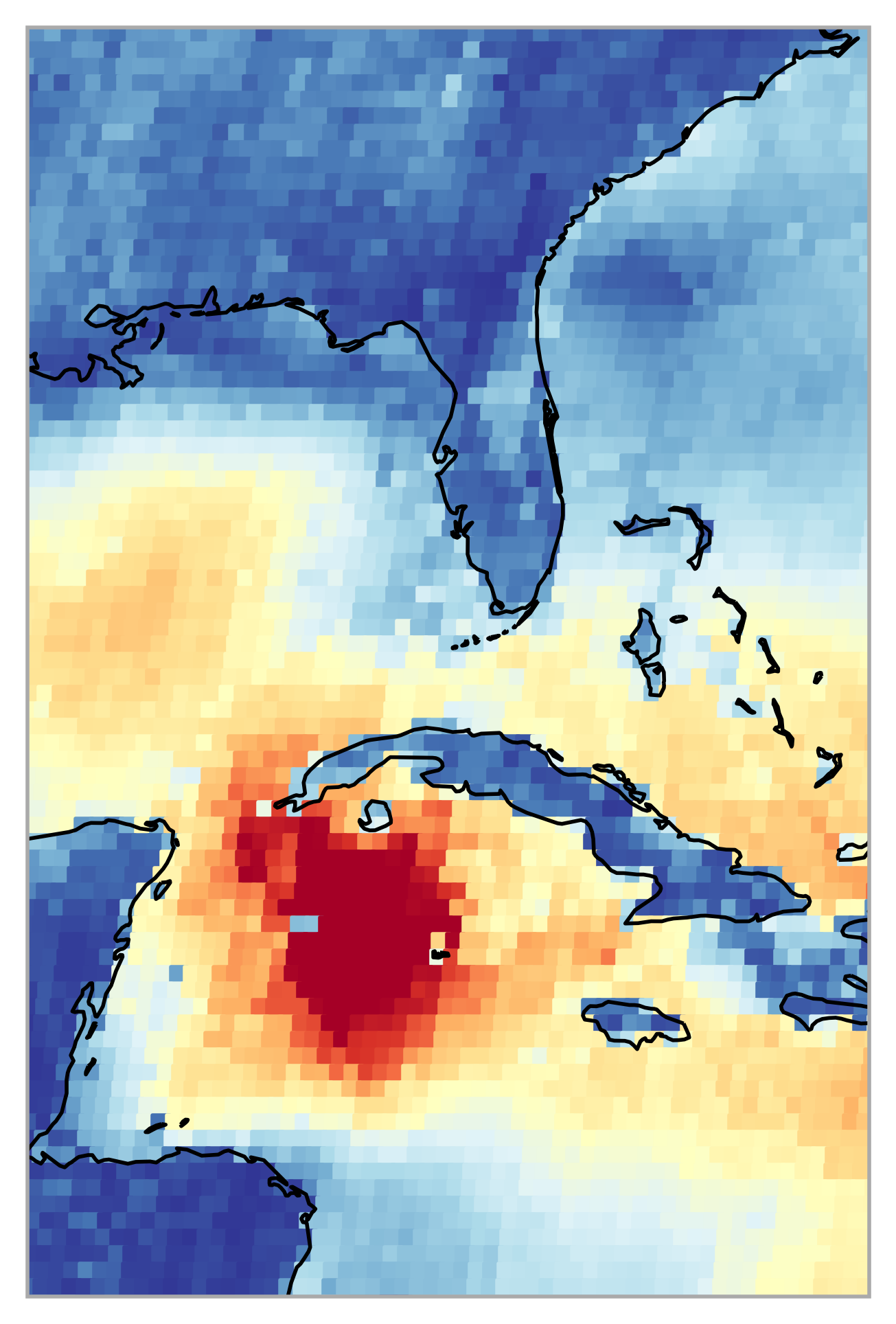}
        \includegraphics[height=.21\linewidth]{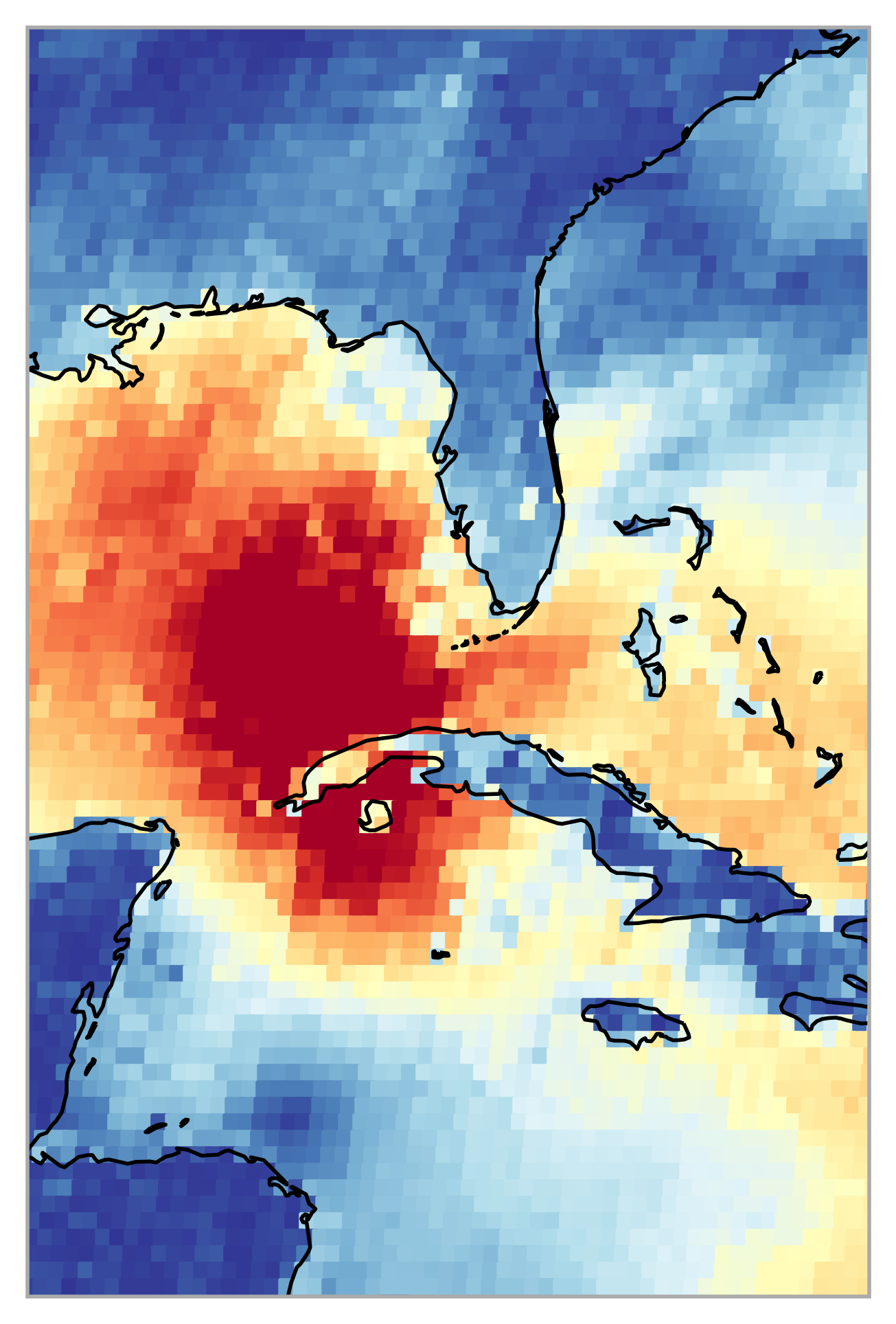}
        \includegraphics[height=.21\linewidth]{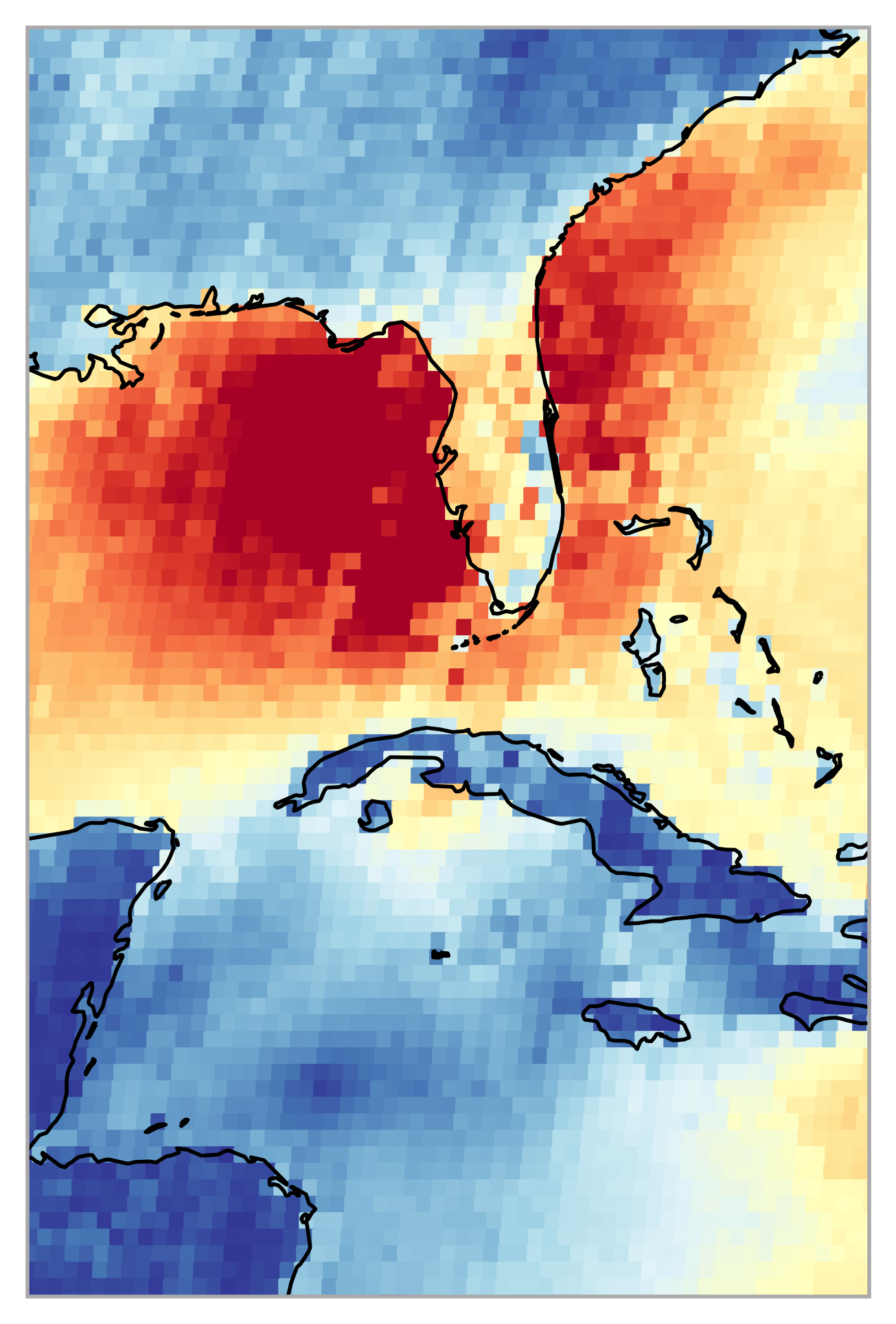}
        \includegraphics[height=.21\linewidth]{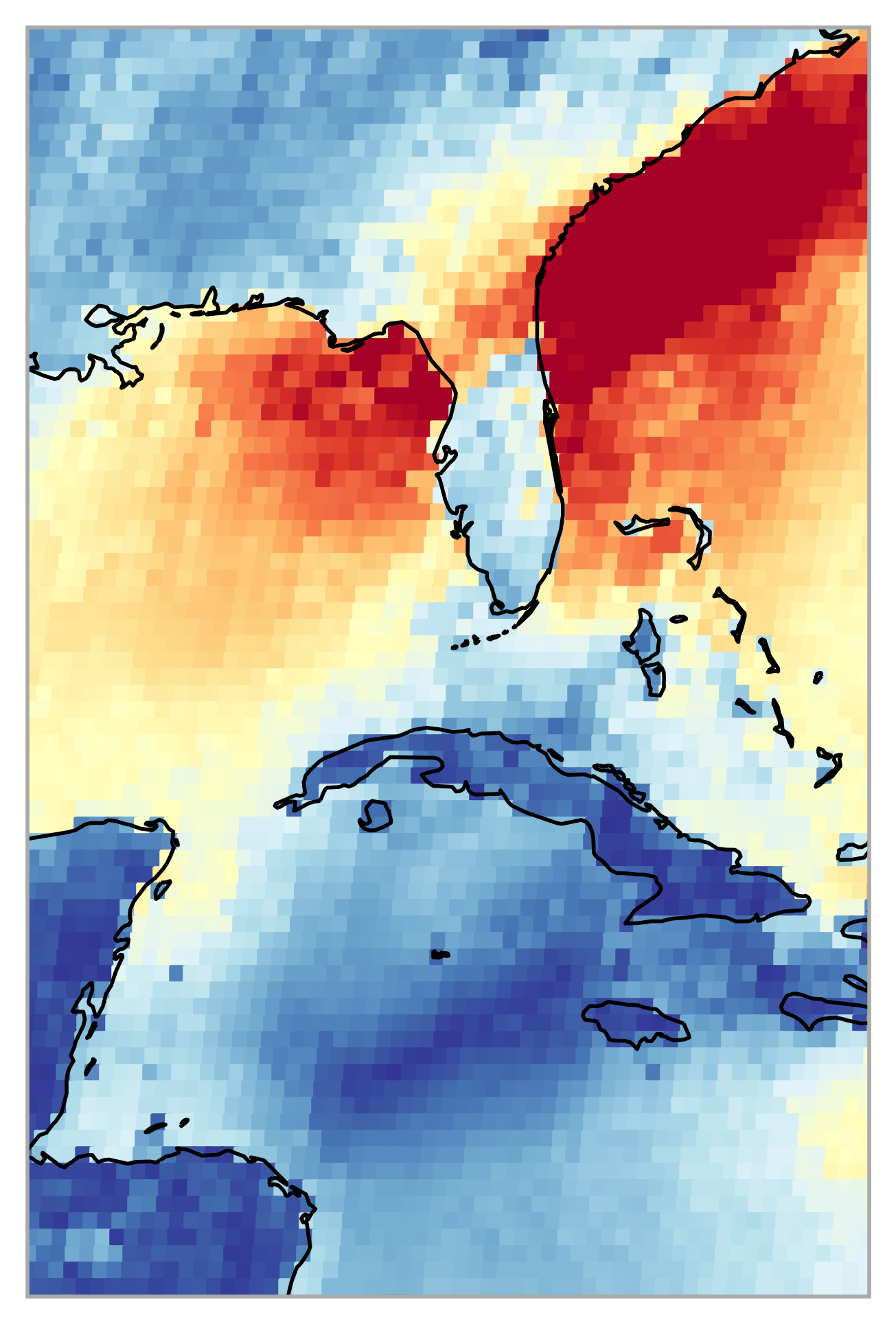}
        \includegraphics[trim=0 6 35 5, clip, height=.21\linewidth]{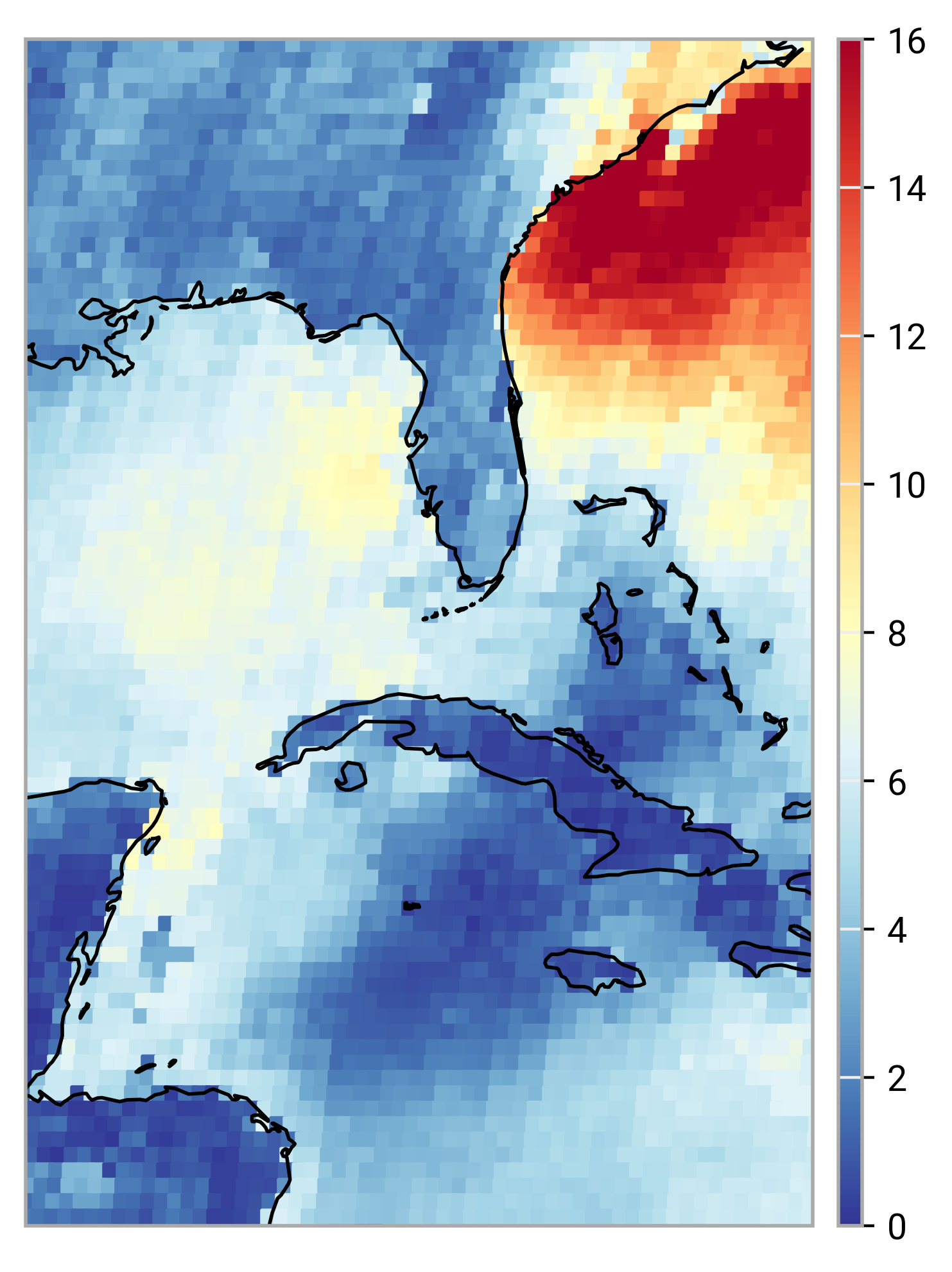}

        \end{minipage}
        &
        \begin{minipage}{0.03\linewidth}
        \includegraphics[height=4.5cm]{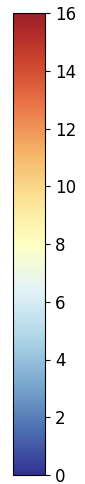}
        \end{minipage}
    \end{tabular}
}

\caption{ERA5 reanalysis (top row of each subfigure) and GraphDOP forecasts (bottom row) for (a) mean sea-level pressure and (b) 10~m wind speed, valid at 00:00~UTC from 26 September to 1 October 2022 (left to right). Forecasts were initialized with observations between 09:00~UTC and 21:00~UTC on 24 September 2022.}
\label{fig:ian}
\end{figure}

A GraphDOP forecast was initialized from 12 hours of observations spanning 24 September 2022, 09:00-21:00~UTC. Forecasts are evaluated at 00:00~UTC daily over the six-day period from 26 September to 1 October. Figures~\ref{fig:ian} and \ref{fig:ian_ocean} show the evolution of key storm variables: mean sea-level pressure (\ref{fig:ian}a) and 10-meter wind speed (\ref{fig:ian}b) for atmospheric states, and significant wave height (\ref{fig:ian_ocean}a), and sea-surface temperature (\ref{fig:ian_ocean}b) for the ocean, comparing ERA5 reanalysis (top rows) and GraphDOP N320 (approximately 0.25 degrees) forecasts (bottom rows). 

\subsection{Atmospheric fields}

The 10-meter wind fields reflect the storm's low-level intensity and structure, while the mean sea-level pressure captures the storm’s central depth and track. These atmospheric variables together reveal how well GraphDOP reconstructs the evolution of Ian’s core and circulation. 

Figures \ref{fig:ian}(a) and (b) show that GraphDOP is able to forecast the general trajectory and evolution of the hurricane's low-pressure centre. The model is also capturing the deepening of the central pressure, although it suffers from substantial underestimation of the intensity throughout. We believe this could be due to the limited number of buoys (that tend to avoid storms) or weather stations reporting surface pressure inside a storm. This highlights the importance of conventional observations for a system such as GraphDOP. Further work includes adding the International Best Track Archive for Climate Stewardship (IBTrACS) central pressure estimates into our surface pressure training datasets. 

Aside from this, finer details are visible in the wind speed forecasts in Figure \ref{fig:ian}(b). The first two panels clearly show the hurricane’s eye as a central minimum (blue pixel) surrounded by high wind speeds, demonstrating the model’s capacity to capture coherent storm structures even several days after initialisation (we recall here that the forecast was initialised 2 days prior to the first panel). 

Beyond the storm core, GraphDOP also captures broad-scale atmospheric flow features, including the sharp contrast between wind speeds over land and ocean. As the storm progresses northward and begins to interact with the US coastline, the model reproduces the weakening of wind speeds inland and maintains stronger gradients along the coastlines. This land–sea contrast is consistent with physical expectations, as surface roughness and frictional effects reduce wind speeds over land. The fact that these patterns are present-even though GraphDOP is not explicitly physics-based-suggests the model has learned to represent surface boundary interactions through observational data alone. 

\begin{figure}[htpb]
\centering
\includegraphics[width=.3\linewidth]{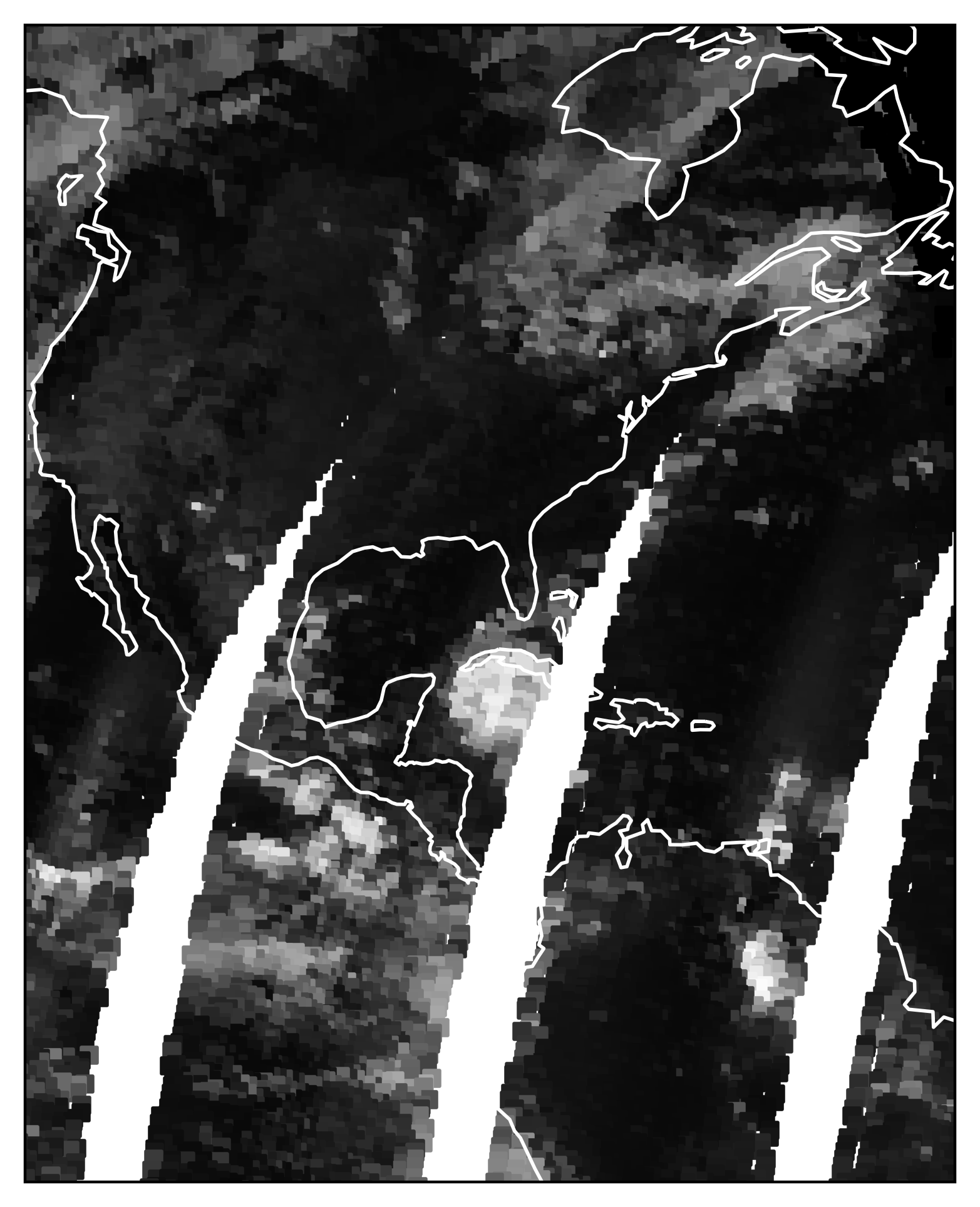}
\includegraphics[width=.3\linewidth]{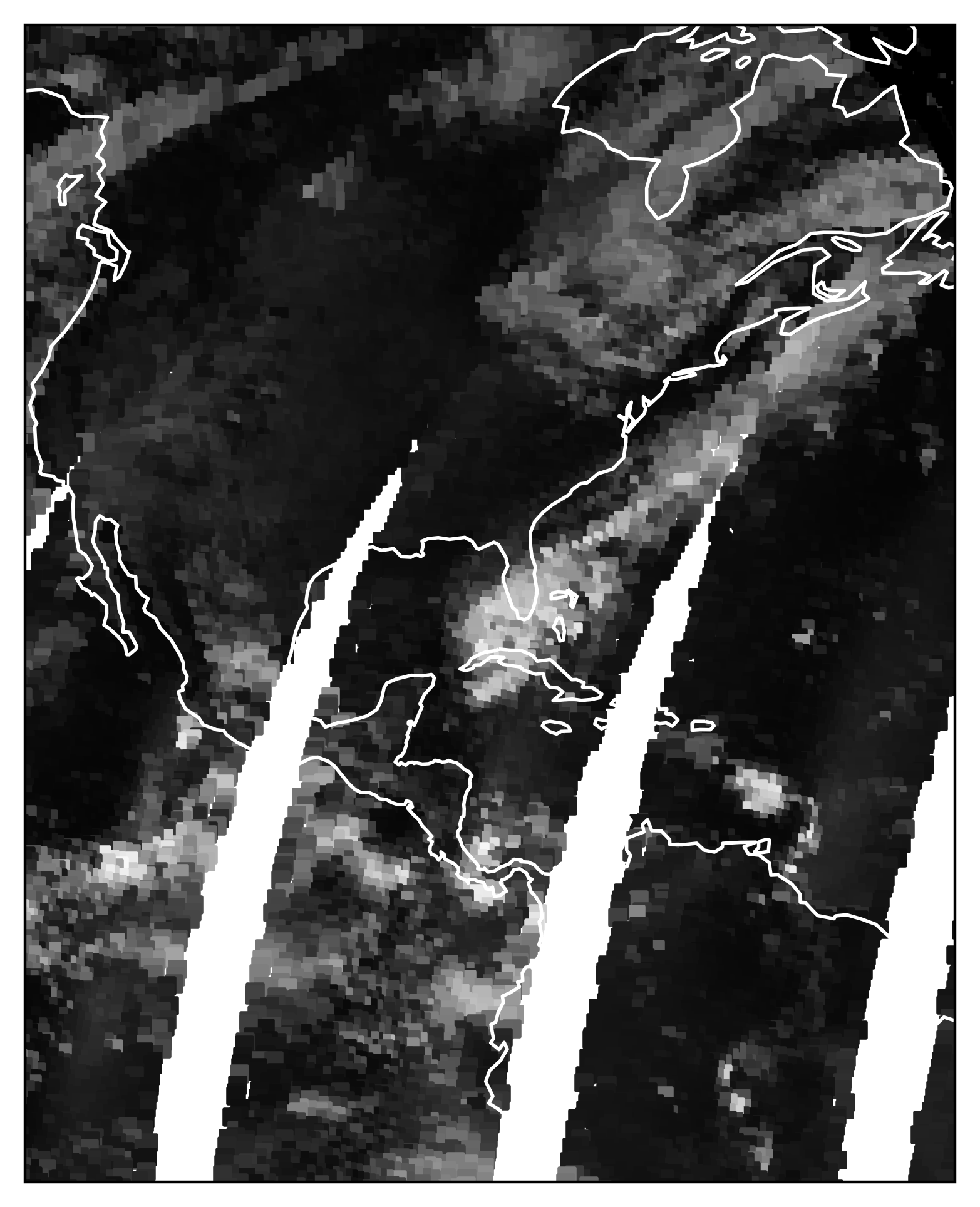}
\includegraphics[width=.3\linewidth]{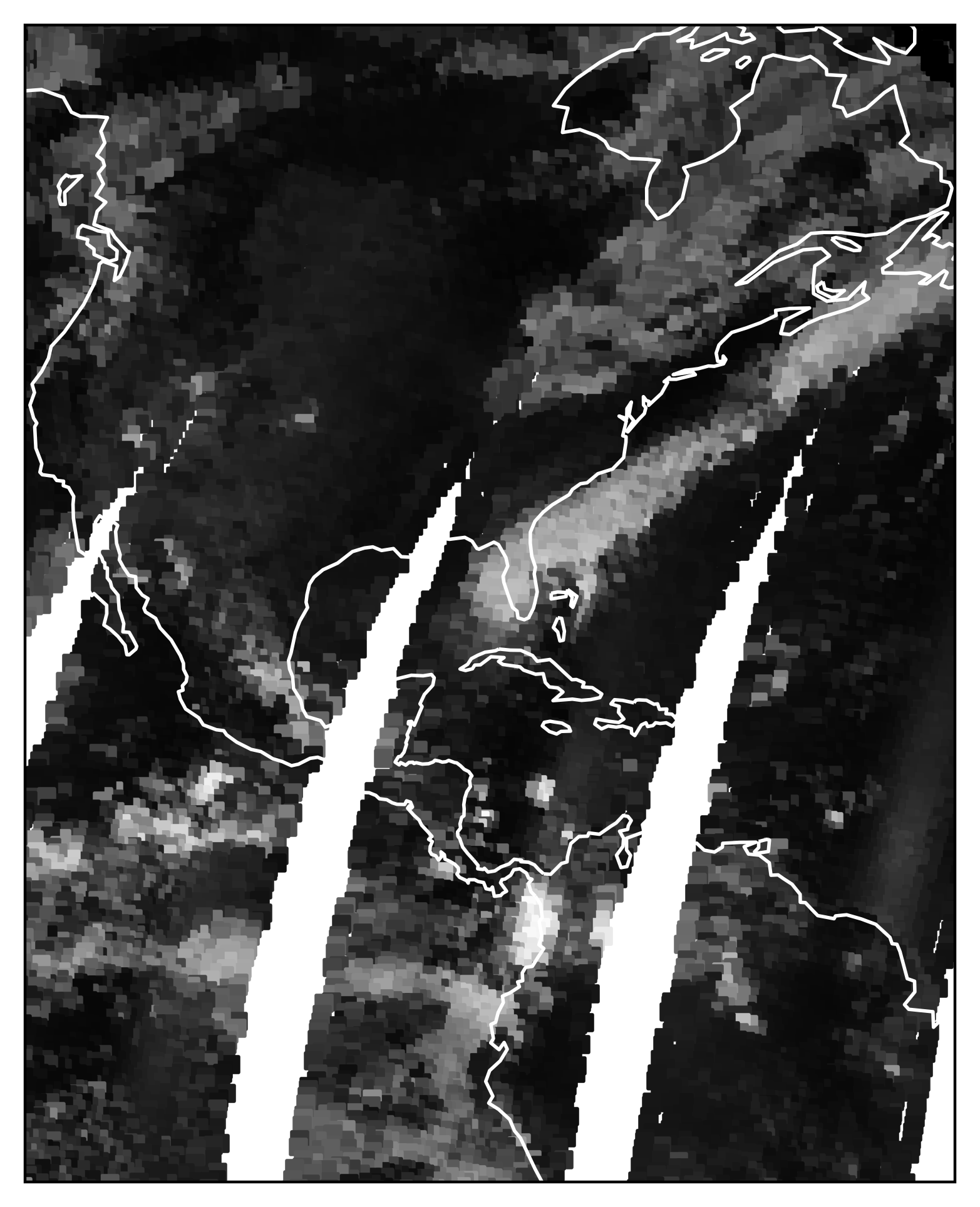} \\

\includegraphics[width=.3\linewidth]{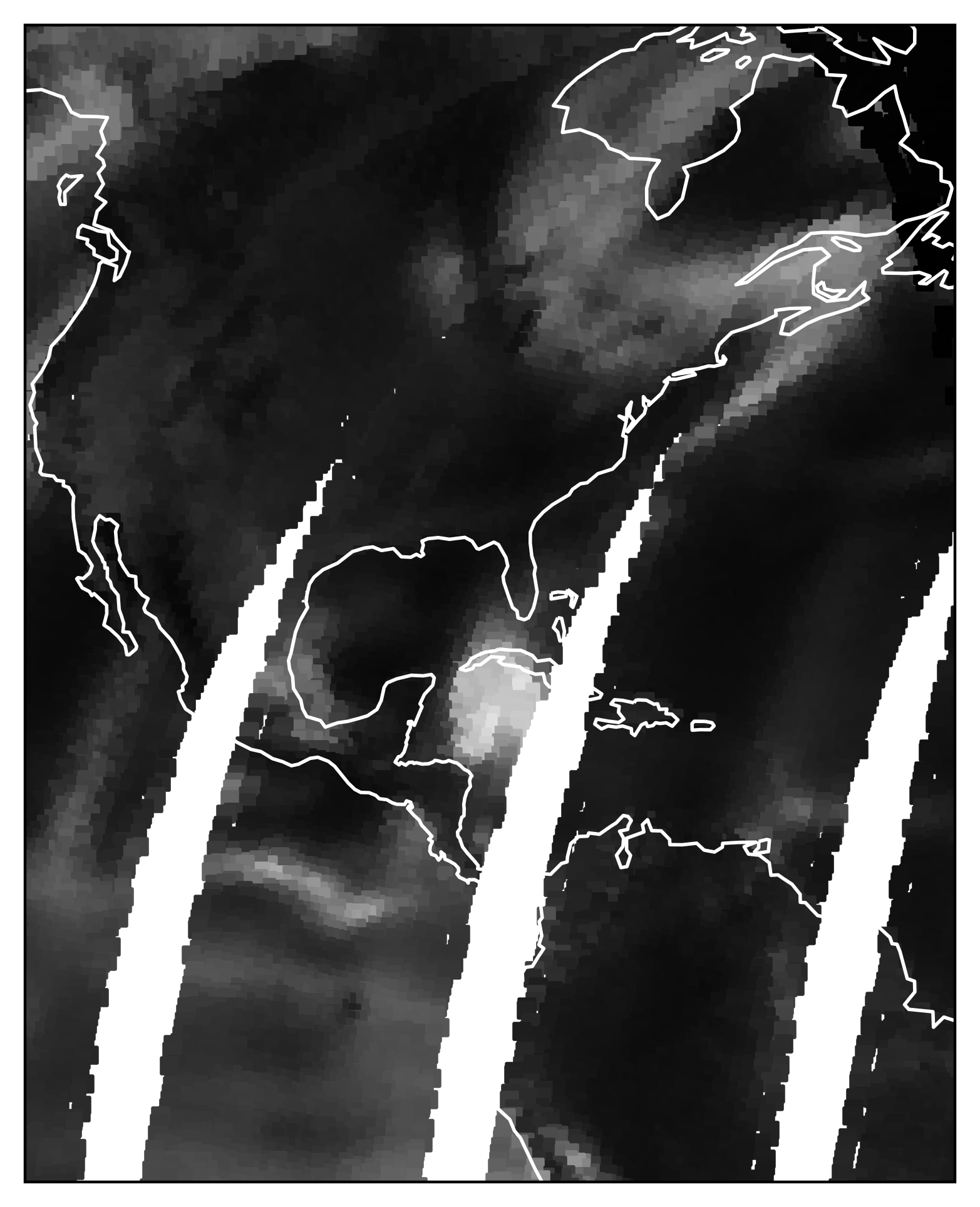}
\includegraphics[width=.3\linewidth]{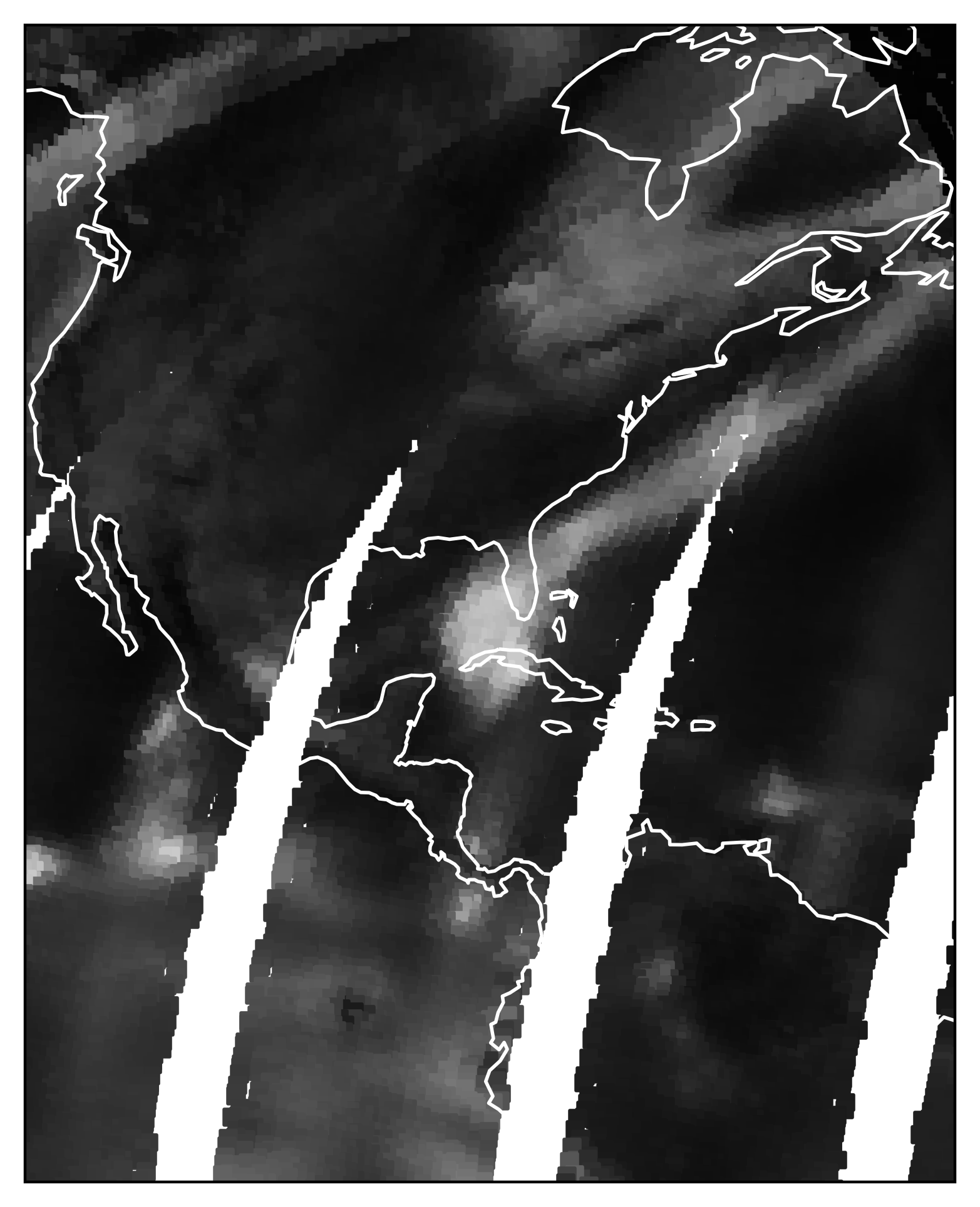}
\includegraphics[width=.3\linewidth]{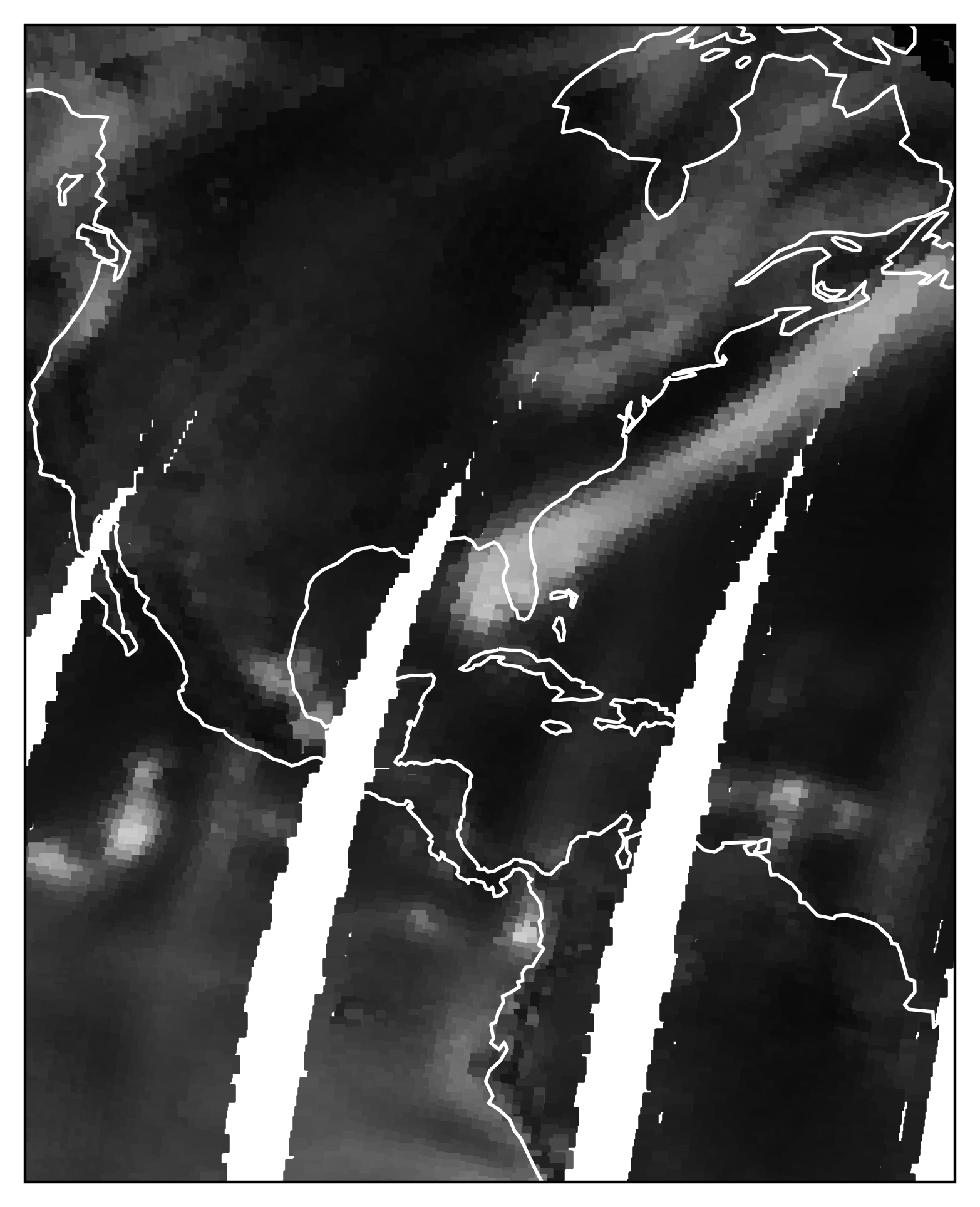}

\caption{Observed (top row) and forecasted (bottom row) AVHRR visible reflectance at 2, 3 and 4-day lead times, valid between 26 and 28 September 2022.}
\label{fig:avhrr}
\end{figure}

Additional evidence supporting the model’s accurate reconstruction of Hurricane Ian’s evolution comes from satellite observations. Figure~\ref{fig:avhrr} presents AVHRR visible channel co-located with IASI pixels (top row), alongside corresponding GraphDOP forecasts (bottom row), over the first three days of the storm (i.e., 26 to 28 September 2022). The observed cloud structures align well with the wind and pressure fields produced by GraphDOP. Importantly, the cloud field advances northward in both observation and forecast, consistent with the trajectory seen in pressure and wind fields. The structural realism in the simulated cloud patterns, despite GraphDOP being trained without explicit radiative or cloud physics, suggests the model has implicitly learned these features from the underlying dynamics captured in the training observations. While in some regions further from the storm, such as over the Pacific, the resemblance to observations is less distinct, the high fidelity achieved in reconstructing Ian’s development illustrates the model’s strong potential for capturing the dynamics of impactful weather systems.

\subsection{Ocean surface and wave response}

\begin{figure}[h!tb]
\centering

\subfigure[Significant wave height (m)]{
    \begin{tabular}{c @{} c} 
        \begin{minipage}{0.9\linewidth}
        \centering
        \includegraphics[trim=0 3 0 0, clip, height=.21\linewidth]{hurricane_ian/era.jpg}
        \includegraphics[height=.21\linewidth]{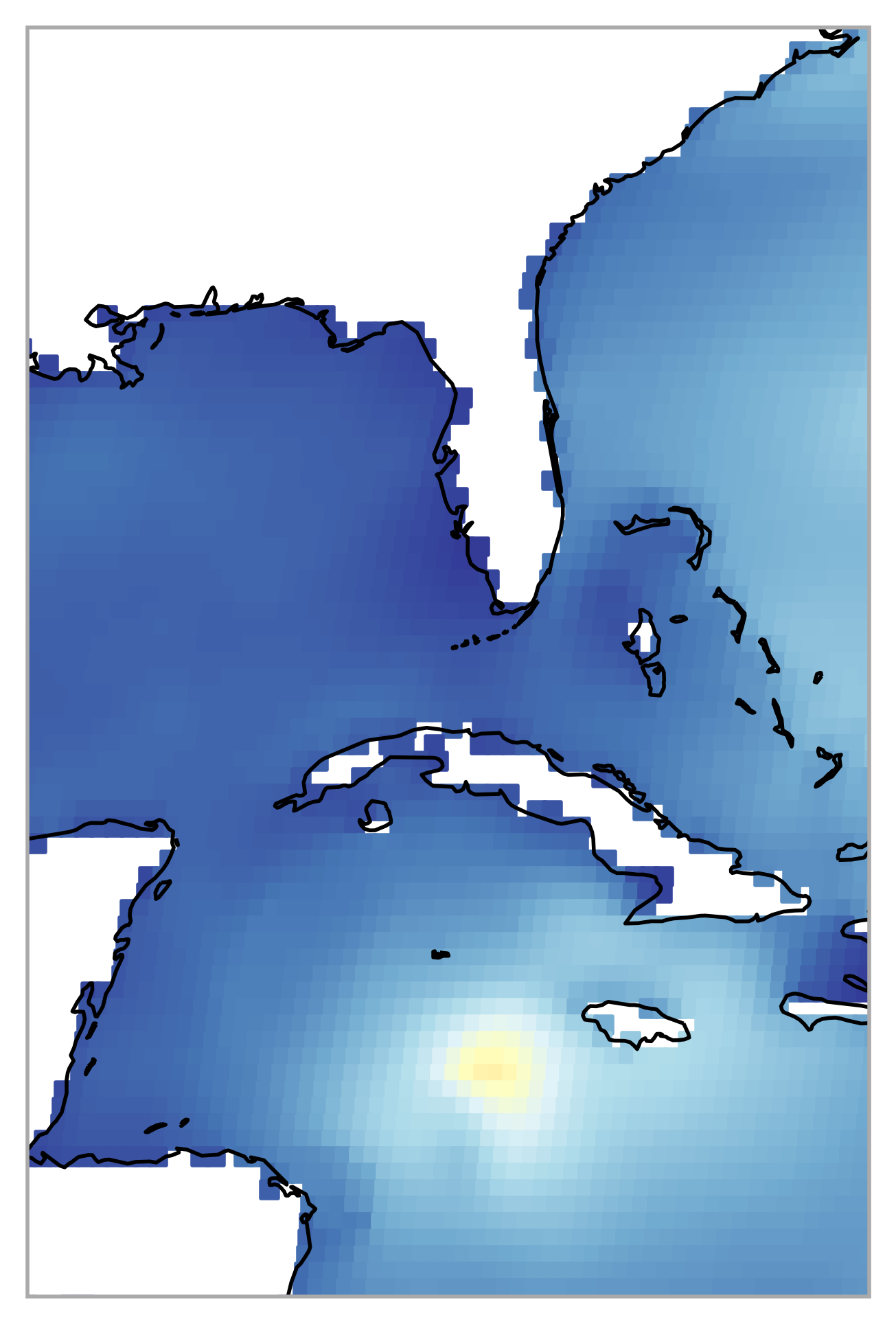}
        \includegraphics[height=.21\linewidth]{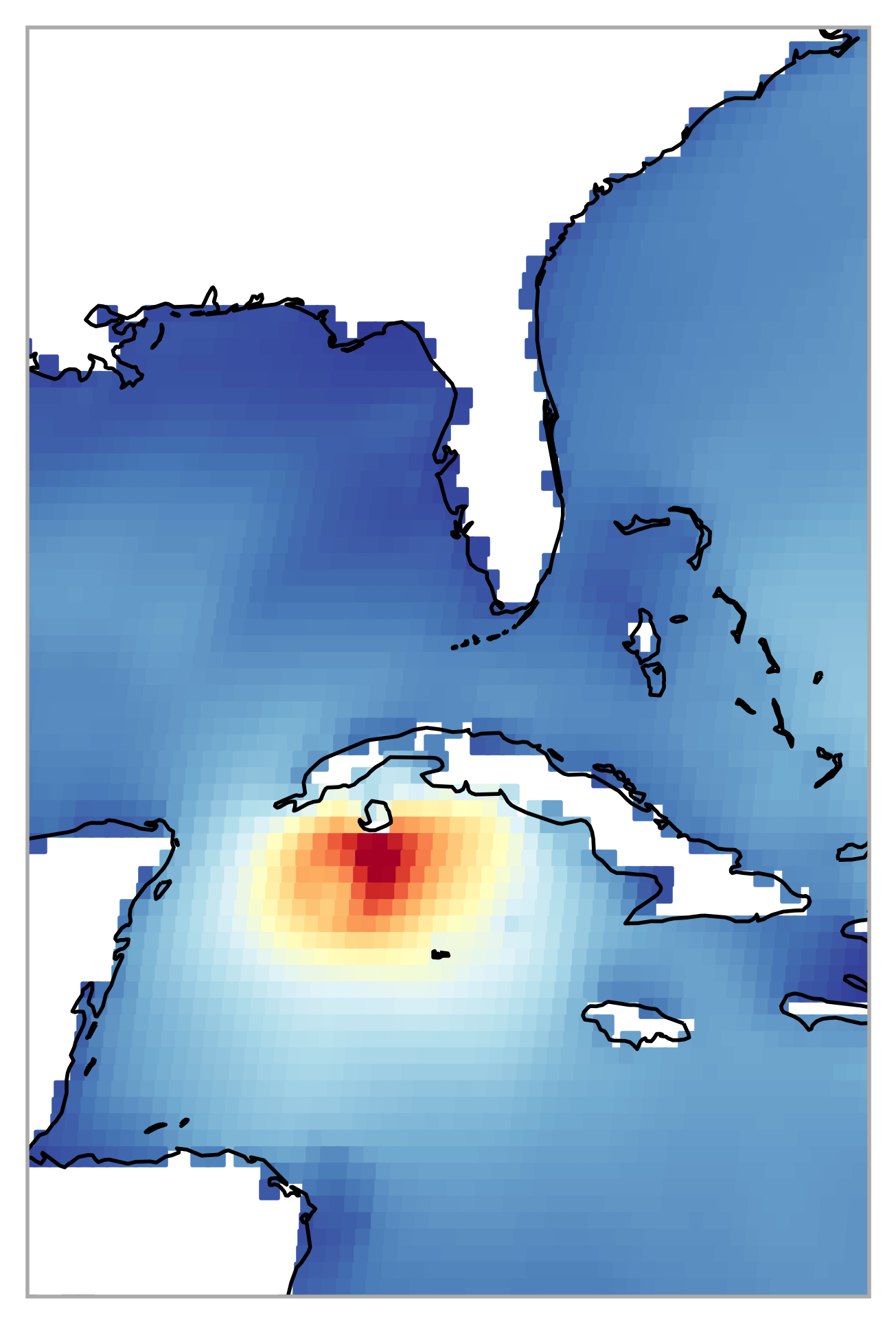}
        \includegraphics[height=.21\linewidth]{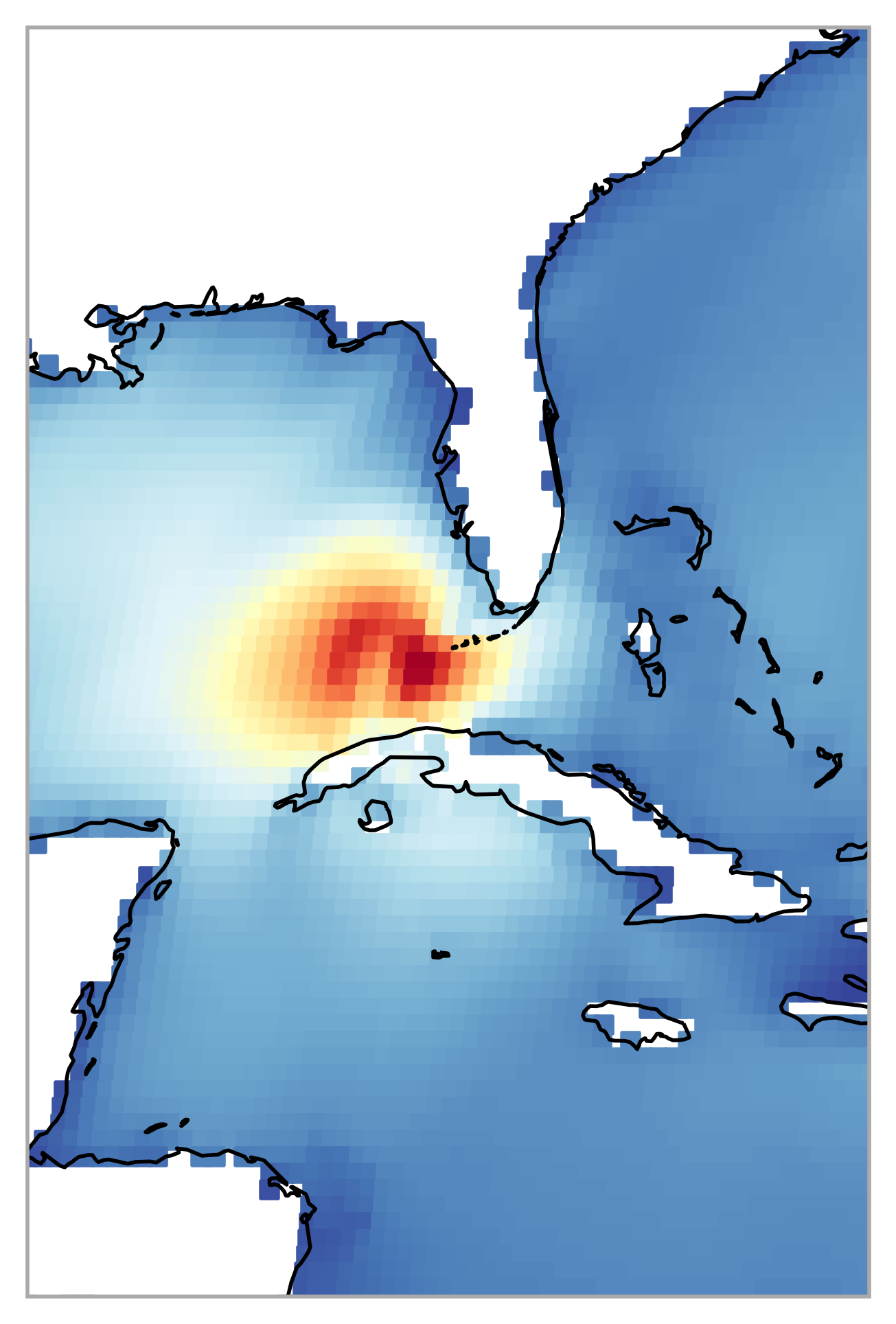}
        \includegraphics[height=.21\linewidth]{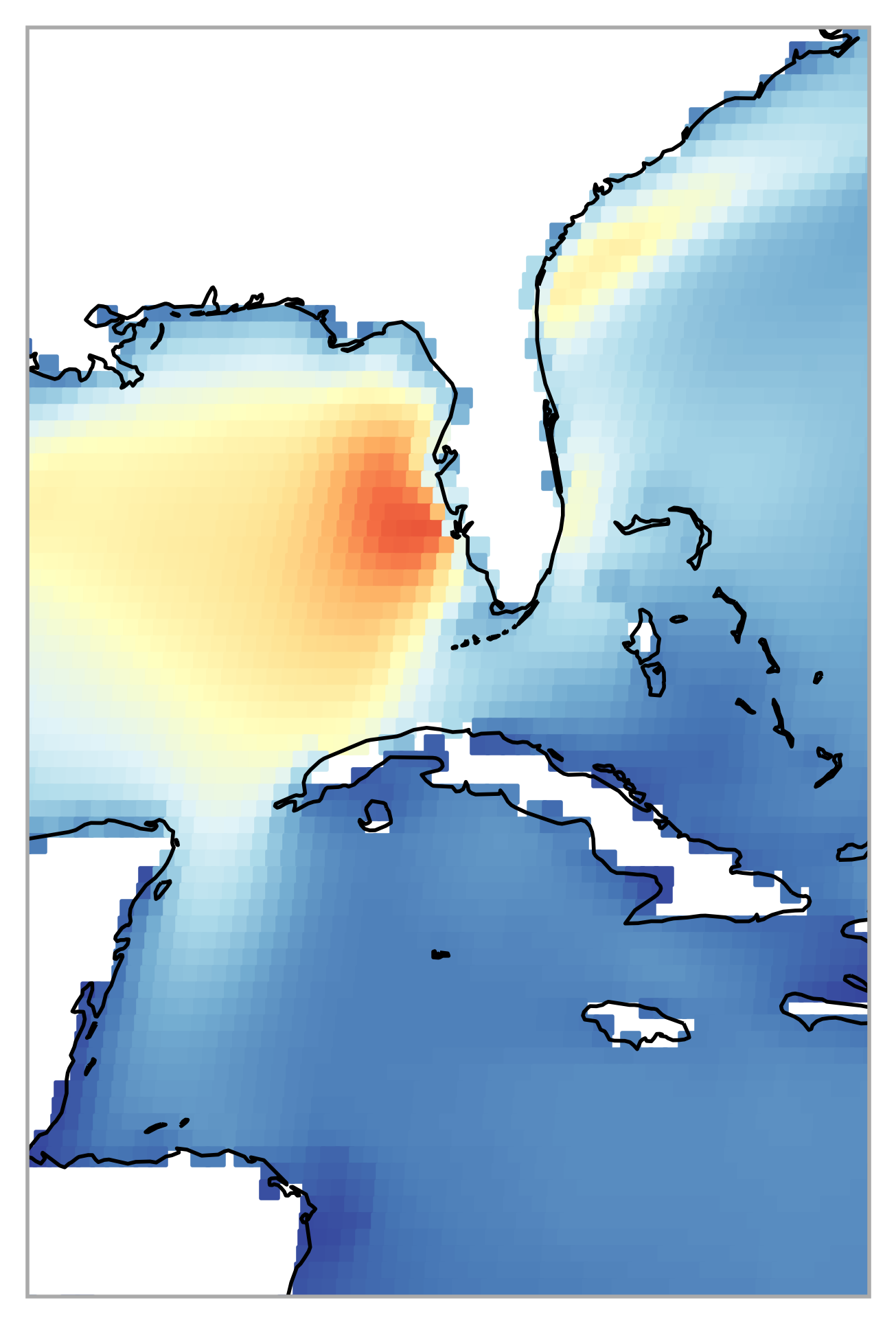}
        \includegraphics[height=.21\linewidth]{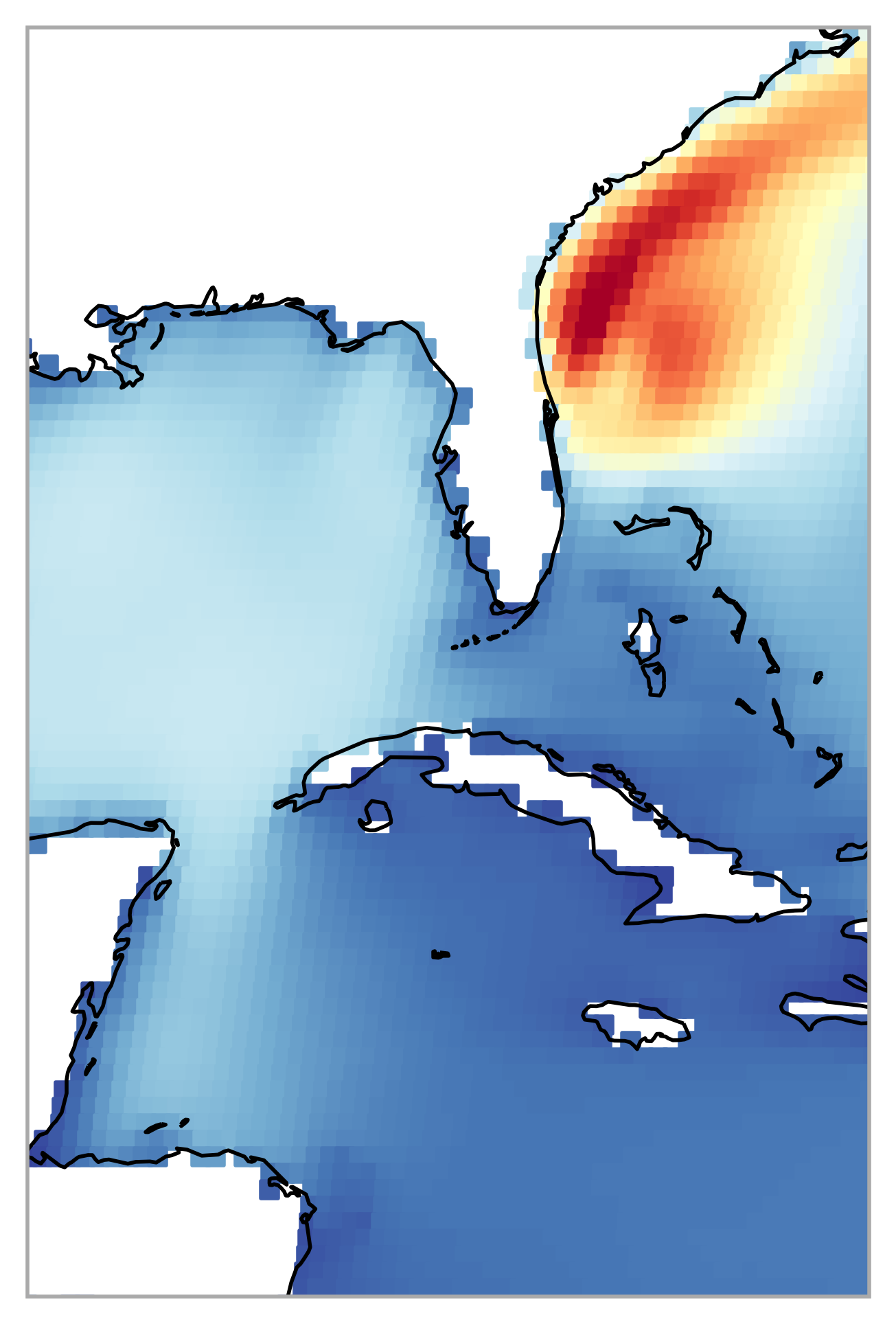}
        \includegraphics[trim=0 4 30 5, clip, height=.21\linewidth]{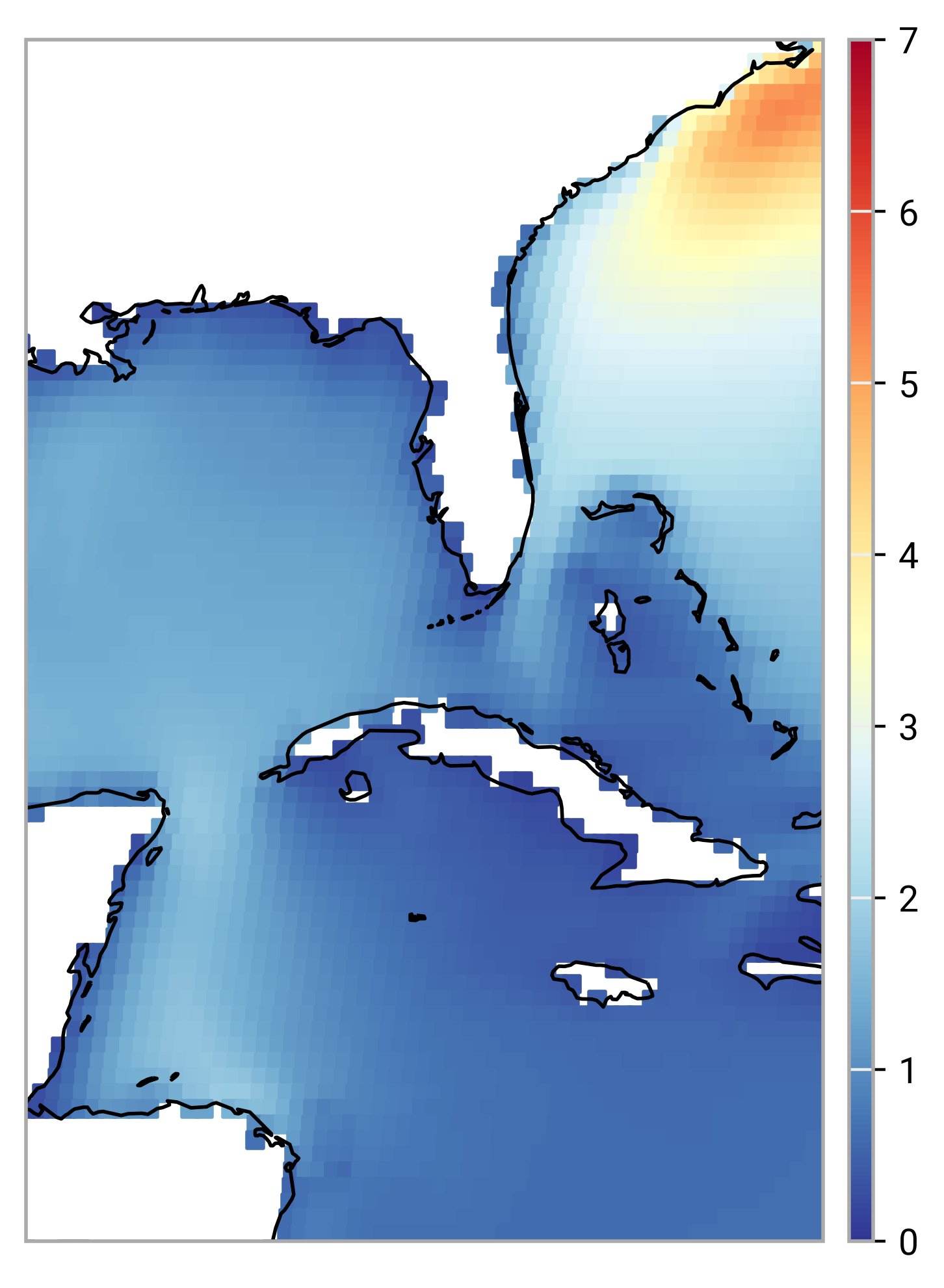}

        \vspace{0.3em}

        \includegraphics[trim=0 3 0 0, clip, height=.21\linewidth]{hurricane_ian/dop.jpg}
        \includegraphics[height=.21\linewidth]{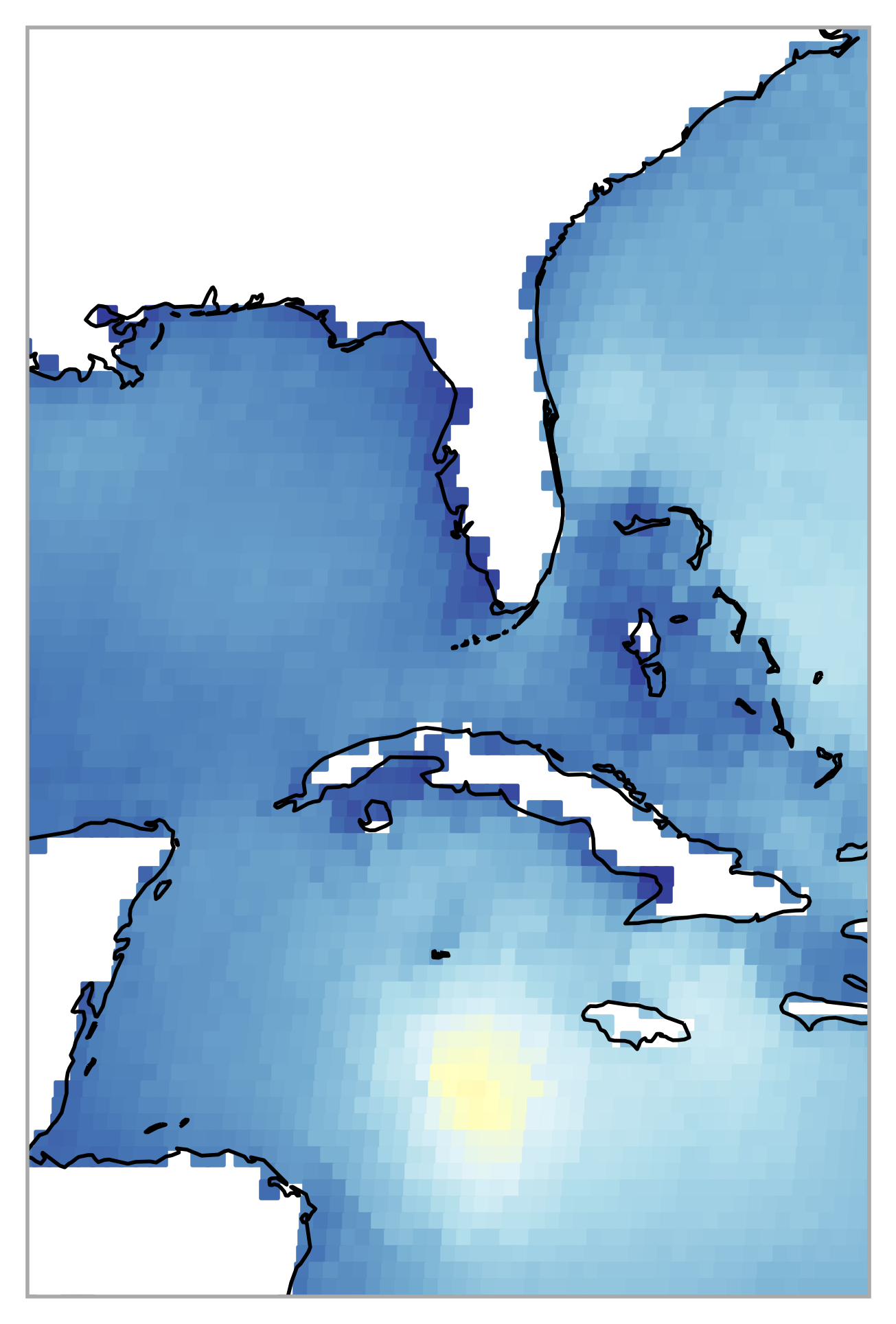}
        \includegraphics[height=.21\linewidth]{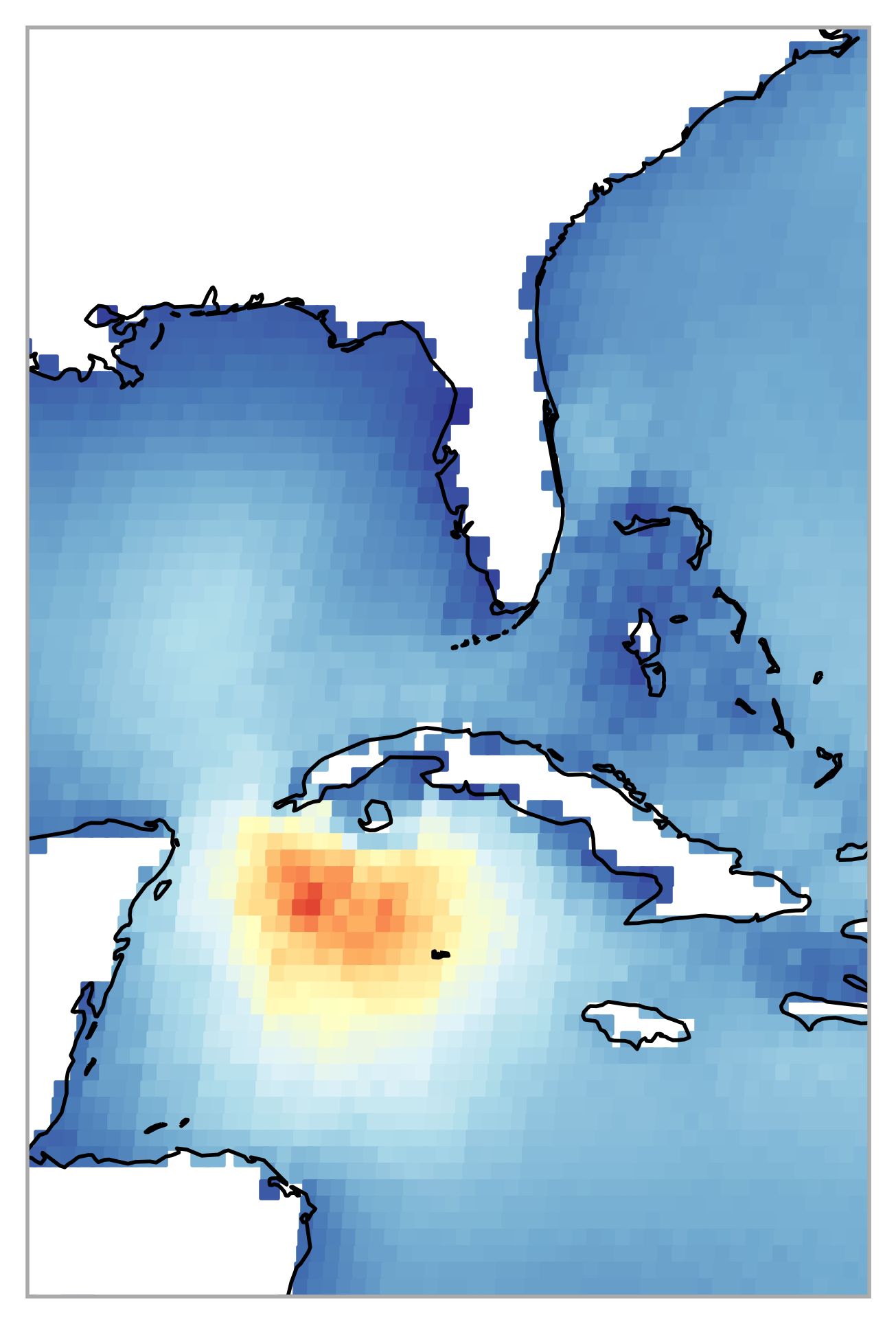}
        \includegraphics[height=.21\linewidth]{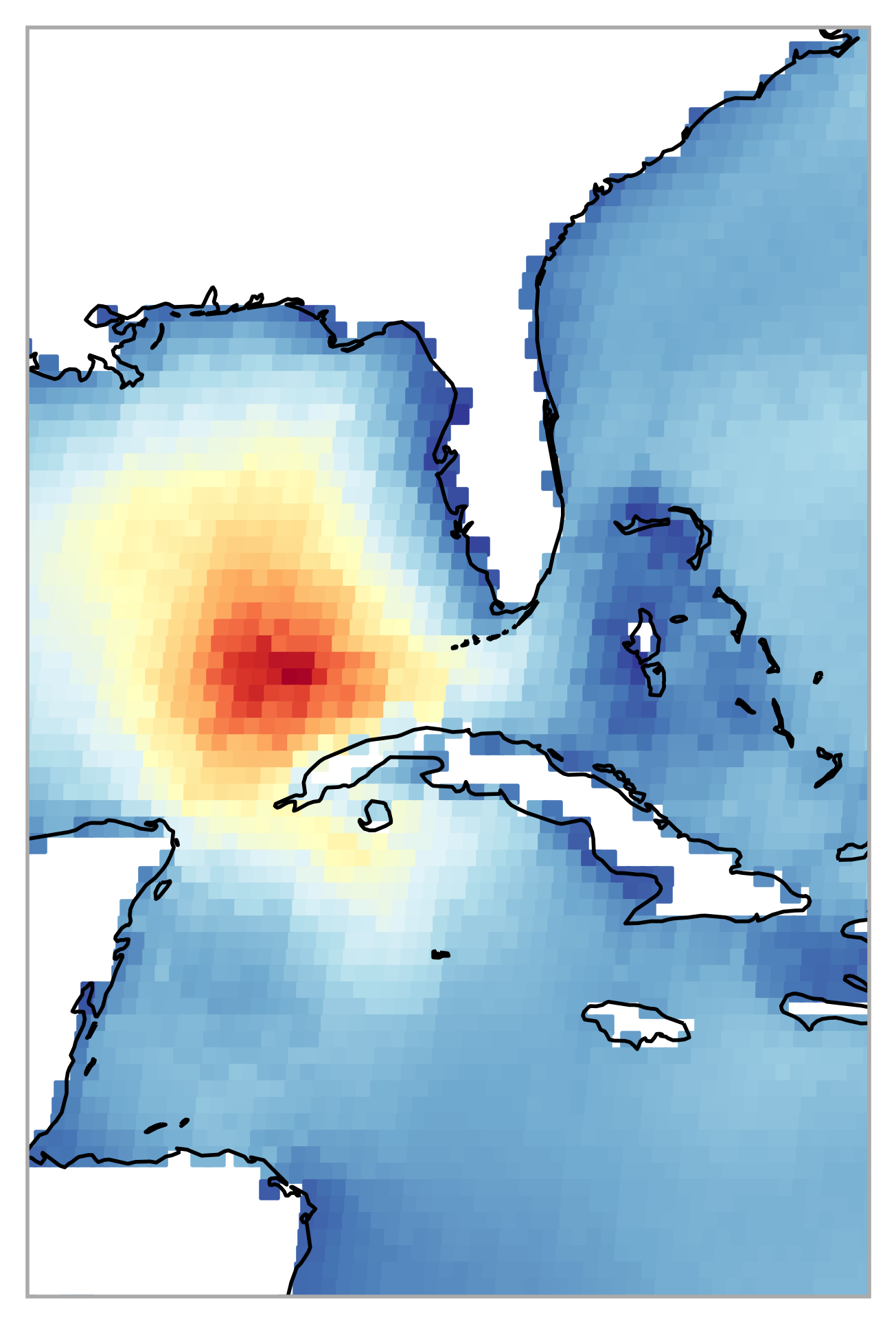}
        \includegraphics[height=.21\linewidth]{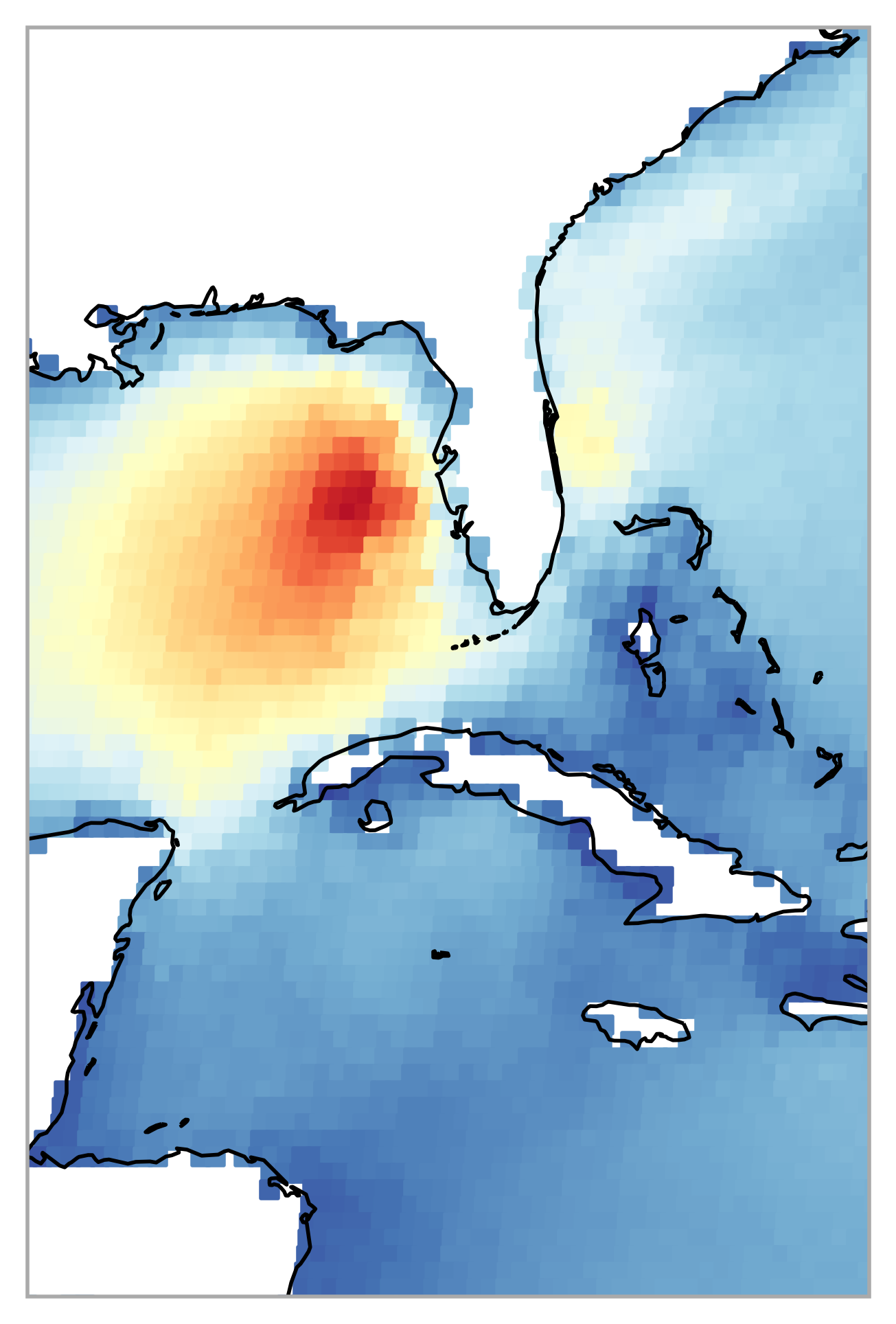}
        \includegraphics[height=.21\linewidth]{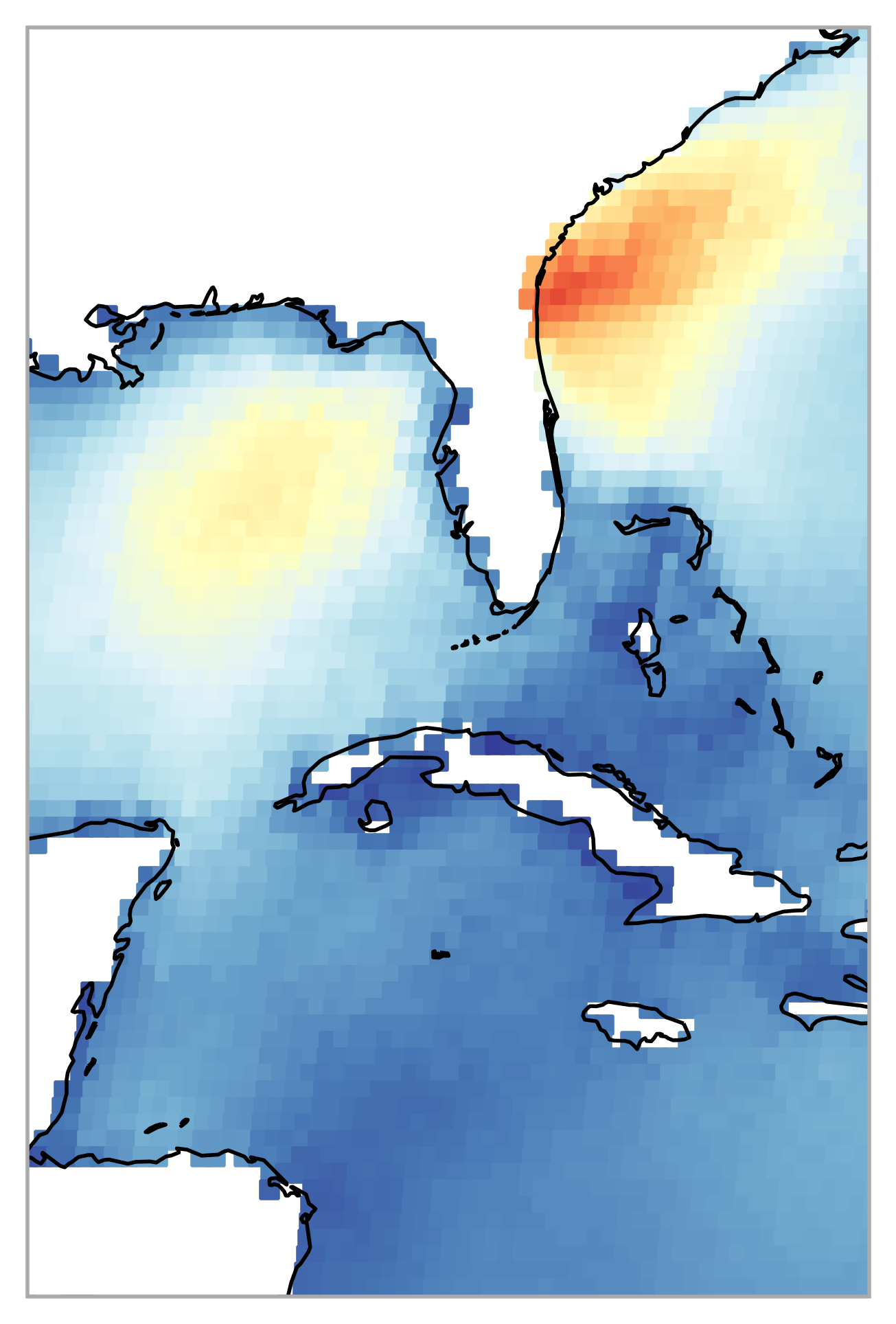}
        \includegraphics[trim=0 4 30 5, clip, height=.21\linewidth]{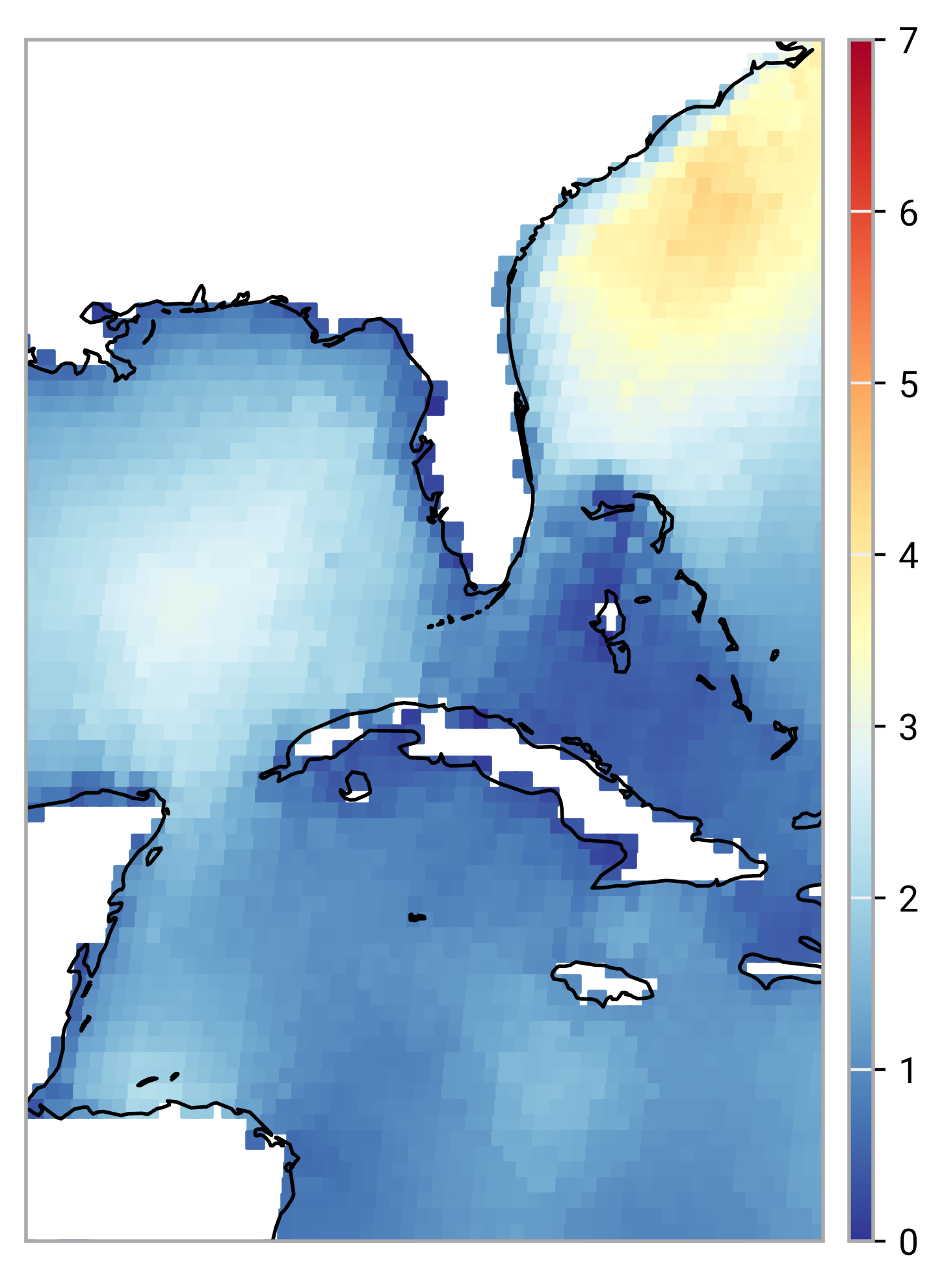}
        \end{minipage}
        &
        \begin{minipage}{0.03\linewidth}
        \includegraphics[height=4.5cm]{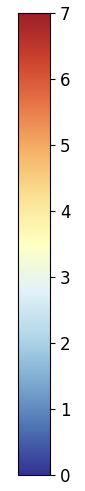}
        \end{minipage}
    \end{tabular}
}

\hspace{1cm} 

\subfigure[Sea surface temperature (K)]{
    \begin{tabular}{c @{} c}
        \begin{minipage}{0.9\linewidth}
        \centering
        \includegraphics[trim=0 3 0 0, clip, height=.21\linewidth]{hurricane_ian/era.jpg}
        \includegraphics[height=.21\linewidth]{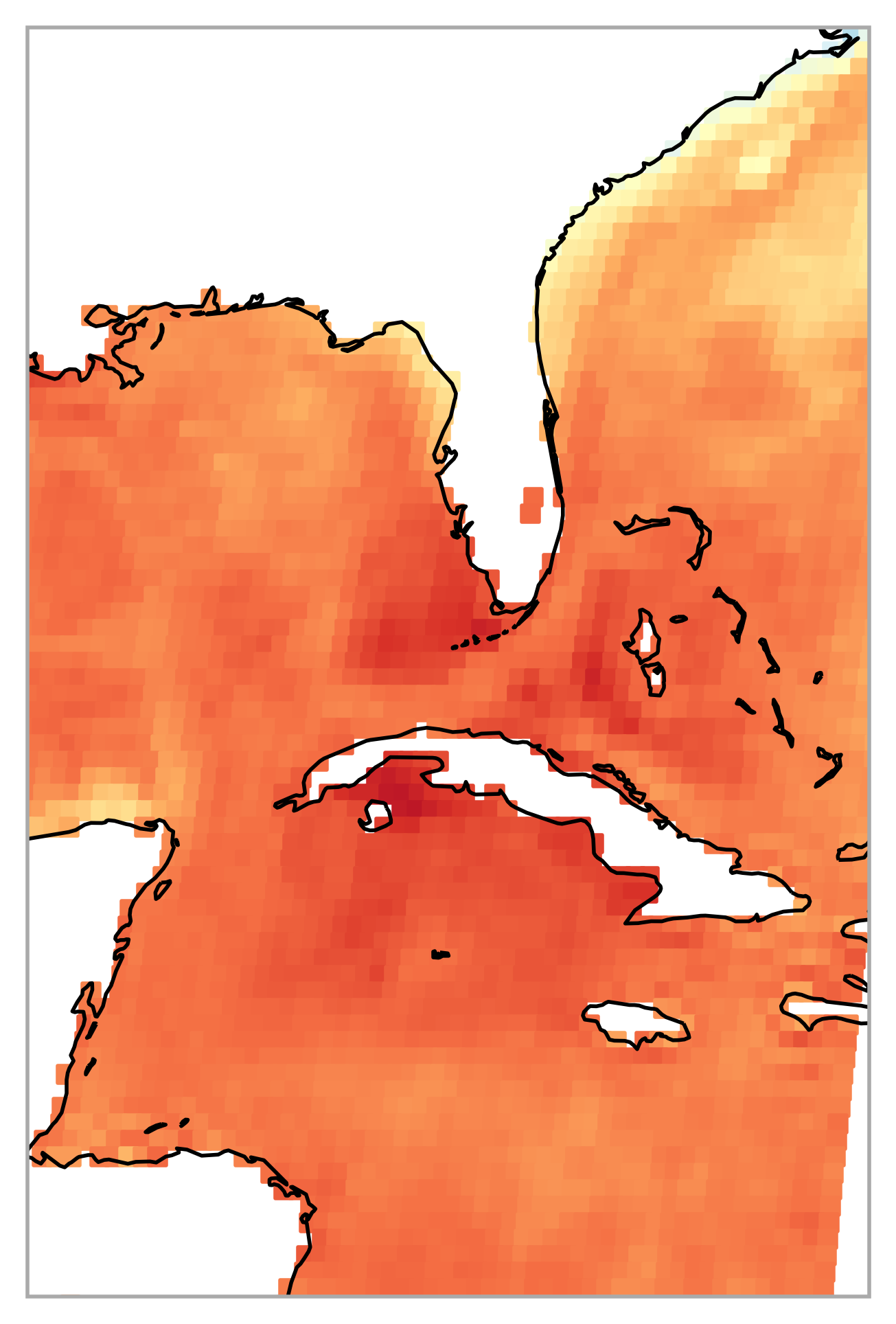}
        \includegraphics[height=.21\linewidth]{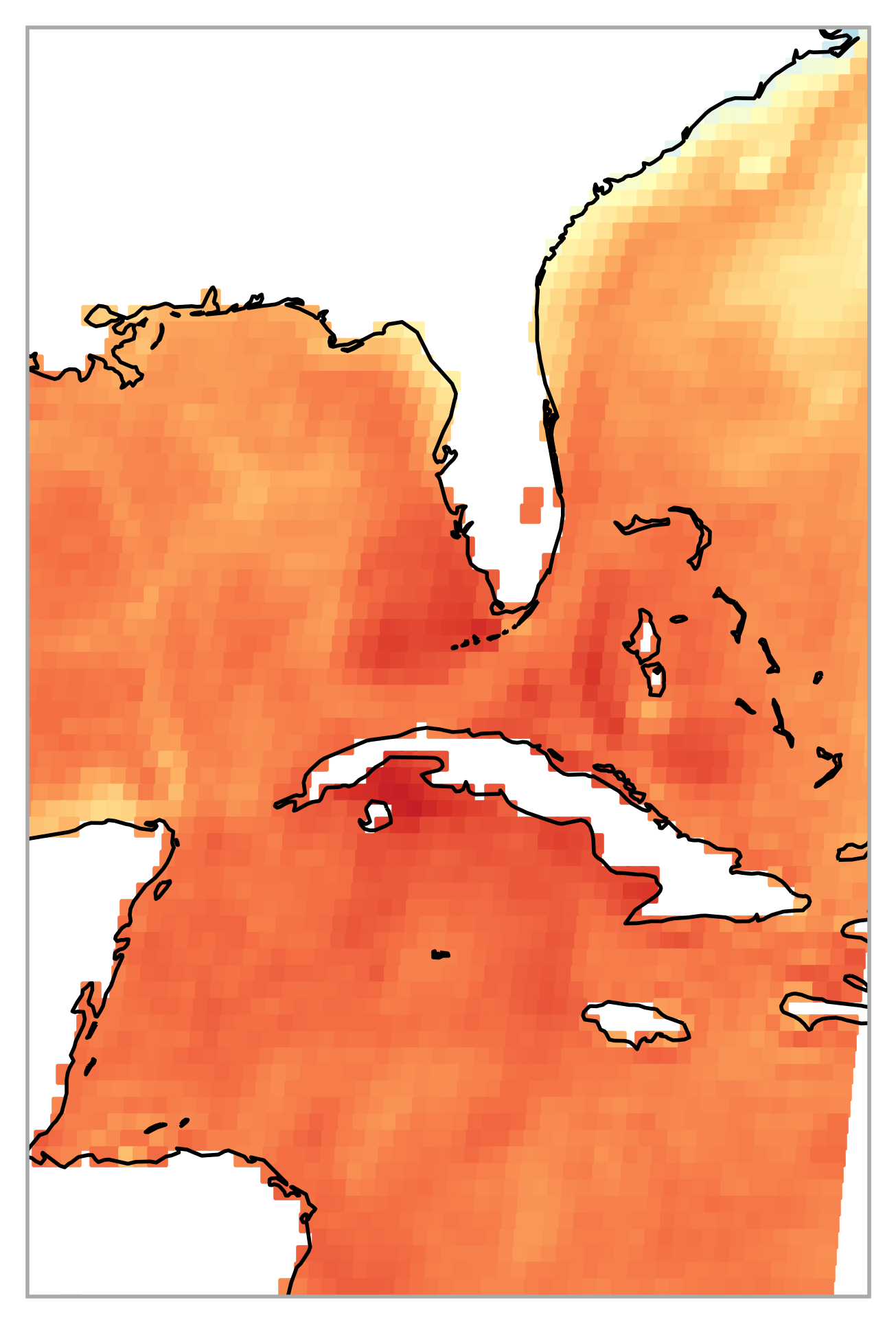}
        \includegraphics[height=.21\linewidth]{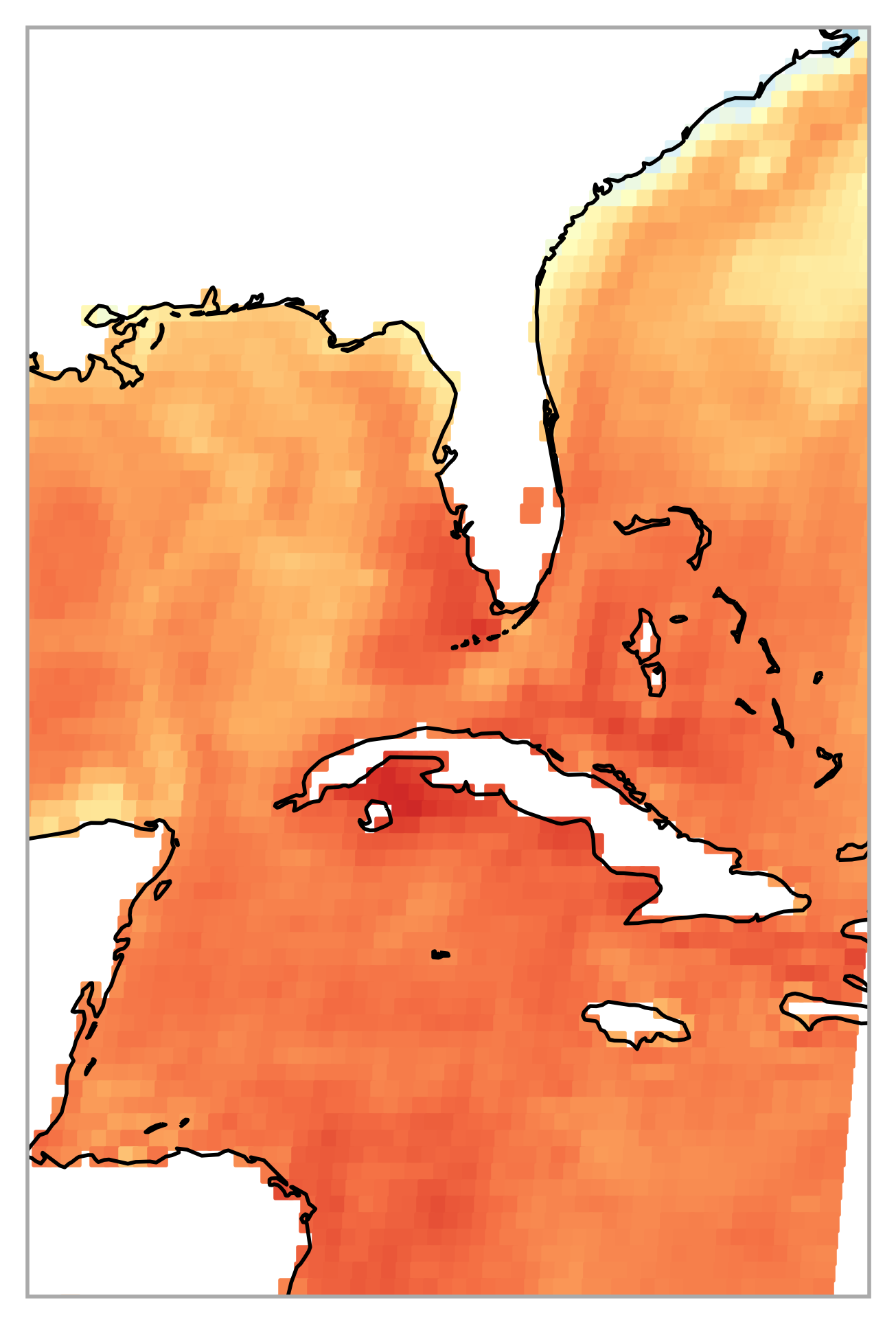}
        \includegraphics[height=.21\linewidth]{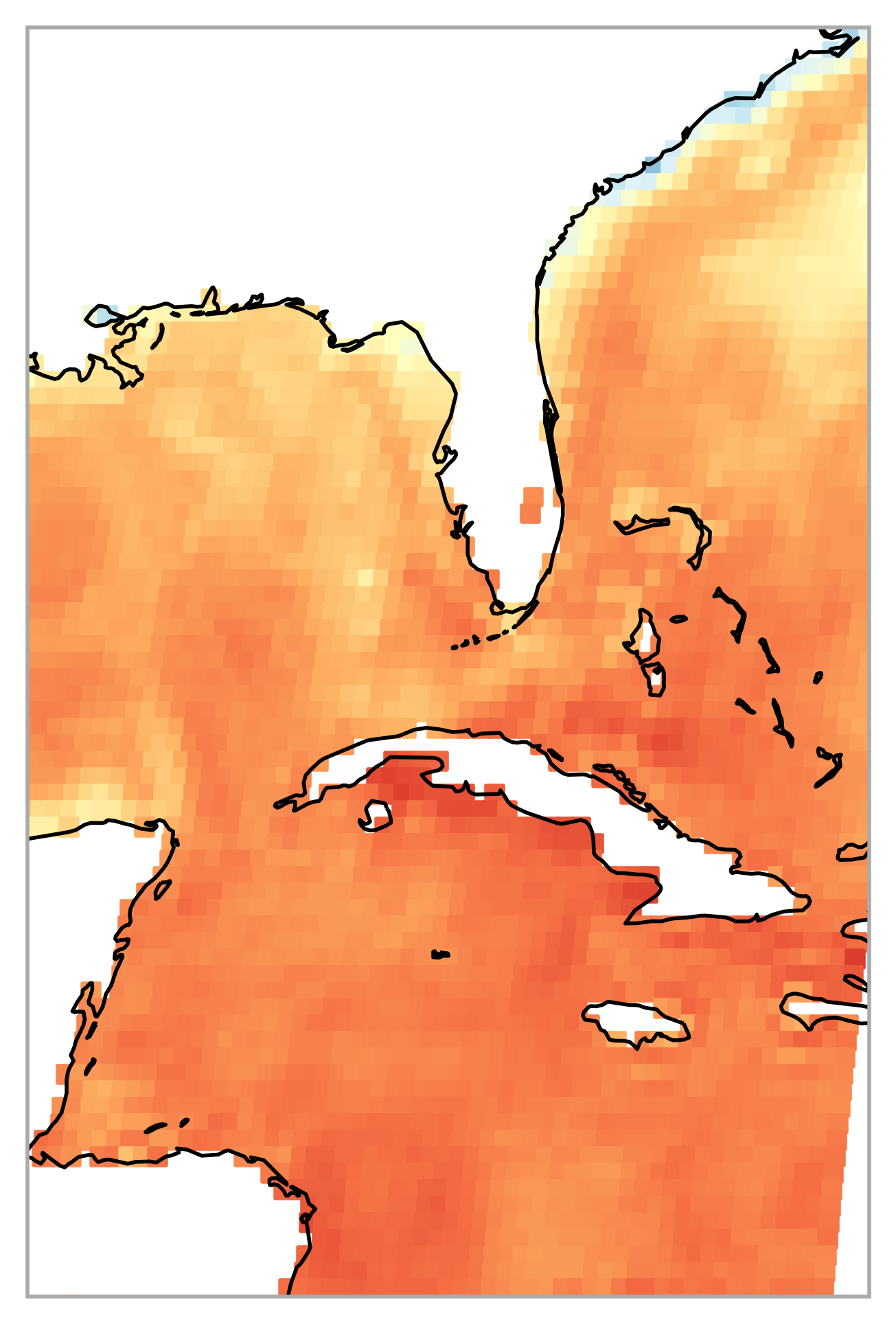}
        \includegraphics[height=.21\linewidth]{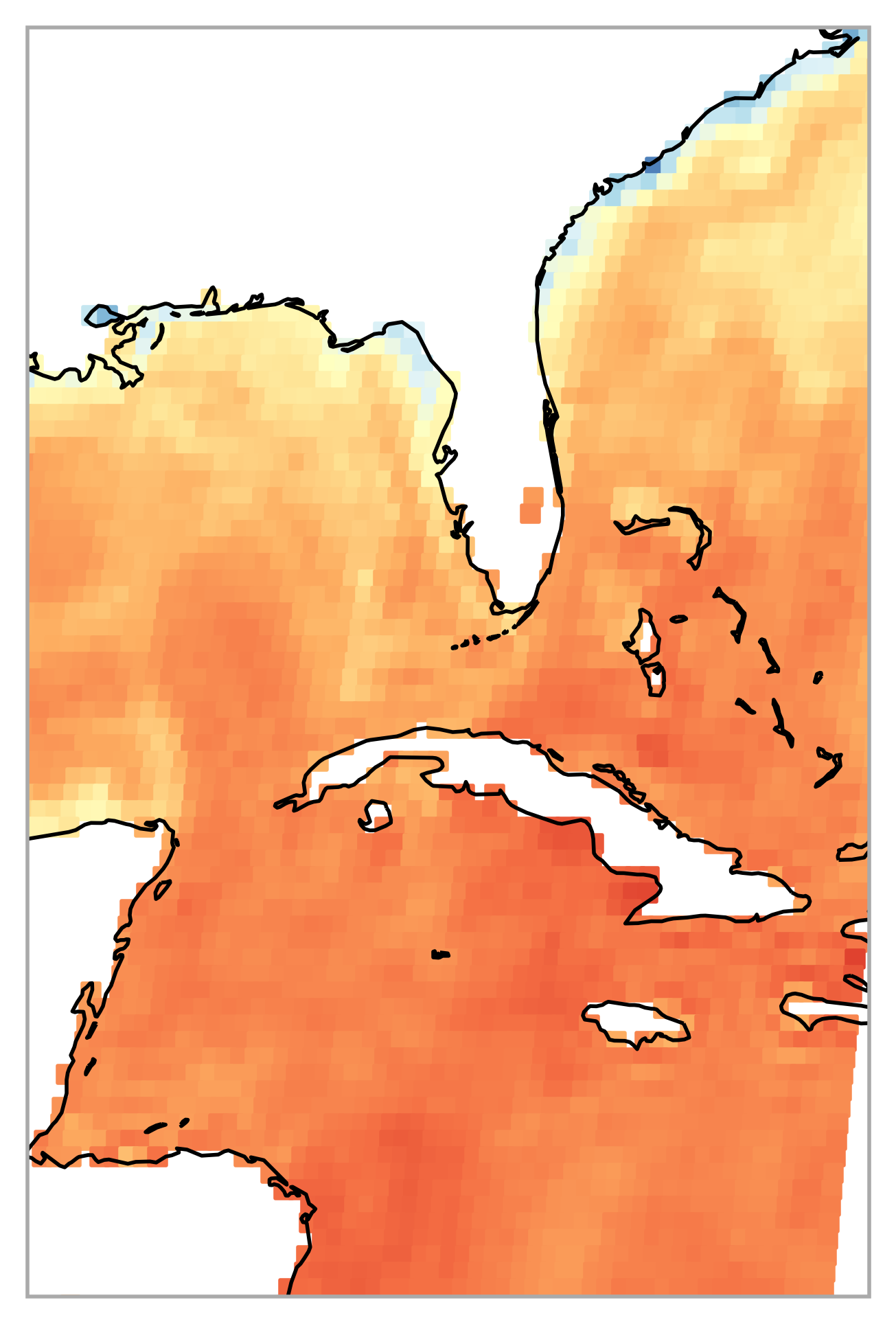}
        \includegraphics[trim=0 0 40 0, clip, height=.211\linewidth]{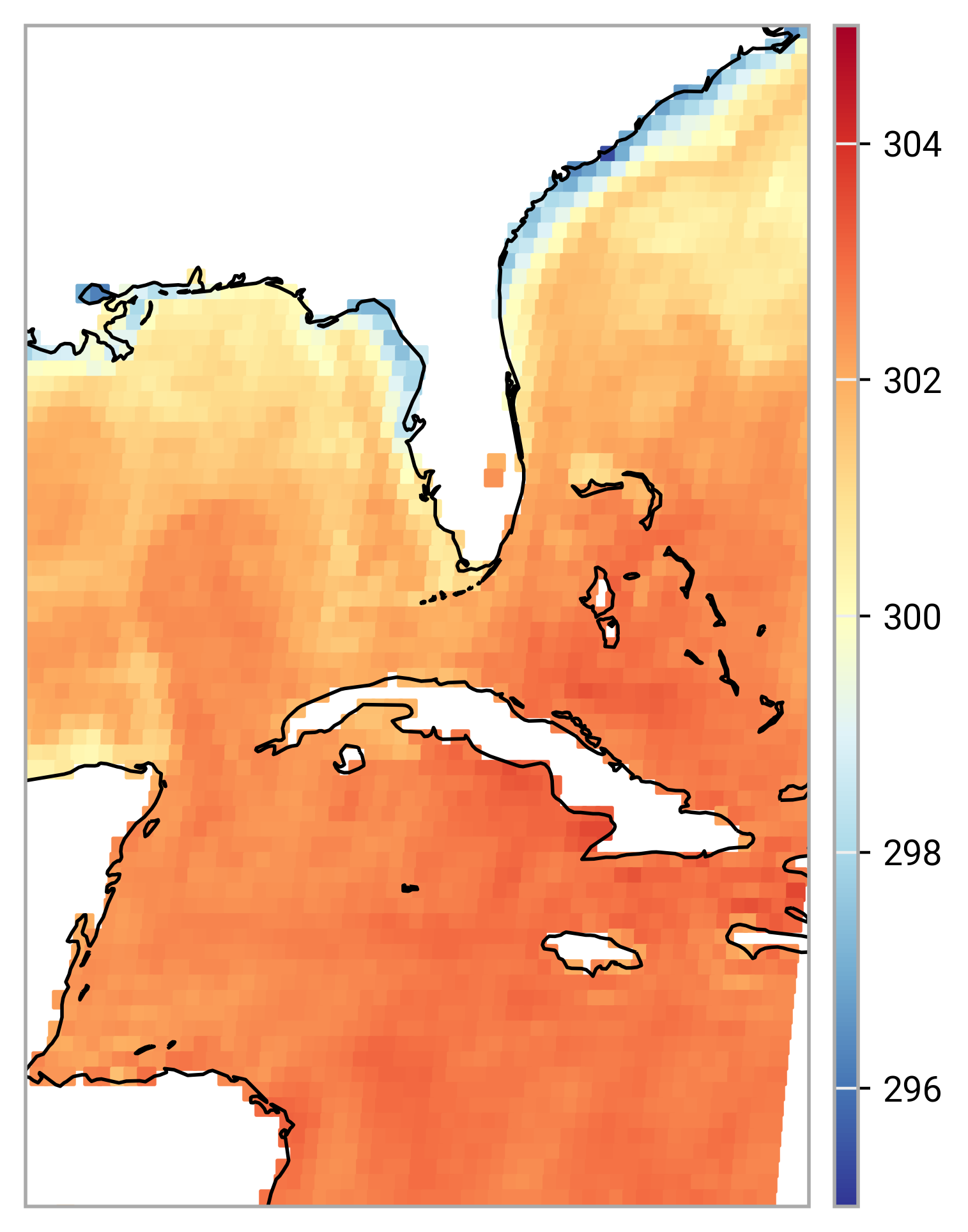}
        
        \vspace{0.3em}

        \includegraphics[trim=0 3 0 0, clip, height=.21\linewidth]{hurricane_ian/dop.jpg}
        \includegraphics[height=.21\linewidth]{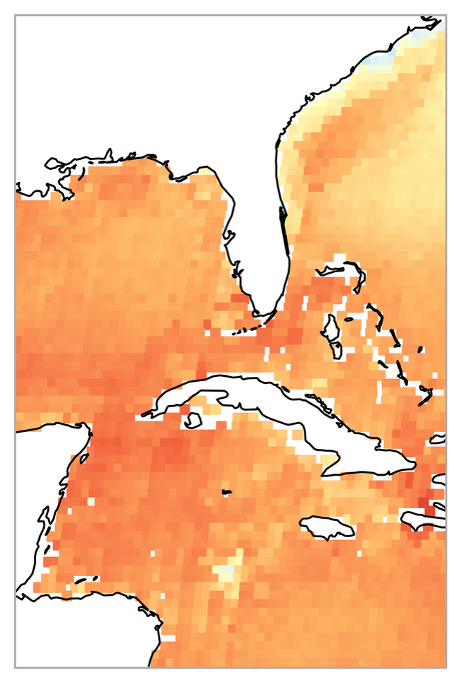}
        \includegraphics[height=.21\linewidth]{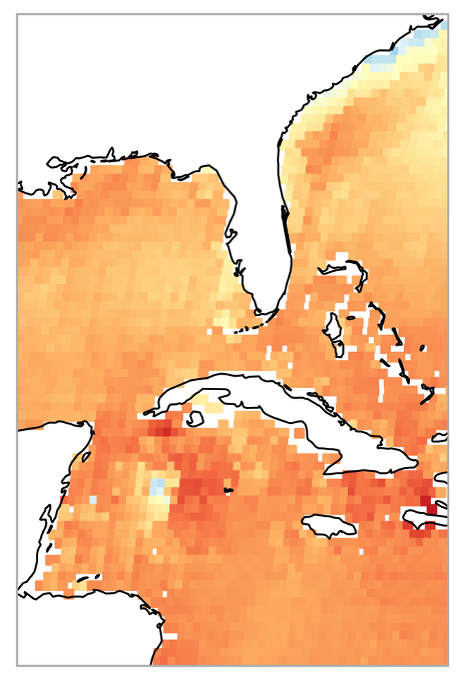}
        \includegraphics[height=.21\linewidth]{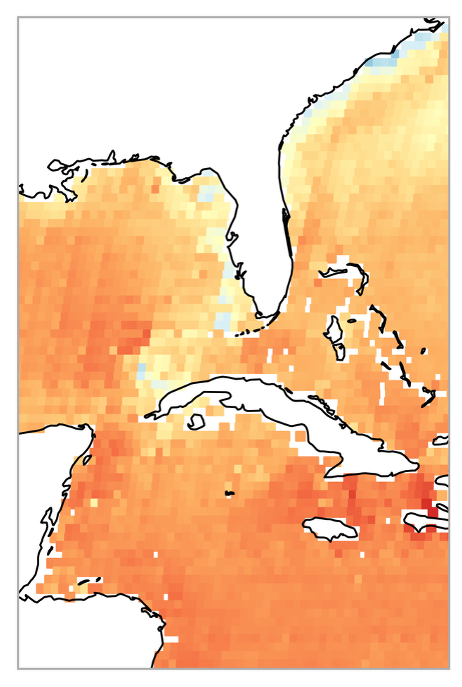}
        \includegraphics[height=.21\linewidth]{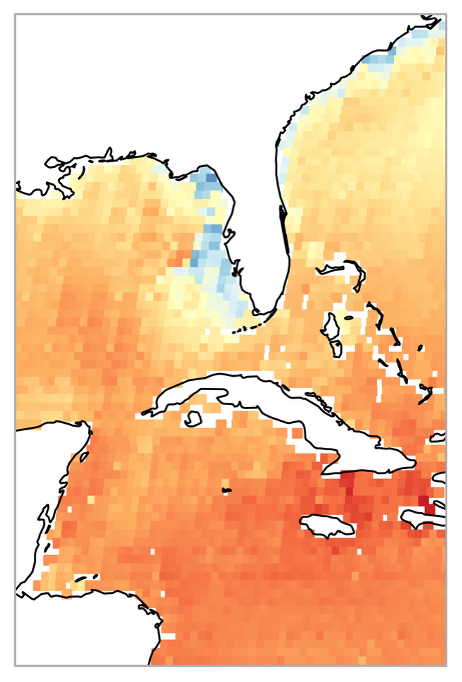}
        \includegraphics[height=.21\linewidth]{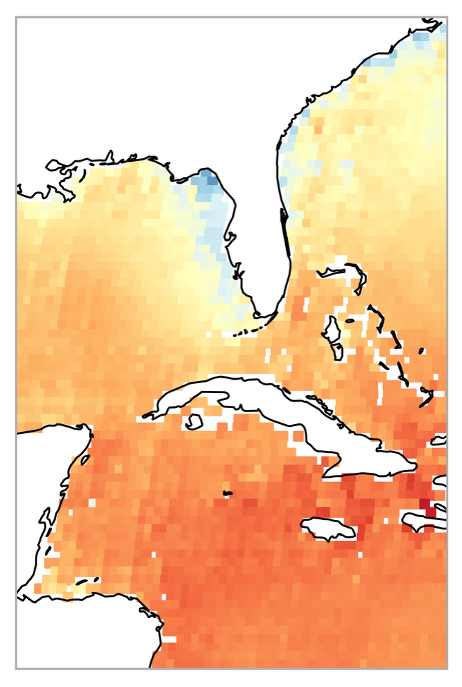}
        \includegraphics[trim=0 0 22 0, clip, height=.211\linewidth]{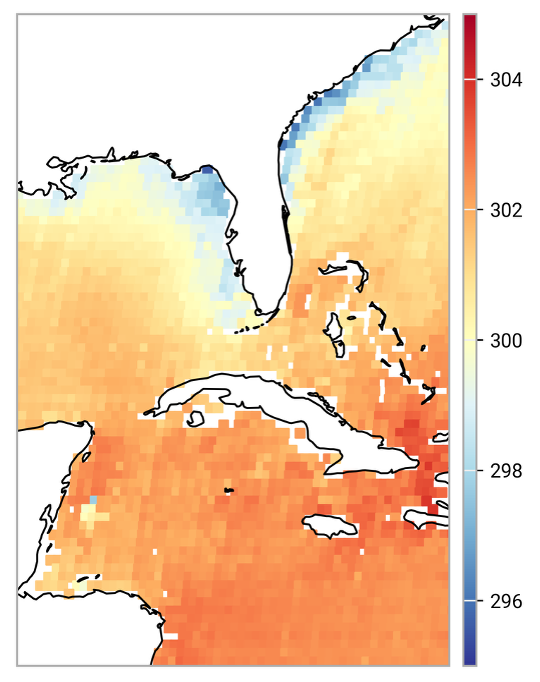}
        
        \settowidth{\figAwidth}{\includegraphics[height=.21\linewidth]{hurricane_ian/swh_era5_day1.png}}
        \settowidth{\figBwidth}{\includegraphics[trim=0 3 0 0, clip, height=.21\linewidth]{hurricane_ian/dop.jpg}}
        
        \makebox[\figBwidth]{\includegraphics[height=.21\linewidth]{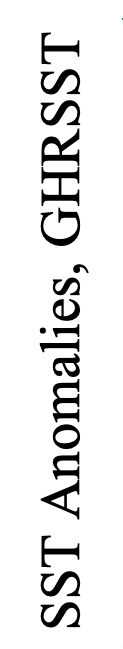}}
        \includegraphics[height=.21\linewidth]{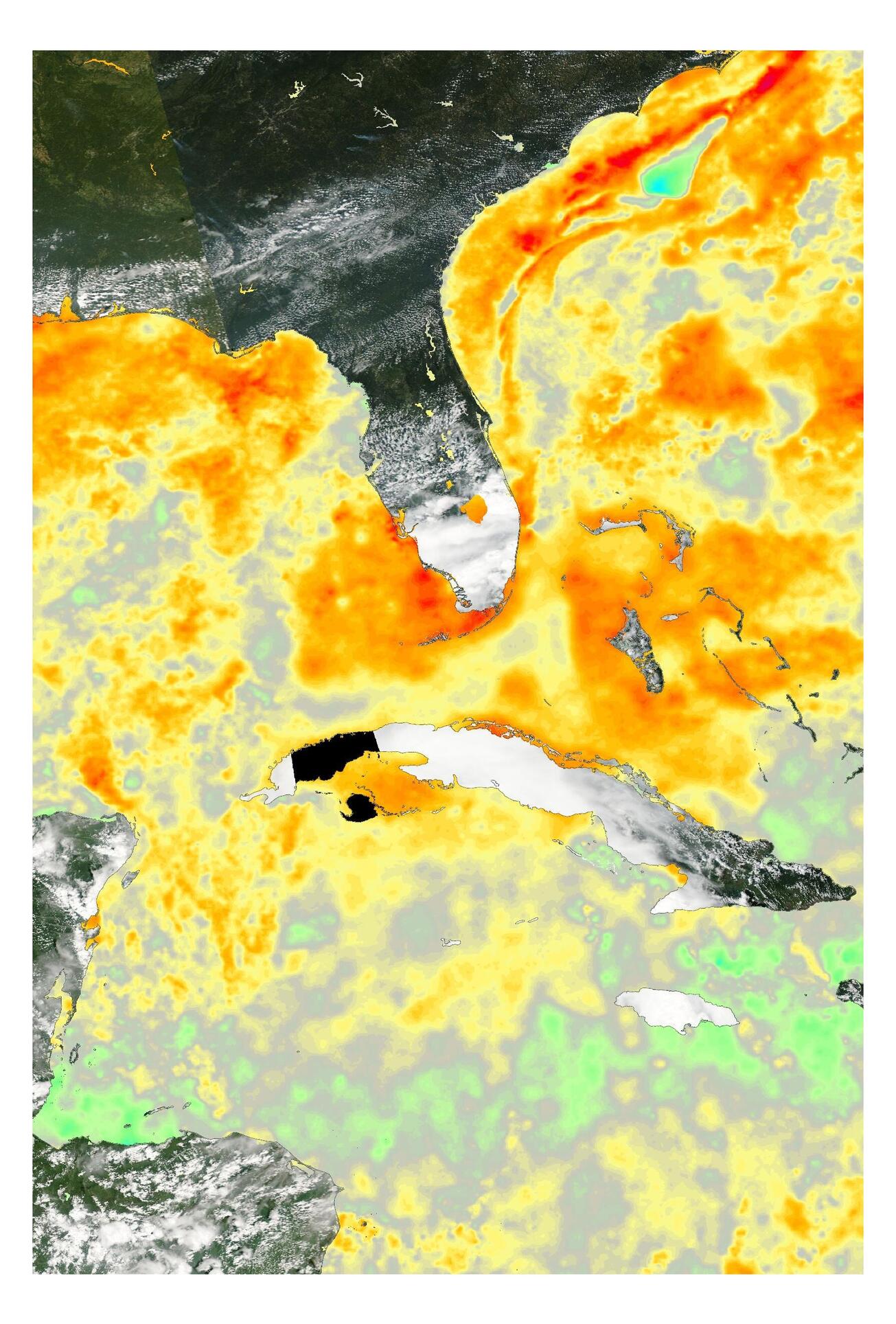}
        \includegraphics[height=.21\linewidth]{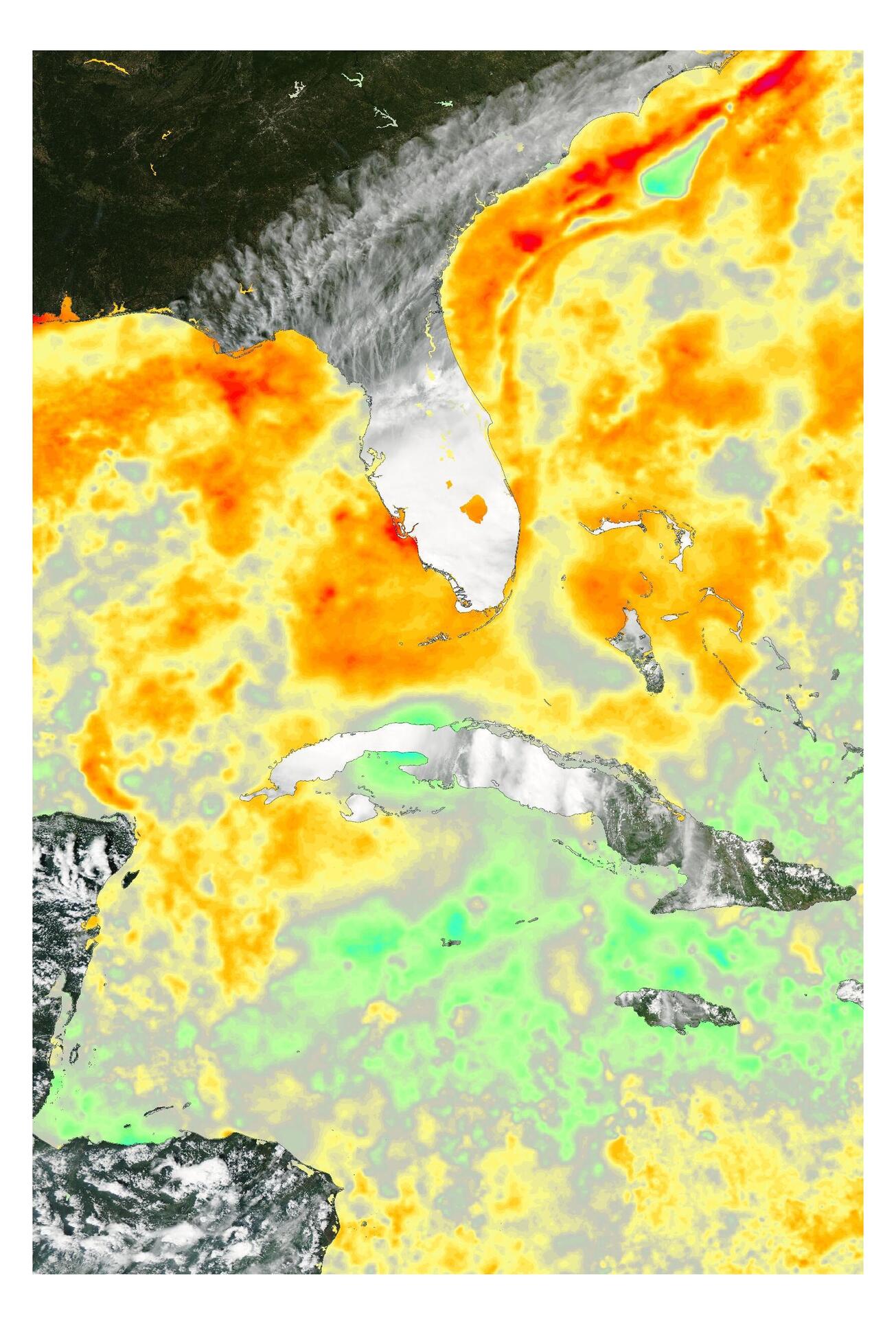}
        \includegraphics[height=.21\linewidth]{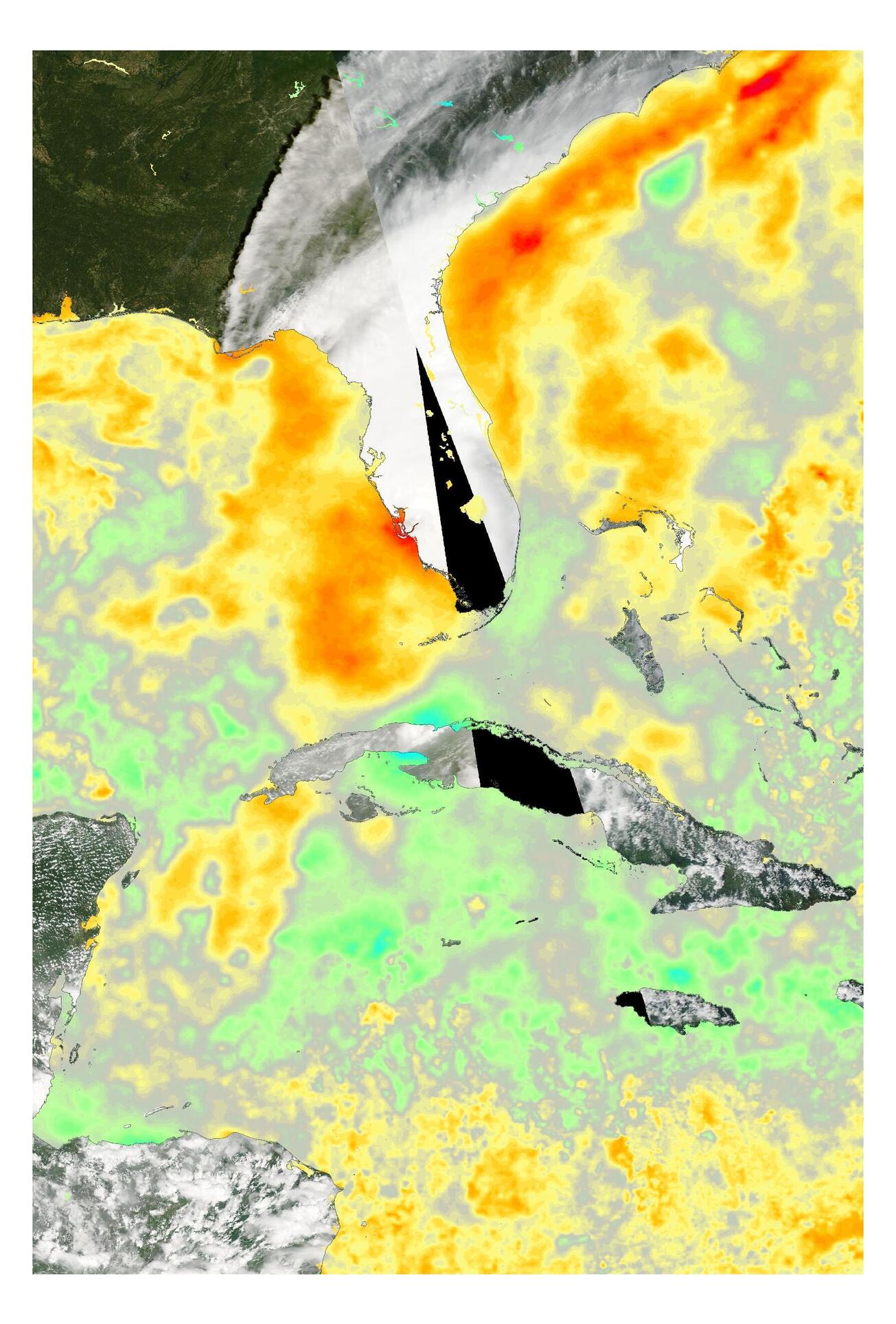}
        \includegraphics[height=.21\linewidth]{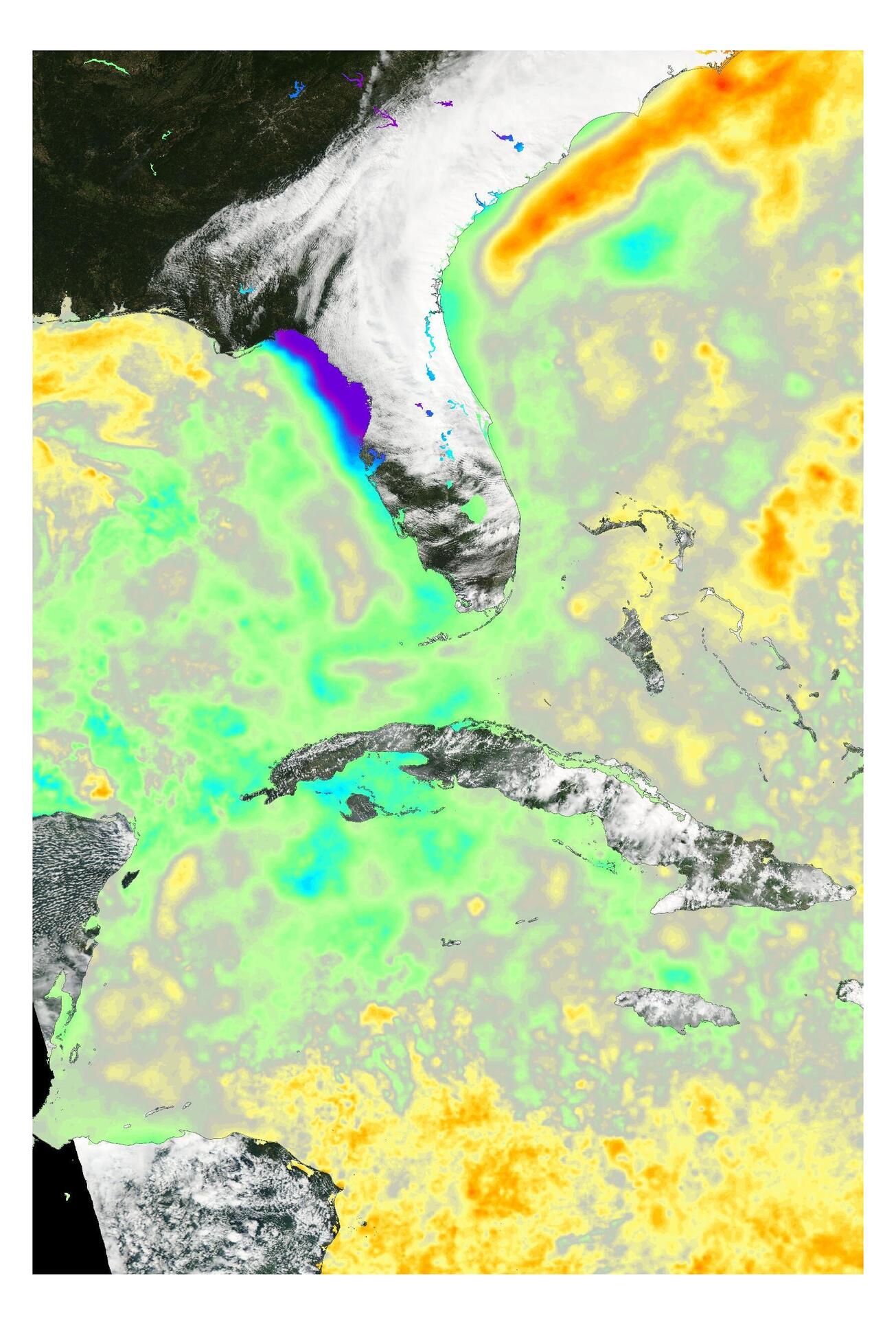}
        \includegraphics[height=.21\linewidth]{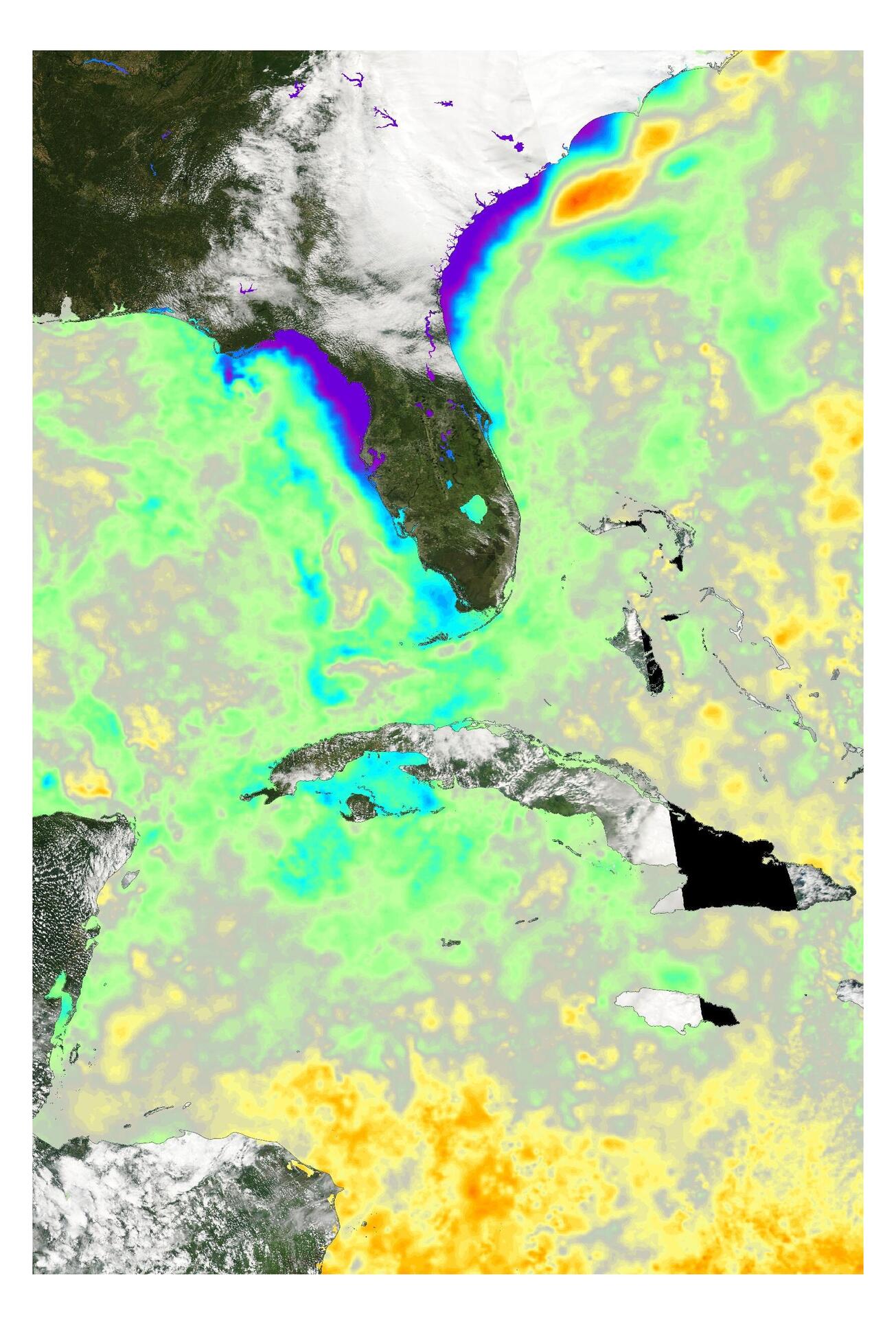}
        \includegraphics[trim=0 0 22 0, clip, height=.21\linewidth]{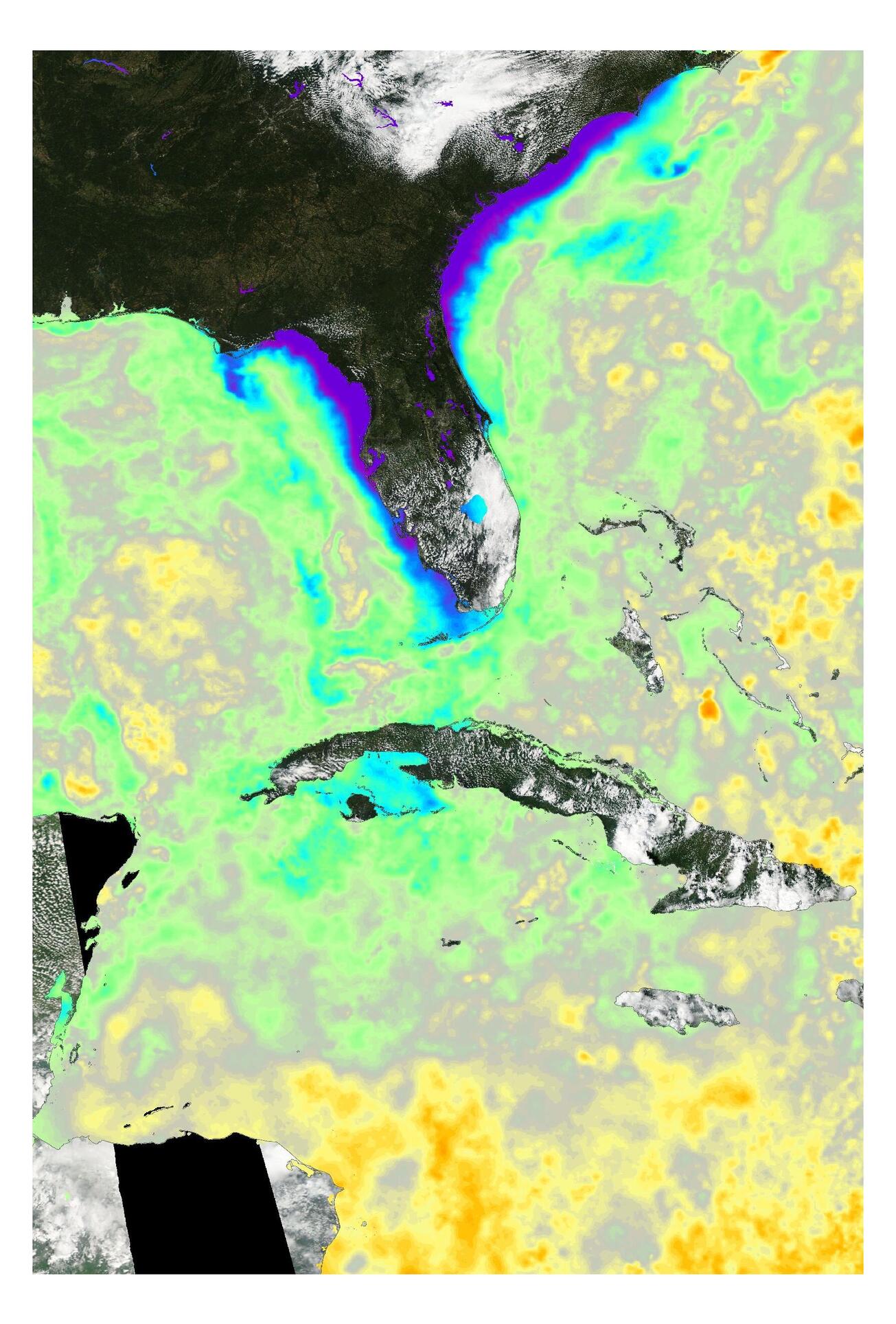}
    
        \end{minipage}
        &
        \begin{minipage}[c]{0.03\linewidth}
        \vspace*{\fill} %
        \includegraphics[trim=0 0 0 0, clip, width=1.2cm]{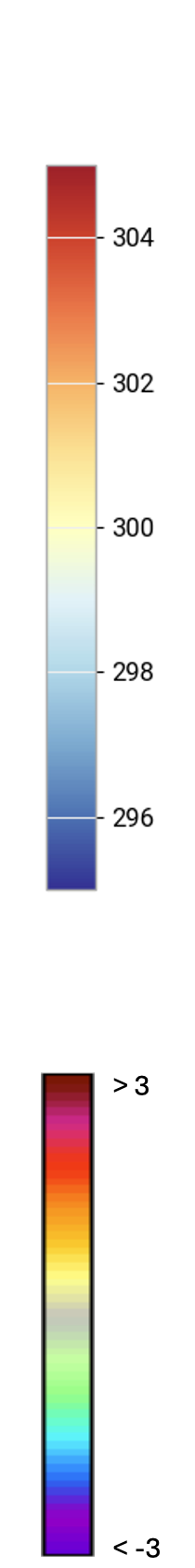}
        \end{minipage}
    \end{tabular}
}

\caption{ERA5 reanalysis (top row of each subfigure) and GraphDOP forecasts (bottom row) for (a) significant wave height and (b) sea surface temperature, valid at 00:00~UTC from 26 September to 1 October 2022 (left to right). The SST Anomalies Level 4 product derived from GHRSST (in °C) is also shown. Forecasts were initialized with observations between 09:00~UTC and 21:00~UTC on 24 September 2022.}
\label{fig:ian_ocean}
\end{figure}

Although not explicitly coupled to an ocean model, GraphDOP also produces forecasts of sea surface temperature (SST; coming from the buoy decoder) and wave conditions (by the radar altimeter decoder), as illustrated Figure \ref{fig:ian_ocean}. The significant wave height field (Figure \ref{fig:ian_ocean}(a)), driven primarily by near-surface winds, provides an indirect but valuable consistency check with the wind forecasts. Similarly, SST forecasts (Figure \ref{fig:ian_ocean}(b)), offer insight into how the model captures cooling patterns typically induced by storm-induced ocean mixing. Together, these variables test GraphDOP’s ability to recover the broader Earth System response to the hurricane, despite the lack of explicitly resolved feedbacks.

As shown in Figure \ref{fig:ian_ocean}(a), GraphDOP successfully captures the large-scale evolution of significant wave height associated with Hurricane Ian. The model reproduces the spatial extent and general location of peak wave activity, including the translation of the wave field as the storm progresses toward Florida and then into the Atlantic. Notably, peak values and wave structure in GraphDOP remain coherent across several forecast days, indicating that the near-surface wind field driving these waves is well represented. Although the predicted wave heights are slightly under or overestimated compared to ERA5, GraphDOP still reflects the expected alignment of the wave field with the hurricane’s trajectory. This is in line with the wind forecasts seen in Figure \ref{fig:ian}, and thus supports the view that GraphDOP captures the internal representation of the storm and produces consistent forecasts across parameters. An example of this can be seen on the fourth panel, where GraphDOP overestimates the wave height at landfall. This behaviour is consistent with the location of the strong wind forecasts seen in Figure~\ref{fig:ian}(b), where the model maintains intense winds for slightly longer than observed. The overestimation of waves in these areas therefore reflects a physically coherent response to the lingering wind field.

The sea surface temperature (SST) forecasts in Figure \ref{fig:ian_ocean}(b) show a particularly compelling result. Along Hurricane Ian’s track, GraphDOP captures a clear surface cooling signal, with temperatures dropping consistently with storm-induced mixing. 

Tropical cyclones commonly leave behind a cold wake (i.e., a band of anomalously cool SSTs along the storm's path). This wake forms as the cyclone’s strong winds and surface waves induce vigorous vertical mixing of the upper ocean and drive Ekman pumping, which brings colder subsurface water toward the surface. These processes, in conjunction with enhanced surface evaporation, can depress SSTs in the wake by several degrees. Although the ocean surface gradually rewarms due to air–sea heat fluxes, the cooling anomaly can persist for days to weeks, influencing subsequent storm development or atmospheric conditions \citep[e.g.][]{vincent2012ocean, pasquero2021cold}. 

Accurately representing this oceanic feedback requires coupled modelling. As the cyclone extracts heat from the ocean, the resulting SST reduction feeds back on the storm by limiting the energy available to sustain or intensify it. While simple slab ocean models can simulate surface heat loss, they are insufficient to capture vertical mixing or subsurface processes such as upwelling. Simulating all three—surface fluxes, mixing, and upwelling—requires a fully three-dimensional ocean model. This coupling is especially critical for slow-moving cyclones, where the duration of wind forcing allows deeper mixing and stronger cold wake formation. 

Traditional NWP systems have made substantial progress in representing atmosphere–ocean interactions, yet forecasting cold wakes remains a challenging task \citep{TC_Mogensen, TC_Polichtchouk}. These challenges arise from the limited resolution needed to resolve mesoscale ocean processes, simplified or absent ocean coupling in some configurations, and uncertainties in the initial ocean state. Accurate cold wake prediction depends on well-initialized SST, mixed-layer depth, and subsurface temperature structure—variables that are often difficult to constrain in tropical cyclone–prone regions due to sparse \textit{in-situ} observations (e.g., ARGO floats, drifting buoys). Furthermore, many widely used SST analyses, while robust for large-scale monitoring, can smooth or lag storm-induced anomalies when based on daily mean products.

ERA5 relies on daily-averaged OSTIA SSTs \citep{donlon2012operational}, which often delay the capture of rapid SST changes due to storm passage. In the upcoming IFS Cycle 50r1 update, these constraints are expected to lessen with the planned adoption of NEMO4-SI3 as the ocean component of all operational configurations \cite{NEMO4}. This new coupling framework is designed to enhance the representation of upper-ocean variability and improve the timeliness and realism of SST responses to extreme events such as tropical cyclones. Against this backdrop, GraphDOP produces a sharper and more localized cooling signal than ERA5, especially from panel three onwards. This behavior is consistent with previous case studies of Hurricane Ian’s wake, such as that of \cite{Shi03082023}, who reported a pronounced drop in SST across the Gulf of Mexico and along the U.S. southeast Atlantic shelf beginning on 28 September (third panel of Figure \ref{fig:ian_ocean}(b)) and intensifying over 29–30 September (fourth and fifth panels) as Ian crossed Florida. A similar pattern is seen in the GHRSST Level 4 SST anomalies\footnote{Accessed from https://worldview.earthdata.nasa.gov} \citep{GHRSST}, shown in the third row of Figure \ref{fig:ian_ocean}(b). Taken together, both GHRSST and the findings of \cite{Shi03082023} support the GraphDOP forecasts, reinforcing its ability to forecast the timing and intensity of hurricane-induced SST cooling.

Collectively, this underscores the capacity of GraphDOP not only to reconstruct the atmospheric structure of a major hurricane but also to reproduce key features of the ocean surface response. That GraphDOP, trained solely on observational data and without explicit coupling to an ocean model, can capture cold wake dynamics suggests that it has implicitly learned robust relationships between atmospheric forcing and ocean surface evolution. This capability indicates the potential of data-driven models like GraphDOP to internalize and reconstruct complex cross-domain processes that emerge during extreme events. This is something that has also been seen when training joint ML models from NWP reanalysis, as shown in \cite{ecmwf2025aifswaves}.

\section{European heatwave case study: land surfaces}
\label{sec:heatwave}

Two-meter temperature (T2m) is a key meteorological variable that responds quickly to surface–atmosphere interactions. It is also widely used over land as an indicator of thermal stress for ecosystems and animals. Unlike upper-air temperature, which reflects broader synoptic-scale conditions, T2m is strongly modulated by surface processes, including, over land surfaces, soil moisture availability, evapotranspiration, and surface radiative fluxes. During heatwave events, these processes act to amplify or damp anomalies, making T2m a sensitive tracer of the coupling between the surface and the atmospheric boundary layer. 

In the IFS, T2m is not a prognostic model variable but is instead derived diagnostically (using vertical interpolation) from surface temperature and the lowermost model levels temperatures using Monin-Obukhov similarity theory. This means that its fidelity depends critically on the representation of surface fluxes and boundary-layer exchange, which in turn are strongly influenced by land-surface model formulations and soil moisture initialization \citep{coupled_da_3, Rosnay22}. Errors in soil moisture, vegetation parameters, or deficiencies in surface-atmosphere coupling processes and surface energy partitioning can therefore propagate directly into biases in T2m forecasts, particularly under persistent high-pressure regimes typical of European summer heatwaves.

Prior to IFS cycle 49r1 (that became operational at ECMWF in 2024), the assimilation of SYNOP T2m observations was only done in the separate land-surface analysis \cite{Rosnay22}, meaning that information from near-surface observations was feeding back via soil moisture and soil temperature interaction processes with the atmosphere. In recent years, important progress has been made to fully exploit the information contained in T2m observations within physics-based systems, with the activation of T2m assimilation in the atmospheric 4D-Var from cycle 49r1 \cite{land_coupling_4941}, and with strongly coupled land–atmosphere data assimilation expected to be implemented in the near future. Nevertheless, even with these advances, the degree of constraint remains sensitive to parametrization choices and error covariance specification. 

The GraphDOP approach does not require separate treatment of land and atmosphere. Instead, T2m observations are embedded into a shared latent representation together with other surface and atmospheric variables. This means that correlations between T2m, soil moisture-sensitive observations, radiation, and near-surface winds are learned directly, without the need for explicit parametrizations of land-surface exchange. For example, hot and dry anomalies associated with depleted soil moisture are represented in the same latent space as the corresponding atmospheric circulation patterns, enabling the model to internalize land–atmosphere coupling pathways. In practice, this allows GraphDOP to exploit the information content of T2m observations more fully, since they contribute both to constraining the land-surface state and to shaping the broader atmospheric evolution.

The summer of 2022 marked one of the hottest seasons in recent European history, with a series of prolonged periods of anomalously high near-surface temperatures. This had widespread impacts on ecosystems, including severe drought and increased fire activity, but also on public health \cite{heatwave1, heatwave2, heatwave3}. It thus provides a stringent test case for understanding extreme climate processes. In particular, it challenges the ability of forecasting systems to capture the propagation of temperature and wind anomalies across the continent, as well as the role of land–atmosphere feedbacks in sustaining extreme heat. It is worth noting that the IFS, benefitting from the recent coupling developments described above, demonstrated notably good predictability of this heatwave, capturing both its onset and large-scale structure with high fidelity \citep{heatwave_newsletter}.

In what follows, we focus on the episode of extreme heat that struck Western Europe between 17 and 20 July 2022, during which both the United Kingdom and France registered unprecedented temperature records. A GraphDOP forecast was initialized from 12 hours of observations spanning 16 July 2022, 09:00-
21:00 UTC. Forecasts are evaluated at 12:00 UTC daily between 17 and 20 July 2022. 

\begin{figure}
\centering
\begin{tabular}{c c c}
\raisebox{.8cm}{\rotatebox{90}{17/07/2022 12:00}} & 
\includegraphics[width=15cm]{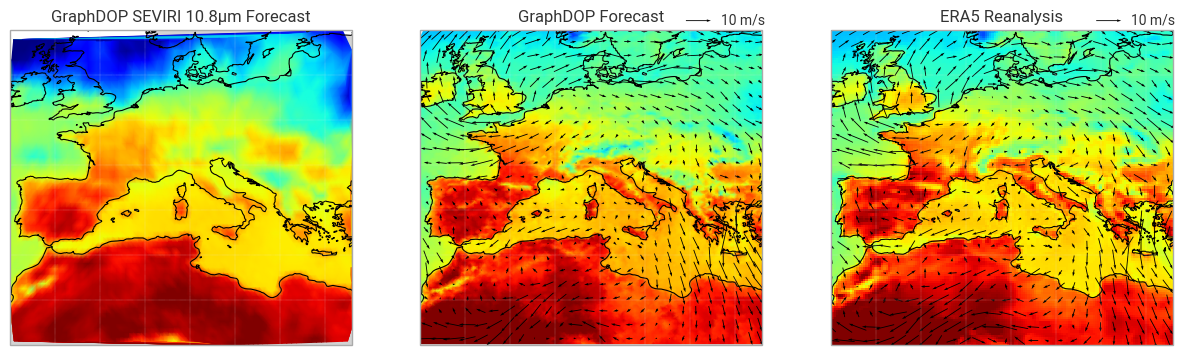} \\

\raisebox{.8cm}{\rotatebox{90}{18/07/2022 12:00}} &
\includegraphics[width=15cm]{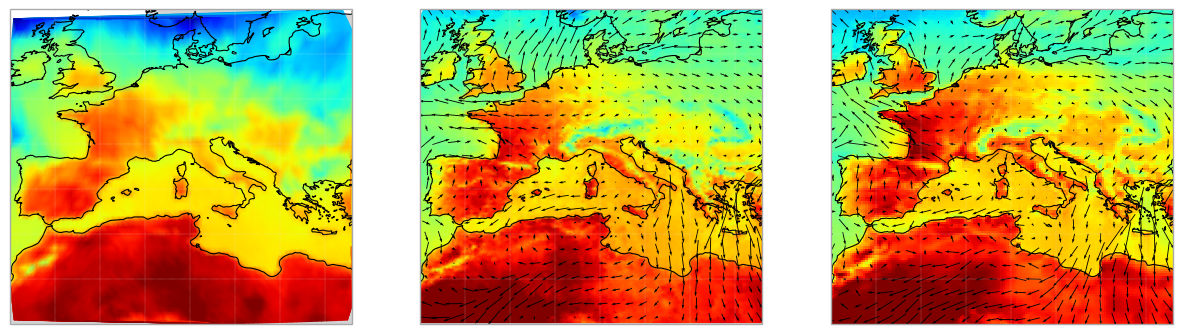} \\

\raisebox{.8cm}{\rotatebox{90}{19/07/2022 12:00}} &
\includegraphics[width=15cm]{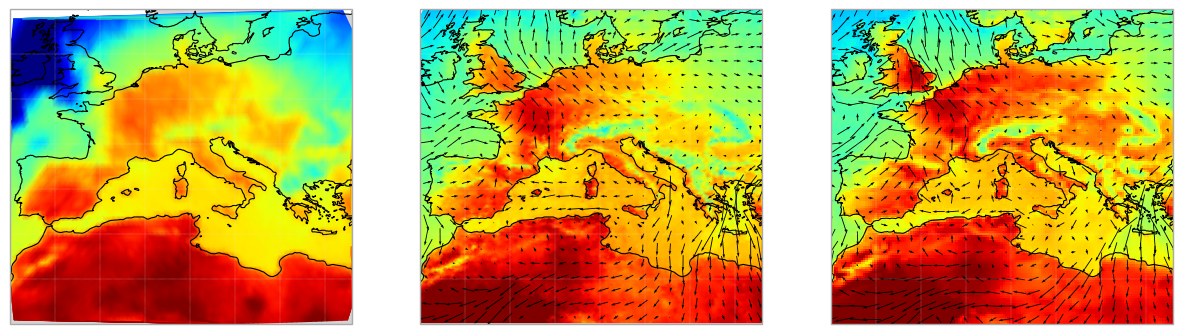} \\

\raisebox{.8cm}{\rotatebox{90}{20/07/2022 12:00}} &
\includegraphics[width=15cm]{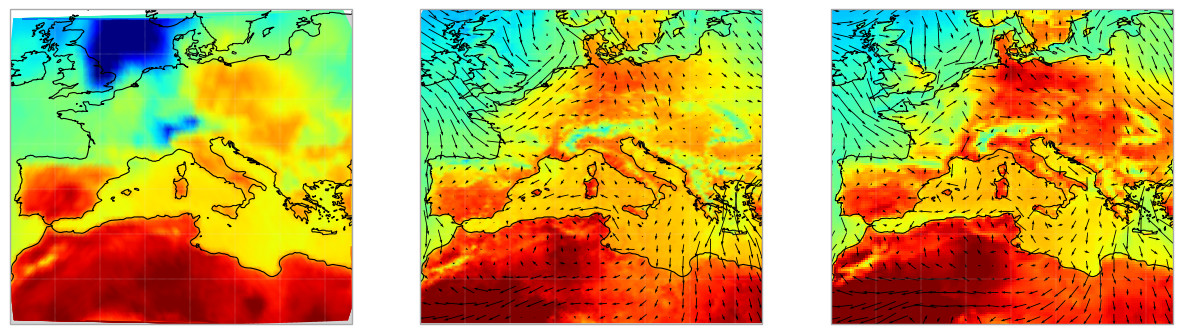} \\
 & 
\includegraphics[width=15.9cm]{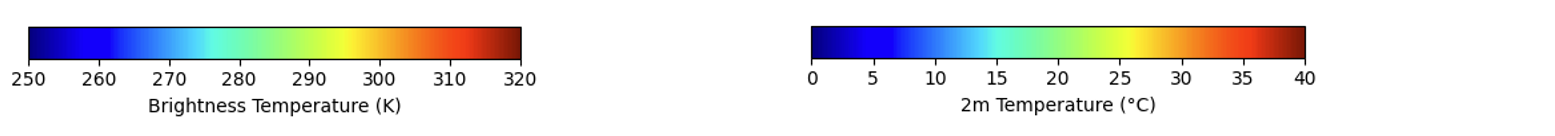} \\ 
\end{tabular}

\caption{SEVIRI brightness temperature (10.8 µm, window channel) GraphDOP forecast (left), near-surface atmospheric conditions (10\,m wind (m/s) and 2\,m temperature ($^{\circ}\mathrm{C}$))) from the GraphDOP forecast (centre), and the ERA5 reanalysis (right) at 1–4 day lead times between 17 and 20 July 2022, 12 UTC.}
\label{fig:heatwave_near_surface}
\end{figure}

Figure \ref{fig:heatwave_near_surface} compares the GraphDOP forecast of near-surface fields (i.e. 2~m temperature and 10~m winds) with ERA5. In the first days of the forecast (17-19 July), the large-scale structure of the heatwave is realistically captured. GraphDOP reproduces the pronounced northward extension of anomalously warm conditions from the Iberian Peninsula into France and central Europe, in close agreement with ERA5. The correspondence between the 2~m temperature field and the associated low-level south westerly winds is particularly encouraging, indicating that the model’s representation of the advective transport is dynamically consistent. 

Whilst the advancement and spatial patterns of the heatwave are very well captured, the GraphDOP forecast underestimates the intensity of the heatwave (i.e. T2m is too low). This is particularly evident over the United Kingdom, where ERA5 shows a sharp amplification of the heat signal across southern and eastern England. Although this is a recurrent problem with data-driven forecasting models that tend to underestimate extremes \citep{boucherextremes,lang2024aifs}, the corresponding GraphDOP forecast of SMOS observations (not shown) highlights that the model overestimated the soil moisture in those same areas. This could partly explain the damped temperature response, as overly moist soils tend to reduce sensible heat flux and limit the build-up of near-surface temperature anomalies. Therefore, the forecast is physically consistent, in the sense that overly moist soils tend to reduce sensible heat flux and limit the build-up of near-surface temperature anomalies. The poorer representation of soil conditions in GraphDOP may reflect that comparatively few observations are currently included in GraphDOP that allow a representation of soil or vegetation conditions. Incorporating more surface and vegetation-sensitive observations, such as ASCAT backscatter over land could help better constrain the land–atmosphere coupling and thus improve the representation of heat extremes. Similarly, extending the SMOS dataset used for training, or complementing it with additional missions such as SMAP, would provide a more comprehensive characterization of soil moisture variability. These additions could supply the model with richer information on land surface conditions, ultimately supporting a more realistic simulation of temperature amplification during heatwave events.

The leftmost column of Figure \ref{fig:heatwave_near_surface} shows, for the same times, the GraphDOP forecasts of SEVIRI 10.8 $\mu$m brightness temperature. This channel is sensitive to land-surface temperature in clear-sky conditions. The close alignment (especially between 17 and 19 July) between the brightness temperatures and 2~m temperature forecasts demonstrates that the model’s latent representation effectively captures land–atmosphere interactions. Large discrepancies, for instance on 20 July, arise from clouds obstructing the surface, appearing as colder brightness temperatures. 

Overall, GraphDOP demonstrated its ability to capture the onset and evolution of the heatwave, particularly over France and western continental Europe, but struggles to reproduce the extreme intensities further north once errors in the circulation accumulate. Nevertheless, the co-evolution of the 2~m temperature forecast with the near-surface winds highlights the model’s ability to maintain physically consistent surface fields, even when regional details diverge from the reanalysis.

\section{Discussion and conclusion}
This paper has evaluated GraphDOP, a fully observation-driven Earth System forecasting model originally presented in \cite{alexe2024graphdopskilfuldatadrivenmediumrange}, through targeted case studies designed to probe its ability to represent cross-domain coupling in the Earth System. Unlike most existing ML forecasting systems that are trained on reanalysis or model output, GraphDOP learns directly from satellite and \textit{in-situ} observations, allowing relationships between Earth System components to emerge naturally from the data. Results presented in this paper and summarised below align well with recent work presented in \cite{lean2025learningnatureinsightsgraphdops}, which argues that representations of the physical state and processes emerge in purely observation-driven models.

The rapid Arctic sea ice freeze-up case study (Section \ref{sec:seaice}) illustrates how GraphDOP can reproduce key signatures of sea ice formation in microwave brightness temperatures, despite the absence of explicit sea ice labels in the training data. The model not only captured the large-scale expansion of ice cover but also represented its impact on other components of the system, such as wave attenuation (not shown). These results suggest that GraphDOP has learned an internal latent representation of sea ice that extends beyond simple correlations. 

The Hurricane Ian case study (Section \ref{sec:hurricane}) further stresses the coupled capabilities of GraphDOP. The model tracked the storm’s trajectory and reproduced coherent pressure, wind, wave, and cloud structures. Particularly notable was its ability to simulate storm-induced sea surface cooling (cold wake) despite no explicit ocean model coupling. This indicates that GraphDOP can internalize feedbacks between the atmosphere and ocean directly from observational constraints.

Finally, the European heatwave case study (Section \ref{sec:heatwave}) highlights GraphDOP’s ability to reproduce the evolution of persistent near-surface anomalies. The model realistically captured the northward advection of hot air masses from the Iberian Peninsula into France and the British Isles, as well as the associated low-level wind patterns, demonstrating dynamically consistent transport processes. Importantly, the brightness temperature forecasts from the SEVIRI 10.8~$\mu$m channel exhibited spatial and temporal structures that are also consistent with the heatwave event.

Across all case studies, GraphDOP demonstrated an ability to propagate cross-domain dependencies even in data-sparse regions, such as the marginal ice zone or the open ocean. Its graph-based architecture and shared latent representation appear well-suited to leveraging heterogeneous observations, enabling the model to generalize coupling patterns beyond the regions with dense coverage. This highlights a central strength of GraphDOP: the flexibility to incorporate new or unconventional observations without the need for predefined observation operators or hand-tuned error models. GraphDOP already makes use of some interface observations that are currently too difficult to exploit successfully in physics-based systems.

Nonetheless, important challenges remain. First, forecast fidelity decreases in situations where observational constraints are limited. For instance, structured errors in regions such as the Kara Sea and Baffin Bay highlight inaccuracies in atmospheric circulation forecasts that propagate into biases in sea ice predictions. This highlights the dependence of observation-driven systems on the quality of initial conditions (observations). Also, the intensity of Hurricane Ian was systematically underestimated, which we attribute to the scarcity of \textit{in-situ} pressure observations within the storm core. This points to an important aspect of observation-only models: unlike traditional DA, they do not make use of a forecast model-based prior and necessarily rely on (1) the learnt dynamics and processes of the latent space, and (2) the constraints stemming from the initializing observations. A further possible contributing factor is the smoothing tendency of AI models trained against a MSE objective, which can damp extremes \citep[see e.g.,][]{lam2022graphcast,lang2024aifs, blogaifs3}.

Without sufficient coverage, especially in extreme weather regimes, skill will remain limited. While GraphDOP demonstrates strong skill in capturing upper-ocean processes like cold wake formation, extending this capability to the deeper ocean presents a larger challenge. Observations at depth are far sparser than at the surface, with ARGO profiling floats providing limited temporal and spatial coverage below 2000~m and satellite remote sensing offering no direct access. As a result, the deep ocean lacks the dense and diverse observational constraints that have enabled data-driven models to succeed at the surface.
 
Looking forward, progress can be made along two main complementary lines. On the observational side, expanding the range of inputs particularly through more dense and diverse ocean, land, and cryosphere observations would provide stronger constraints on coupled dynamics. Incorporating additional storm-focused data, such as central pressure estimates from best-track archives, could mitigate intensity underestimation in tropical cyclones. On the methodological side, advances such as recurrent input cycling, latent priors, and improved treatment of observational biases could extend model memory (currently limited by the 12-hour input window) helping the model to represent slowly evolving processes that drive medium-to-longer-range predictability). Based on previous experience, we expect probabilistic training to encourage realistic variability in the forecasts \citep{lang2024aifscrpsensembleforecastingusing}.

GraphDOP exemplifies a shift in paradigm for Earth System prediction. Traditional NWP systems rely on explicit physical models and carefully engineered coupling schemes, while most ML surrogates replicate the dynamics of such models by training on reanalysis. GraphDOP instead learns directly from the observational record, bypassing both. Its ability to recover physically consistent coupled behaviour without any coupling infrastructure suggests that future forecasting systems may increasingly emerge from the observational data streams themselves. While challenges remain, especially in handling extremes and extending skill to poorly observed domains, the results presented here provide a compelling demonstration that end-to-end observation-driven forecasting is both viable and scientifically valuable.

\section*{Acknowledgements}
The development of AI-DOP at ECMWF has benefited from the support and input of many colleagues. In particular, we wish to thank Christian Lessig, Matthew Chantry, Peter Dueben, Chris Burrows, Josh Kousal, Rachel Furner, Sara Hahner, Sean Healy, Mike Rennie, Florian Pinault, Mario Santa Cruz, Cathal O’Brien, Jan Polster, Aaron Hopkinson, Marcin Chrust, Tobias Necker and Alan Geer for their valuable contributions. We acknowledge the Partnership for Advanced Computing in Europe (PRACE) for awarding us access to Leonardo, CINECA, Italy, and the EuroHPC Joint Undertaking (JU) for awarding us project access to the EuroHPC supercomputer MareNostrum5, hosted by the Barcelona Supercomputing Centre, through a EuroHPC JU Special Access call.

\bibliographystyle{unsrtnat}
\bibliography{dop}

\end{document}